\documentclass{article}
\usepackage[active]{srcltx}
\usepackage{amsmath}
\usepackage{bbm}
\usepackage{amsthm}
\usepackage{amsfonts}
\usepackage{leftidx}
\usepackage{mathrsfs}
\usepackage{amssymb}
\usepackage{amscd}
\usepackage{cases}
\usepackage{a4wide}
\usepackage{bm}
\usepackage{latexsym}
\usepackage[usenames,dvipsnames]{xcolor}
\usepackage{dsfont}
\usepackage{indentfirst}
\usepackage[latin1]{inputenc}

\def\qed{\hfill $\square$}

\newtheorem{definition}{Definition}[section]
\newtheorem{proposition}[definition]{Proposition}
\newtheorem{theorem}[definition]{Theorem}
\newtheorem{lema}[definition]{Lemma}
\newtheorem{corollary}[definition]{Corollary}
\newtheorem{remark}[definition]{Remark}

\theoremstyle{definition}

\renewcommand{\thedefinition}{\arabic{section}.\arabic{definition}}
\numberwithin{equation}{section}

\DeclareMathAlphabet{\mathpzc}{OT1}{pzc}{m}{it}

\begin{document}
\begin{center}
{\Large{ \textbf{Rigorous investigation of the reduced density matrix for the ideal Bose gas in harmonic traps by a loop-gas-like approach.}}}

\medskip

\today
\end{center}

\begin{center}
\small{Mathieu Beau\footnote{Dublin Institute for Advanced Studies
School of Theoretical Physics, 10 Burlington Road, Dublin 4, Ireland; e-mail:
    mbeau@stp.dias.ie}, Baptiste Savoie\footnote{Department of Mathematical Sciences, University of Aarhus, Ny Munkegade, Building 1530, DK-8000 Aarhus C, Denmark; e-mail: baptiste.savoie@gmail.com .}}

\end{center}
\vspace{0.5cm}

\begin{abstract}
In this paper, we rigorously investigate the reduced density matrix (RDM) associated to the ideal Bose gas in harmonic traps. We present a method based on a sum-decomposition of the RDM allowing to treat not only the isotropic trap, but also general anisotropic traps. When focusing on the isotropic trap, the method is analogous to the \textit{loop-gas approach} developed by W.J. Mullin in \cite{Mullin1}. Turning to the case of anisotropic traps, we examine the RDM for some anisotropic trap models corresponding to some quasi-1D and quasi-2D regimes. For such models, we bring out an additional contribution in the local density of particles which arises from the mesoscopic loops. The close connection with the occurrence of \textit{generalized-BEC} is discussed. Our loop-gas-like approach provides relevant information which can help guide numerical investigations on highly anisotropic systems based on the Path Integral Monte Carlo (PIMC) method.
\end{abstract}
\vspace{0.5cm}

\noindent
\textbf{PACS-2010 number}: 05.30.-d, 05.30.Jp, 67.85.Bc, 67.85.Hj, 67.85.Jk, 03.75.Hh

\medskip

\noindent
\textbf{MSC-2010 number}:  82B10, 82B21, 82D05

\medskip

\medskip

\tableofcontents
\medskip


\section{Introduction.}
\label{intro}

\subsection{Bose-Einstein condensation (BEC) in dilute cold alkali atoms gases.}

BEC was for the first time observed in 1995 in a series of experiments on dilute cold alkali atoms gases, such as Rubidium $^{87}\mathrm{Rb}$ \cite{C-al}, Sodium $^{23}\mathrm{Na}$ \cite{K-al} and Lithium $^{7}\mathrm{Li}$ \cite{BSTH}. Although the first theoretical predictions go back to the 1920s \cite{Bose,Einstein} and were made for the ideal Bose gas in isotropic cubic boxes \cite{Einstein}, these recent experiments were realized in a magnetic-optical trap.\\
\indent Let us give the two-key principles of these experiments. The first step consists in pre-cooling the atoms by the \textit{laser cooling method}. The dilute atoms gas is confined in a vacuum chamber and is cooled by two lasers facing each other in each direction at a frequency slightly lower than the resonance frequency of the atoms so that the moving atoms are slowed by Doppler effect. A temperature of the order of $10^{-4}$K can be reached. The second step consists in lowering the temperature by the \textit{magnetic evaporative cooling method}. An inhomogeneous magnetic field is introduced to trap the atoms. After switching off the laser beams, the magnetic evaporation allows to remove the high-energy atoms. The temperature is of the order of $10^{-6}$K with about $10^{4}-10^{6}$ atoms in the magnetic trap. The temperature of the gas can be adjusted by moving the energy cutoff of the evaporating process, and then it can be below the predicted critical temperature
$T_C\approx 10^{-6}K$ at the center of
the trap. Note that this critical temperature is a good approximation for a dilute gas, see e.g. \cite{Phys1,Phys2,Phys3}. To observe the BEC, by an absorbing image technics, one can measure the spatial density profile of the atomic cloud. At high temperature (or at low density), one can observe a widely spread spatial distribution. At low temperature (or at high density), one can observe a \textit{spatial condensation} through a peak of density.\\
\indent Since 1995, a very large number of experiments have been realized to study more precisely the features of BEC created by cold alkali atoms gases and, naturally there is a huge amount of literature on this topic. We refer the readers to the modern references \cite{Phys1,Phys2,Phys3}. In relation with the problem treated in this review, we point out that many experiments have been realized for anisotropic traps in \cite{GVL,Aspect,Dalibard} and have revealed that the Bose gas may manifest singular and unusual phenomena, see e.g. \cite{Ketterle,Ketterle2}.

\subsection{Investigating the features of BEC and the reduced density matrix.}

\begin{itemize}
\item \textbf{A review of different approaches.}
\end{itemize}

To figure out at first stage BEC phenomenon created by cold alkali atom gases, the most widespread model in literature is a $d$-dimensional ideal Bose gas trapped in an isotropic harmonic potential $V_{\mathrm{osc}}(\bold{x}):=\frac{m}{2}\omega^2 \vert \bold{x}\vert^2$ (here, $\omega$ stands for the angular frequency of the harmonic oscillator). Such a system only has discrete energy levels whose the ground-state
energy is $E_{0}:= \frac{d}{2} \hbar \omega$. When investigating the thermodynamics of the $d$-dimensional trapped ideal Bose gas in the grand-canonical situation, the usual method consists in approximating the sum over the energy levels involved in the thermodynamic functions (such as the average number of particles, total energy, etc.) by an integral in the semiclassical regime $\hbar\omega\beta\ll 1$, $\beta:=(k_B T)^{-1}$. It amounts to approximating the density of states by its high-energy value, i.e., when $E \gg \hbar \omega$. We refer the reader to the founding articles \cite{GHS,BPK}, and
also to the reference textbooks \cite{Phys1,Phys3,Phys2}. This procedure is related to the thermodynamic limit concept, and because of the inhomogeneity arising from the trap, the role of canonical parameter is given to an effective number of particles $N \omega^{d}$, see e.g.
\cite[Sec. 10.4]{Phys1}. Actually, this procedure turns out to be necessary to make appear a critical average number of particles $N_c$ when the chemical potential tends to $E_{0}$ (similarly to the standard critical density $\rho_c$ for homogeneous systems) for $d\geq 2$. When this critical number of particles is attained, the number of particles computed by the integral is then saturated. In the semiclassical regime $\hbar\omega\beta \ll 1$, if the total number of particles in the trap $N$ is greater than $N_c$, it is usually assumed that the excess number of particles $N_0=N-N_c$ has to fall in the ground state, in accordance with the Einstein criterion of BEC. We mention that the necessity of such a semiclassical approximation to compute the thermodynamic functions has been discussed in \cite{KT1,2HR,GH,KT2}.\\
\indent A crucial ingredient to study precisely the features of BEC is the local density function defined as the diagonal part of the reduced density matrix. Detailed information on the local density of particles allows to 'draw' a density profile. This is actually what is measured in the experiments to demonstrate the occurrence of BEC, see e.g. \cite{Phys2}. When $N>N_{c}$, it is usually assumed that the 3-dimensional local density of particles is divided into two parts: a term corresponding to the condensate plus a term corresponding to the non-condensate (often referred to as thermal gas):
\begin{equation}
\label{Intro1}
\rho(\bold{x})= \rho^{\mathrm{(BEC)}}(\bold{x}) + \rho^{\mathrm{(therm)}}(\bold{x}),
\end{equation}
where:
\begin{equation}
\label{Intro11}
\rho^{\mathrm{(BEC)}}(\bold{x})= N_0 \left\vert\Psi^{(0)}(\bold{x})\right\vert^2.
\end{equation}
Here, $N_{0} = N- N_{c}>0$ is the number of particles in the ground-state and $\Psi^{(0)}$ the ground-state eigenvector. Since $N_{c}= \mathcal{O}((\hbar \omega \beta)^{-3})$ and $\vert\Psi^{(0)}(\bold{x})\vert^{2} = \mathcal{O}((\hbar \omega \beta)^{\frac{3}{2}})$ when $\hbar \omega \beta \ll 1$, then it is found that $\rho^{\mathrm{(BEC)}}(\bold{x}) = \mathcal{O}((\hbar \omega \beta)^{-\frac{3}{2}})$ when $\hbar \omega \beta \ll 1$. As for the non-condensate part, the expression that is generally given reads as:
\begin{equation}
\label{Intro12}
\rho^{\mathrm{(therm)}}(\bold{x}) = \int_{\mathbb{R}^3}\frac{\mathrm{d}\bold{p}}{(2\pi\hbar)^3} \frac{1}{\mathrm{e}^{\beta \left(\frac{\vert\bold{p}\vert^2}{2m}+V_{\mathrm{osc}}(\bold{x})\right)}-1}.
\end{equation}
\eqref{Intro12} is obtained by considering the semiclassical limit $\hbar\downarrow0$ ($\hbar$ being seen as a parameter). This is justified by the fact that the semiclassical regime $\hbar\omega\beta\ll1$ is assumed. Note that if one considers the limit of zero angular frequency of the harmonic trap, the expression for $\rho^{\mathrm{(therm)}}(\bold{x})$ is obtained from \eqref{Intro12} by setting $V_{\mathrm{osc}}(\bold{x}) = 0$. We stress the point that the two-terms decomposition in \eqref{Intro1} relies on the Einstein criterion of BEC, and \textit{a priori}, it only holds for the isotropic harmonic traps. Indeed, when considering highly anisotropic traps, we expect \eqref{Intro1} to be modified because of the possible occurrence of generalized-BEC (g-bec), see \cite{BZ,Mullin} and also \cite{Beau,BeauThese} and references therein.\\

Since our article mainly deals with the reduced density matrix (RDM) associated to the ideal Bose gas in harmonic traps, let us discuss some of the methods encountered in literature used to examine the RDM. Some attempts have been made to study the RDM from its representation by the sum involving directly the Hermite functions (see formula \eqref{rdm}), see \cite{Mullin2} and \cite{BeauThese} respectively for the 2-dimensional and 3-dimensional ideal Bose gas trapped in the isotropic trap. In \cite{Mullin2}, it is found that the non-condensate part of the 2-dimensional local density function at $\bold{x}=\bold{0}$ diverges as $\ln(1/\omega)$ in the regime of weak angular frequencies. In \cite{BeauThese}, a decomposition of type \eqref{Intro1} has been recovered in the semiclassical regime, and the formula given for $\rho^{\mathrm{(therm)}}$ corresponds to the zero angular frequency regime. To make this approach rigorous, the main difficulty is to have a good control on the behavior of the Hermite polynomials 
associated to high eigenvalues which oscillate rapidly in the regime of weak angular frequencies. At the same time, another approach based on a path integral representation, originally introduced by R.P.
Feynman in \cite{Feynman} for the Bose gas in boxes with periodic boundary conditions, has been developed. This is the so-called \textit{loop-gas approach (or cycles permutation)}, see \cite{Mullin1,Mullin12} for the ideal Bose gas in 3-dimensional harmonic traps. Following the Feynman original idea, this approach consists in representing the RDM as a sum of reduced density matrices associated to the loops of size $l$, which is given by the Mehler kernel at a scaled inverse-temperature $l\beta$. In \cite{Mullin1,Mullin12}, W. Mullin revisited the 3-dimensional isotropic trap and recovered the well-known results related to the localization of the condensate and thermal gas. Moreover, he identified the sizes of the loops in the scale of $\hbar\omega\beta$: the condensate comes from the large loops corresponding to $l > (\hbar \omega \beta)^{-1}$, and the non-condensate comes from the small loops. This work is connected to the Path Integral Monte Carlo (PIMC) numerical method, see e.g. \cite{Ceperley,Krauth,
Krauth1,Holzmann,Chevallier}. Finally, we mention that there exists also a stochastic approach based on the theory of random point fields. Considering a model of the mean-field interacting boson gas trapped by a weak harmonic potential, Tamura \textit{et al.} proved in \cite{TZ} the existence of two phases distinguished by the boson condensation and by a different behavior of the local particle density in  \textit{weak harmonic trap limit} (WHT-limit) mimicking the regime of weak angular frequency. The properties of the system are derived from the generating functional in the WHT-limit whose the detailed study is the main subject of \cite{TZ}. The same method has been used to investigate the perfect Bose gas in exponential- and polynomial-anisotropic boxes in \cite{TZ2}.

\begin{itemize}
\item \textbf{Our approach: a loop-gas-like approach.}
\end{itemize}

The purpose of this paper is to introduce a rigorous method allowing to derive in the semiclassical regime $\hbar \omega \beta \ll 1$, accurate information on the RDM associated to a $d$-dimensional ideal Bose gas in harmonic traps. Our starting-point is the representation of the RDM by the sum involving the Mehler's kernel in which we introduce a dilatation of the angular frequency  by a dimensionless parameter $\kappa$. The regime of small values of $\kappa$ mimics the regime of weak angular frequencies of the trap. In the regimes in which BEC occurs, our method consists in a suitable sum-decomposition of the RDM. The bounds of the sums are well-chosen monotone increasing functions of $\kappa$ when $\kappa$ approaches zero. Performing the limit $\kappa \downarrow 0$ (the so-called \textit{open-trap limit}) allows to identify the parts of the decomposition from which arise the condensate and non-condensate contributions. Our approach is similar to the so-called \textit{loop-gas approach} developed in \cite{
Mullin1} for the ideal Bose gas in isotropic traps in that sense that, our sum-decomposition resembles to the loops-decomposition. To introduce the method, we first focus on the isotropic harmonic trap. From our \textit{loop-gas-like approach}, we investigate the RDM in open-trap limit which mimics the regime of weak angular frequencies of the trap. By a suitable rescaling of the spatial variables, the semiclassical regime corresponding to small values of $\hbar$ (seen as a parameter) is also investigated. All the results stated in the literature are recovered, and we provide accurate information on the localization of the condensate/thermal gas via the local density function. Regarding the loop-gas approach in \cite{Mullin1}, we give accurate information on the length scale of the loops from which arise the different contributions involved in the RDM. We refer the reader to Remark \ref{Remloop}.\\
\indent The loop-gas-like approach that we develop extends to the case of anisotropic traps. Given a model of anisotropic trap, our method allows an accurate study of the RDM in the regime of weak angular frequencies (ensuring the semiclassical regime). For illustrative purposes, we treat two particular models of 3-dimensional anisotropic traps: an exponential-quasi-1D ($\hbar\omega_1\beta\ll\hbar\omega_\perp\beta\ll1$) and an exponential-quasi-2D ($\hbar\omega_\perp\beta\ll\hbar\omega_1\beta\ll1$) model. We refer the reader to Sec. \ref{quasi1D} and \ref{quasi2D} respectively for a precise definition of these models. Remarks \ref{RemQ1D} and \ref{RemQ2D1} provide guidance on how to prepare experimentally such systems. Investigating the RDM for such models is relevant because of the following. For the quasi-1D model that we consider, it has been shown in \cite{BZ} that the ideal Bose gas can manifest both BEC and generalized-BEC (g-BEC) in a suitable regime corresponding to a second kind of transition. 
Therefore,
 we expect the RDM to exhibit a non-usual behavior arising from the presence of g-BEC. As for the quasi-2D model that we consider, it actually allows to mimic the properties of the 2-dimensional isotropic trap in a suitable regime. Ergo, we expect the 'extra-dimension' to regularize the logarithmic divergence occurring in the $2$-dimensional isotropic trap. In particular, we show for both models that the local density of particles has the form:
\begin{equation}\label{Intro2}
\rho(\bold{x})=\rho^{\mathrm{(BEC)}}(\bold{x})+\rho^{\mathrm{(add)}}(\bold{x})+\rho^{\mathrm{(therm)}}(\bold{x}).
\end{equation}
The first and third contribution is the counterpart of the first and second term in \eqref{Intro1}. In the regime of weak angular frequencies, the additional contribution $\rho^{\mathrm{(add)}}(\bold{x})$ exhibits  different behaviors according to the models: it is divergent for the exponential-quasi-1D model, whereas it is a $\bold{x}$-independent constant for the exponential-quasi-2D model. Our loop-gas-like approach allows to bring out that the additional term comes from the loops of mesoscopic size, see Remarks \ref{RemQ1D3} and \ref{RemQ2D3} for further details. Due to this feature, the additional term for the exponential-quasi-1D and exponential-quasi-2D model can be interpreted as a \textit{g-BEC contribution} and \textit{local g-BEC contribution} respectively. Further investigations on these models will be made in a companion paper.\\

To conclude this introduction, we stress the point that our method is related to the Path Integral Monte Carlo (PIMC) numerical computations, see e.g. \cite{Ceperley,Krauth,Krauth1,Holzmann,Chevallier}. Since it permits to treat anisotropic harmonic traps, then it could be useful for future numerical investigations. We believe that PIMC numerical simulations might serve to exhibit the additional term appearing in \eqref{Intro2}, and also might make the connections with the presence of g-BEC in anisotropic traps.

\section{The setup \& The main results.}
\label{Sec2}

\subsection{The single-particle Hamiltonian and related operators.}

Consider a $d$-dimensional ($d=1,2,3$) ideal quantum gas composed of a large number of non-relativistic spin-0 identical particles with rest mass $m>0$, and obeying the Bose-Einstein statistics. The gas is confined in a box given by $\Lambda_{L}^{d}:= \{\bold{x} \in \mathbb{R}^{d}: -\frac{L}{2}< x_{l}<\frac{L}{2},\, l=1,\ldots,d\}$ with $L>0$, and trapped in an external isotropic harmonic potential whose the angular frequency is given by $\omega_{\kappa} := \omega_{0} \kappa$, with $\omega_{0}>0$ and $\kappa>0$ being a dimensionless parameter. The interactions between particles are neglected, and the system is at equilibrium with a thermal and particles bath.\\
\indent Introduce  the one-particle Hamiltonian. On $\mathcal{C}_{0}^{\infty}(\Lambda_{L}^{d})$, define $\forall \kappa>0$ the family of operators:
\begin{equation}
\label{HL}
H_{L,\kappa} := \frac{1}{2m}(- i \hbar\nabla_{\bold{x}})^{2} + \frac{1}{2} (\omega_{0}\kappa)^{2} \vert \bold{x}\vert^{2}.
\end{equation}
Here and hereafter, $m>0$ and $\omega_{0}>0$ are kept fixed. By standard arguments, \eqref{HL} extends $\forall\kappa>0$ to a family of self-adjoint and bounded from below operators for any $L \in (0,\infty)$, denoted again by $H_{L,\kappa}$, with domain $D(H_{L,\kappa}) = W_{0}^{1,2}(\Lambda_{L}^{d}) \cap W^{2,2}(\Lambda_{L}^{d})$. This definition corresponds to choose Dirichlet boundary conditions on the boundary $\partial \Lambda_{L}^{d}$. Since the inclusion $W_{0}^{1,2}(\Lambda_{L}^{d}) \hookrightarrow L^{2}(\Lambda_{L}^{d})$ is compact, then $H_{L,\kappa}$ has a purely discrete spectrum with an accumulation point at infinity. In the case of $d=1$, we denote by $\{\epsilon_{L,\kappa}^{(s)}\}_{s \in \mathbb{N}}$ the set of eigenvalues counting multiplicities and in increasing order. Due to the property of separation of variables, the eigenvalues of the  multidimensional case are related to those of the one-dimensional case by: $E_{L,\kappa}^{(\bold{s})} := \sum_{j=1}^{d} \epsilon_{L,\kappa}^{(s_{j})}$,
$\bold{s}=\{s_{j}\}_{j=1}^{d} \in \mathbb{N}^{d}$. Here and hereafter, $\mathbb{N}$ denotes the set of non-negative integers and $\mathbb{N}^{*}$ the set of strictly positive integers.\\
\indent When $\Lambda_{L}^{d}$ fills the whole space (i.e., when $L \uparrow \infty$), define $\forall \kappa>0$ on $\mathcal{C}_{0}^{\infty}(\mathbb{R}^{d})$ the operator:
\begin{equation}
\label{Hinfini}
H_{\infty,\kappa}:= \frac{1}{2m}(- i \hbar \nabla_{\bold{x}})^{2} + \frac{1}{2} (\omega_{0} \kappa)^{2} \vert \bold{x}\vert^{2}.
\end{equation}
From \cite[Thm. X.28]{RS2}, $\forall \kappa>0$ \eqref{Hinfini} is essentially self-adjoint and its self-adjoint extension, denoted again by $H_{\infty,\kappa}$, is bounded from below. By \cite[Thm. XIII.16]{RS4}, the spectrum of $H_{\infty,\kappa}$ is purely discrete with eigenvalues increasing to infinity. From the one-dimensional problem, the eigenvalues and eigenfunctions of the multidimensional case can be written down explicitly. The eigenvalues of the one-dimensional problem are all non-degenerate and given by, see e.g. \cite[Sec. 1.8]{BS}:
\begin{equation}
\label{vap1}
\epsilon_{\infty,\kappa}^{(s)} := \hbar \omega_{0} \kappa \left(s + \frac{1}{2}\right),\quad s \in \mathbb{N}.
\end{equation}
The corresponding eigenfunctions, which form an orthonormal basis in $L^{2}(\mathbb{R})$, read as:
\begin{equation}
\label{funcp1}
\forall x \in \mathbb{R},\quad \psi_{\infty,\kappa}^{(s)}(x) := \frac{1}{\sqrt{2^{s} s!}} \left(\frac{m \omega_{0} \kappa}{\pi \hbar}\right)^{\frac{1}{4}} \exp\left(- \frac{m \omega_{0} \kappa}{\hbar} \frac{x^{2}}{2} \right) \mathpzc{H}_{s}\left(\sqrt{\frac{m\omega_{0}\kappa}{\hbar}} x\right), \quad s \in \mathbb{N},
\end{equation}
where $\mathpzc{H}_{s}$, $s \in \mathbb{N}$ are the Hermite polynomials defined on $\mathbb{R}$ by:
$\mathpzc{H}_{s}(x) := (-1)^{s} \exp(x^{2}) \frac{\mathrm{d}^{s}}{\mathrm{d} x^{s}}\exp(-x^{2})$.\\
The eigenvalues and eigenfunctions of the  multidimensional case (i.e., $d=2,3$) are respectively related to those of the one-dimensional case by:
\begin{gather}
\label{vapd}
E_{\infty,\kappa}^{(\bold{s})} := \sum_{j=1}^{d} \epsilon_{\infty,\kappa}^{(s_{j})} = \hbar \omega_{0} \kappa \sum_{j=1}^{d} \left(s_{j} + \frac{1}{2}\right),\quad \bold{s}=\{s_{j}\}_{j=1}^{d} \in \mathbb{N}^{d},\\
\label{funcpd}
\Psi_{\infty,\kappa}^{(\bold{s})}(\bold{x}) := \prod_{j=1}^{d} \psi_{\infty,\kappa}^{(s_{j})}(x_{j}),\quad \bold{x}= \{x_{j}\}_{j=1}^{d} \in \mathbb{R}^{d}.
\end{gather}
From \eqref{vap1}-\eqref{vapd} and by the use of the min-max principle, one has for any $L \in (0,\infty)$:
\begin{equation}
\label{infspect}
\forall \kappa>0,\quad \inf \sigma(H_{L,\kappa}) \geq \inf \sigma (H_{\infty,\kappa}) = E_{\infty,\kappa}^{(\bold{0})} = d \epsilon_{\infty,\kappa}^{(0)}>0.
\end{equation}

For the need of the following section, let us introduce the one-parameter semigroup generated by $H_{\infty,\kappa}$ in \eqref{Hinfini}, and by $H_{\infty,0}:=  \frac{1}{2m} (- i \hbar \nabla)^{2}$ whose the self-adjointness domain is $W^{2,2}(\mathbb{R}^{d})$. For any $\kappa \geq 0$, the one-parameter semigroup is defined by $\{G_{\infty,\kappa}(t) := \mathrm{e}^{- t H_{\infty,\kappa}} : L^{2}(\mathbb{R}^{d}) \rightarrow L^{2}(\mathbb{R}^{d})\}_{t \geq 0}$. It is strongly continuous and it is a self-adjoint and positive operator by the spectral theorem and the functional calculus. For any $\kappa \geq 0$, $\{G_{\infty,\kappa}(t)\}_{t>0}$ is an integral operator whose the integral kernel, denoted by $G_{\infty,\kappa}^{(d)}(\cdot\,,\cdot\,;t)$, is jointly continuous in $(\bold{x},\bold{y},t) \in \mathbb{R}^{d}\times\mathbb{R}^{d} \times (0,\infty)$, see e.g. \cite[Sec. A]{BSa}. If $\kappa=0$, it corresponds to the so-called heat kernel which reads for $d=1$ as:
\begin{equation}
\label{heatk}
\forall(x,y)\in \mathbb{R}^{2},\,\forall t>0,\quad G_{\infty,0}^{(d=1)}(x,y;t) := \sqrt{\frac{m}{2\pi \hbar^{2}}} \frac{\mathrm{e}^{- \frac{m}{\hbar^{2}} \frac{(x-y)^{2}}{2 t}}}{\sqrt{t}}.
\end{equation}
If $\kappa>0$, the one-dimensional kernel is given by the so-called Mehler's formula, see
\cite[pp. 176]{Ku}:
\begin{multline}
\label{Mehler}
\forall(x,y)\in \mathbb{R}^{2},\,\forall t>0,\quad G_{\infty,\kappa}^{(d=1)}(x,y;t)  = \\ \sqrt{\frac{m \omega_{0}\kappa}{2 \pi \hbar \sinh(\hbar \omega_{0} \kappa t)}} \mathrm{e}^{-\frac{m \omega_{0} \kappa}{4 \hbar} \left[(x+y)^{2} \tanh\left(\frac{\hbar \omega_{0} \kappa t}{2} \right) + (x-y)^{2} \coth\left(\frac{\hbar \omega_{0} \kappa t}{2} \right)\right]}.
\end{multline}
Note that the multidimensional kernel (i.e., $d=2,3$) is directly obtained from \eqref{heatk} or \eqref{Mehler} by:
\begin{equation}
\label{multd}
\forall \kappa\geq 0,\quad G_{\infty,\kappa}^{(d)}(\bold{x},\bold{y};t) := \prod_{j=1}^{d} G_{\infty,\kappa}^{(d=1)}(x_{j},y_{j};t), \quad \bold{x}:=\{x_{j}\}_{j=1}^{d},\, \bold{y}:=\{y_{j}\}_{j=1}^{d}.
\end{equation}
From \eqref{heatk}-\eqref{Mehler}-\eqref{multd}, one gets $\forall (\bold{x},\bold{y},t) \in \mathbb{R}^{d}\times \mathbb{R}^{d}\times (0,\infty)$ the following inequalities:
\begin{gather}
\label{roughes}
\forall \kappa\geq0,\quad G_{\infty,\kappa}^{(d)}(\bold{x},\bold{y};t) \leq
G_{\infty,0}^{(d)}(\bold{x},\bold{y};t) \leq m^{\frac{d}{2}} (2 \pi \hbar^{2} t)^{-\frac{d}{2}},\\
\label{morepre}
\forall \kappa > 0,\quad G_{\infty,\kappa}^{(d)}(\bold{x},\bold{y};t) \leq \left(\frac{m \omega_{0} \kappa}{\pi \hbar}\right)^{\frac{d}{2}} \frac{\mathrm{e}^{- E_{\infty,\kappa}^{(\bold{0})} t}}{\left(1 - \mathrm{e}^{-2 \hbar \omega_{0} \kappa t}\right)^{\frac{d}{2}}},\quad  E_{\infty,\kappa}^{(\bold{0})}= \frac{d}{2} \hbar \omega_{0} \kappa.
\end{gather}
From the foregoing, for any $\kappa>0$ the semigroup $\{G_{\infty,\kappa}(t)\}_{t>0}$ is a trace-class operator on $L^{2}(\mathbb{R}^{d})$:
\begin{equation}
\label{trace}
\mathrm{Tr}_{L^{2}(\mathbb{R}^{d})}\left\{G_{\infty,\kappa}(t)\right\} = \left(2 \sinh\left(\frac{\hbar \omega_{0} \kappa t}{2} \right)\right)^{-d} = \mathrm{e}^{- E_{\infty,\kappa}^{(\bold{0})} t}\left(1 - \mathrm{e}^{- \hbar \omega_{0} \kappa t}\right)^{-d}.
\end{equation}
\indent Subsequently, we need to introduce a particular function of the semigroup generated by $H_{\infty,\kappa}$, $\kappa>0$. Define $\forall d \in \{1,2,3\}$, $\forall \kappa>0$, $\forall \beta>0$ and $\forall \mu < E_{\infty,\kappa}^{(\bold{0})}$ the operator on $L^{2}(\mathbb{R}^{d})$:
\begin{equation}
\label{fBE}
\mathfrak{f}_{BE}(\beta,\mu;H_{\infty,\kappa}) := \left( \mathrm{e}^{\beta \left( H_{\infty,\kappa} - \mu\right)} - 1\right)^{-1},
\end{equation}
where $\mathfrak{f}_{BE}(\beta,\mu;\cdot\,)$ stands for the Bose-Einstein distribution function. \eqref{fBE} is defined via the Dunford functional calculus as bounded operator on $L^{2}(\mathbb{R}^{d})$, see e.g.
\cite[Sec. VII.9]{DS}. By expanding $x \mapsto (1 - x)^{-1}$, $\vert x\vert<1$ in power series, one has under the conditions of \eqref{fBE} the representation:
\begin{equation}
\label{fBE2}
\mathfrak{f}_{BE}(\beta,\mu;H_{\infty,\kappa}) = \sum_{l=1}^{\infty} \mathrm{e}^{l \beta \mu} G_{\infty,\kappa}(l \beta),
\end{equation}
and the series is absolutely convergent in the trace-class operators sense on $L^{2}(\mathbb{R}^{d})$ because of \eqref{trace}. From \eqref{Mehler}-\eqref{multd} with \eqref{morepre}, the operator \eqref{fBE} has a jointly continuous integral kernel on $\mathbb{R}^{2d}$.

\subsection{Preparing the \textit{open-trap limit}--Some bulk statistical quantities.}

We start by introducing some bulk quantities associated to the confined and harmonically trapped Bose gas in the grand-canonical (G-C) situation. By bulk quantities, we mean independent of the boundary effects arising from the confining box $\Lambda_{L}^{d}$. In accordance with the usual rules of quantum statistical mechanics, the quantities are first defined at finite-volume. Subsequently, we investigate the large-volume behavior to write down an expression for the usual thermodynamic limit. The first quantities of interest are the G-C potential and average number of particles.

\subsubsection*{The grand-canonical potential and average number of particles.}

In the G-C ensemble, let $(\beta,z,\vert \Lambda_{L}^{d}\vert)$ be the fixed external parameters, where $\beta := (k_{B}T)^{-1}>0$ is the 'inverse temperature' ($k_{B}$ is the Boltzmann constant), $z := \mathrm{e}^{\beta \mu} \in (0,\mathrm{e}^{\beta E_{L,\kappa}^{(\bold{0})}})$ is the fugacity ($\mu$ stands for the chemical potential) and $\vert \Lambda_{L}^{d}\vert$ denotes the Lebesgue-measure of the box $\Lambda_{L}^{d}$. We recall that $E_{L,\kappa}^{(\bold{0})} := \inf \sigma(H_{L,\kappa}) >0$. For the definitions below, we refer the reader to \cite{Ru,Hu}.\\
The G-C partition function is defined $\forall d \in \{1,2,3\}$, $\forall L>0$, $\forall \kappa>0$, $\forall\beta>0$, $\forall z \in (0,\mathrm{e}^{\beta E_{L,\kappa}^{(\bold{0})}})$ by:
\begin{equation}
\label{partition}
\Xi_{L,\kappa}(\beta,z) := \prod_{\bold{s} \in \mathbb{N}^{d}}\left(1 - z \mathrm{e}^{-\beta E_{L,\kappa}^{(\bold{s})}}\right)^{-1}.
\end{equation}
From \eqref{partition}, the finite-volume G-C potential and finite-volume G-C average number of particles are respectively defined $\forall d \in \{1,2,3\}$, $\forall L >0$, $\forall \kappa>0$, $\forall \beta>0$ and $\forall z \in (0,\mathrm{e}^{\beta E_{L,\kappa}^{(\bold{0})}})$ by:
\begin{gather}
\label{grandpotential}
\Omega_{L,\kappa}(\beta,z) := - \frac{1}{\beta} \ln \left(\Xi_{L,\kappa}(\beta,z)\right) = \frac{1}{\beta} \sum_{\bold{s} \in \mathbb{N}^{d}} \ln\left(1 - z \mathrm{e}^{-\beta E_{L,\kappa}^{(\bold{s})}}\right),\\
\label{averagenum}
\overline{N}_{L,\kappa}(\beta,z) := - \beta z \frac{\partial \Omega_{L,\kappa}}{\partial z}(\beta,z) =
\sum_{\bold{s} \in \mathbb{N}^{d}} \frac{z \mathrm{e}^{- \beta E_{L,\kappa}^{(\bold{s})}}}{ 1 - z \mathrm{e}^{-\beta E_{L,\kappa}^{(\bold{s})}}}.
\end{gather}
The series in \eqref{grandpotential}-\eqref{averagenum} are absolutely convergent since the semigroup $\{\mathrm{e}^{-\beta H_{L,\kappa}}\}_{\beta>0}$ is trace-class on $L^{2}(\Lambda_{L}^{d})$, see \eqref{trace0}. Moreover, $\Omega_{L,\kappa}(\beta,\cdot\,)$ is real analytic on $(0,\mathrm{e}^{\beta E_{L,\kappa}^{(\bold{0})}})$, see Sec. \ref{AppendixA}.\\

\indent Next, we turn to the large-volume behavior of the two G-C quantities in \eqref{grandpotential}-\eqref{averagenum}. Below, we summarize all the needed results. We refer the reader to Sec. \ref{AppendixA} for the proofs. When $\Lambda_{L}$ fills the whole space, the thermodynamic limit of the G-C potential and average number of particles generically exist. Denoting $\forall d\in \{1,2,3\}$, $\forall \kappa>0$, $\forall\beta>0$ and $\forall z \in (0,\mathrm{e}^{\beta E_{\infty,\kappa}^{(\bold{0})}})$, the bulk G-C potential by $\Omega_{\infty,\kappa}(\beta,z) := \lim_{L \uparrow \infty} \Omega_{L,\kappa}(\beta,z)$, then one has the pointwise convergence:
\begin{equation}
\label{lthaveragenum1}
\overline{N}_{\infty,\kappa}(\beta,z) := -\beta z \frac{\partial \Omega_{\infty,\kappa}}{\partial z}(\beta,z) = \lim_{L \uparrow \infty} -\beta z \frac{\partial \Omega_{L,\kappa}}{\partial z}(\beta,z),
\end{equation}
and the convergence is uniform on compact sets w.r.t. $(\kappa,\beta,z)$. Moreover, one has the asymptotic expansions:
\begin{equation*}
\Omega_{L,\kappa}(\beta,z) = \Omega_{\infty,\kappa}(\beta,z) + \mathcal{O}\left(\mathrm{e}^{-c L^{2}}\right),\quad
\overline{N}_{L,\kappa}(\beta,z) = \overline{N}_{\infty,\kappa}(\beta,z) + \mathcal{O}\left(\mathrm{e}^{-c L^{2}}\right),
\end{equation*}
for some $L$-independent constant $c=c(\beta,z)>0$. The thermodynamic limit of the grand-canonical potential and average number of particles can be expressed $\forall d\in \{1,2,3\}$, $\forall \kappa>0$, $\forall\beta>0$ and $\forall z \in (0,\mathrm{e}^{\beta E_{\infty,\kappa}^{(\bold{0})}})$ by a sum involving the eigenvalues $\{E_{\infty,\kappa}^{(\bold{s})}\}_{\bold{s} \in \mathbb{N}^{d}}$ as in \eqref{grandpotential}-\eqref{averagenum}:
\begin{gather}
\label{grandpotentialTL}
\Omega_{\infty,\kappa}(\beta,z) = \frac{1}{\beta} \sum_{\bold{s} \in \mathbb{N}^{d}} \ln\left(1 - z \mathrm{e}^{-\beta E_{\infty,\kappa}^{(\bold{s})}}\right),\\
\label{averagenumTL}
\overline{N}_{\infty,\kappa}(\beta,z) =
\sum_{\bold{s} \in \mathbb{N}^{d}} \frac{z \mathrm{e}^{- \beta E_{\infty,\kappa}^{(\bold{s})}}}{ 1 - z \mathrm{e}^{-\beta E_{\infty,\kappa}^{(\bold{s})}}}.
\end{gather}
By involving the semigroup $\{G_{\infty,\kappa}(t)\}_{t \geq 0}$, \eqref{averagenumTL} can be rewritten under the same conditions as:
\begin{equation}
\label{lthaveragenum2}
\overline{N}_{\infty,\kappa}(\beta,z) =  \sum_{l=1}^{\infty} z^{l} \mathrm{Tr}_{L^{2}(\mathbb{R}^{d})}\{G_{\infty,\kappa}(l \beta)\} = \mathrm{Tr}_{L^{2}(\mathbb{R}^{d})} \{\mathfrak{f}_{BE}(\beta,z;H_{\infty,\kappa})\},
\end{equation}
where $\mathfrak{f}_{BE}(\beta,z;H_{\infty,\kappa})$ is the operator defined by \eqref{fBE}, see also \eqref{fBE2}.

\subsubsection*{The rescaled average number of particles.}

So far, we have dealt with the basic bulk statistical quantities related to the Bose gas in the G-C situation. As reviewed in Sec. \ref{intro}, the experiments demonstrate the BEC in \textit{the semiclassical regime $\hbar \omega_{0} \beta \ll 1$}, i.e., $k_{B}T \gg \hbar \omega_{0}$. In the experiments, $\omega_{0}$ usually is a given parameter and the temperature is adapted so that the above condition is fulfilled.
To investigate the global properties of the gas (such as the average number of particles, total energy,...) from \eqref{grandpotentialTL} in the semiclassical regime, the method usually encountered in  literature consists in approximating the sum over the $\bold{s}$-index by an integral. This boils down to identify the bulk quantity of interest by the leading term of its asymptotic expansion in the semiclassical regime while disregarding the remainder term. Preferring to deal with limits instead of equivalents, we introduce a rescaling of the quantities in \eqref{grandpotentialTL}-\eqref{averagenumTL} by the dimensionless parameter $\kappa^{d}$, so that when performing the limit $\kappa \downarrow 0$, the resulting limit coincides with the leading term of the asymptotic expansion in the semiclassical regime. Note that, the $d$-power on $\kappa$ naturally appears in view of the formula used for the first-order approximation of the density of states. We give the
name of \textit{'open-trap limit'} to the limit $\kappa \downarrow 0$ since it mimics the large-opening regime of the trap (i.e., the regime of weak angular frequencies $\omega_{0}$). This explains the introduction of the dimensionless $\kappa$-parameter in \eqref{HL}-\eqref{Hinfini}, and through this procedure, we do not need to set restrictions on the range of $\beta$. 
We emphasize that, when investigating the \textit{global properties} of the Bose gas, the open-trap limit allows to mimics either the regime of weak angular frequencies $\omega_{0}$, or the 'true' semiclassical regime corresponding to small values of $\hbar$ (here, $\hbar$ is seen as a parameter). In all cases, the results stated in literature under the condition $\hbar \omega_{0} \beta \ll 1$ are recovered. This is discussed in Sec. \ref{methoddis}. Finally, we mention that an analogous procedure is introduced in \cite{TZ}, the so-called \textit{'weak harmonic trap limit'}. This limit has a slightly different meaning since it plays the role of thermodynamic limit.\\
\indent In view of the foregoing, we introduce a $\kappa$-rescaling of the bulk average number of particles:

\begin{definition}
\label{defrescale}
$\forall d \in \{1,2,3\}$, $\forall \kappa>0$, $\forall \beta>0$ and $\forall z \in (0,\mathrm{e}^{\beta  E_{\infty,\kappa}^{(\bold{0})}})$, we define the (bulk) rescaled average number of particles from \eqref{lthaveragenum1} by setting:
\begin{equation}
\label{nuinf}
\nu_{\infty,\kappa}(\beta,z) := \kappa^{d} \overline{N}_{\infty,\kappa}(\beta,z).
\end{equation}
\end{definition}

Note that \eqref{nuinf} has to be seen as the thermodynamic limit of the finite-volume rescaled average number of particles defined as $\nu_{L,\kappa}(\beta,z):= \kappa^{d} \overline{N}_{L,\kappa}(\beta,z)$, with $\overline{N}_{L,\kappa}(\beta,z)$ in \eqref{averagenum}.\\

Next, we switch to canonical-like conditions and we assume that the rescaled average number of particles denoted by $\nu>0$, becomes, in addition with the 'inverse' temperature $\beta>0$, an external parameter. Note that, because of the confining harmonic potential, the density of particles vanishes in thermodynamic limit. Thus, the (rescaled) average number of particles turns out to be the 'right' canonical parameter. Seeing the quantity in \eqref{nuinf} as a function of the $\mu$-variable instead of $z$, then $\forall \beta>0$, $\forall\nu>0$ and $\forall\kappa>0$, let $\overline{\mu}_{\infty,\kappa}=\overline{\mu}_{\infty,\kappa}(\beta,\nu) \in (-\infty, E_{\infty,\kappa}^{(\bold{0})})$ satisfying:
\begin{equation}
\label{musolEq}
\nu = \nu_{\infty,\kappa}\left(\beta,\overline{\mu}_{\infty,\kappa}\right),
\end{equation}
and such a $\overline{\mu}_{\infty,\kappa}$ is unique. The inversion of the relation between the bulk rescaled average number of particles and the chemical potential is ensured by the fact that $\forall \beta>0$, $\forall \nu>0$ and $\forall \kappa >0$, $\mu \mapsto \nu_{\infty,\kappa}(\beta,\mu)$ is a strictly increasing function on $\mathbb{R}$, and actually it defines a $\mathcal{C}^{\infty}$-diffeomorphism of $\mathbb{R}$ into $(0,\infty)$. Getting back to the finite-volume quantities, if $\overline{\mu}_{L,\kappa}=\overline{\mu}_{L,\kappa}(\beta,\nu) \in (-\infty, E_{L,\kappa}^{(\bold{0})})$ denotes the unique solution of $\nu = \nu_{L,\kappa}(\beta,\nu)$, then one can prove, see pp. \pageref{liminfsup}:
\begin{equation}
\label{locmu}
\lim_{L \uparrow \infty} \overline{\mu}_{L,\kappa} = \overline{\mu}_{\infty,\kappa}.
\end{equation}

Motivated by the following rewriting of \eqref{nuinf} obtained from \eqref{averagenumTL}:
\begin{equation}
\label{nukappamu}
\forall \kappa>0,\, \forall \beta>0,\,\forall \mu \in \left(-\infty,E_{\infty,\kappa}^{(\bold{0})}\right),\quad \nu_{\infty,\kappa}(\beta,\mu) = \sum_{\bold{s} \in \mathbb{N}^{d}} \frac{\kappa^{d}}{\mathrm{e}^{\beta\left(E_{\infty,\kappa}^{(\bold{s})} - \mu\right)} - 1},
\end{equation}
we end this section by introducing the (bulk) rescaled average number of particles in the $\bold{s}$-state:

\begin{definition}
\label{def3}
$\forall d \in \{1,2,3\}$, $\forall \kappa>0$, $\forall \beta>0$, $\forall \nu>0$ and $\forall \bold{s} \in \mathbb{N}^{d}$, we define the (bulk) rescaled average number of particles in the $\bold{s}$-state as:
\begin{equation}
\label{nustate}
\nu_{\infty,\kappa}(\beta,\nu;\bold{s}) = \nu_{\infty,\kappa}(\beta, \overline{\mu}_{\infty,\kappa};\bold{s}) := \frac{\kappa^{d}}{\mathrm{e}^{\beta(E_{\infty,\kappa}^{(\bold{s})} - \overline{\mu}_{\infty,\kappa})} - 1},
\end{equation}
where $\overline{\mu}_{\infty,\kappa}=\overline{\mu}_{\infty,\kappa}(\beta,\nu) \in (-\infty,E_{\infty,\kappa}^{(\bold{0})})$ satisfies \eqref{musolEq}.
\end{definition}

\subsubsection*{The reduced density matrix, local density function and their rescaling.}

The \textit{reduced density matrix} concept was initially introduced by Penrose and Onsager in \cite{PO} to formulate a BEC criterion and investigate its features. It turns out to be the crucial tool to investigate the \textit{local properties of BEC}. We start by recalling the definition of the reduced density matrix as well as the local density of particles as they are defined in Physics literature.

\begin{definition}
\label{def5}
$\forall d \in \{1,2,3\}$, $\forall \kappa>0$, $\forall \beta>0$ and $\forall \nu>0$, we define the reduced density matrix as the integral kernel of the operator $\mathfrak{f}_{BE}(\beta,\overline{\mu}_{\infty,\kappa};H_{\infty,\kappa})$ in \eqref{fBE} which reads $\forall (\bold{x},\bold{y}) \in \mathbb{R}^{2d}$ as:
\begin{equation}
\label{rdm}
\rho_{\infty,\kappa}\left(\bold{x},\bold{y};\beta,\nu\right) := \left(\mathfrak{f}_{BE}\left(\beta,\overline{\mu}_{\infty,\kappa};H_{\infty,\kappa}\right)\right) (\bold{x},\bold{y}) = \sum_{\bold{s} \in \mathbb{N}^{d}} \frac{\Psi_{\infty,\kappa}^{(\bold{s})}(\bold{x}) \overline{\Psi_{\infty,\kappa}^{(\bold{s})}(\bold{y})}}{\mathrm{e}^{\beta \left(E_{\infty,\kappa}^{(\bold{s})} - \overline{\mu}_{\infty,\kappa}\right)} - 1},
\end{equation}
where $\overline{\mu}_{\infty,\kappa} = \overline{\mu}_{\infty,\kappa}(\beta,\nu) \in (-\infty,E_{\infty,\kappa}^{(\bold{0})})$ satisfies \eqref{musolEq}. The diagonal part of \eqref{rdm} (obtained by setting $\bold{y}=\bold{x}$) is usually named the local density of particles at the point $\bold{x} \in \mathbb{R}^{d}$.
\end{definition}

The sum in the r.h.s. of \eqref{rdm} comes from the spectral theorem. Since the eigenfunctions in \eqref{funcpd}-\eqref{funcp1} are real-valued functions, then we drop the complex conjugation in the following. Without involving directly the eigenfunctions of $H_{\infty,\kappa}$, one has also from \eqref{fBE2} the representation:
\begin{equation}
\label{otrep}
\forall(\bold{x},\bold{y}) \in \mathbb{R}^{2d},\quad \rho_{\infty,\kappa}(\bold{x},\bold{y};\beta,\nu) = \sum_{l=1}^{\infty} \mathrm{e}^{l \beta \overline{\mu}_{\infty,\kappa}} G_{\infty,\kappa}^{(d)}(\bold{x},\bold{y};l\beta).
\end{equation}
Note that, because of \eqref{morepre}, the above series is absolutely convergent uniformly in $(\bold{x},\bold{y}) \in \mathbb{R}^{2d}$. We point out that the reduced density matrix in \eqref{rdm} has the dimension of a density of particles since by \eqref{funcpd} the product of two wave functions has the dimension of the inverse of a volume. Then, the local density of particles at $\bold{x}\in \mathbb{R}^d$ is interpreted as the number of particles at $\bold{x}\in \mathbb{R}^d$ per unit volume.
From \eqref{rdm}, \eqref{funcpd} and \eqref{funcp1}, the reduced density matrix can be rewritten  as:
\begin{equation}
\label{rdmscalepb}
\forall (\bold{x},\bold{y}) \in \mathbb{R}^{2d},\quad \rho_{\infty,\kappa}(\bold{x},\bold{y};\beta,\nu)= \frac{1}{\kappa^{\frac{d}{2}}} \sum_{\bold{s} \in \mathbb{N}^{d}}
\nu_{\infty,\kappa}\left(\beta,\overline{\mu}_{\infty,\kappa};\bold{s}\right) \Psi_{\infty,1}^{(\bold{s})}\left(\bold{x}\sqrt{\kappa}\right) \Psi_{\infty,1}^{(\bold{s})}\left(\bold{y}\sqrt{\kappa}\right).
\end{equation}

In Sec. \ref{p3}, we investigate the reduced density matrix in \textit{open-trap limit}. Remind that the open-trap limit mimics the large-opening regime of the harmonic trap. We will see that, for certain regime of the rescaled density $\nu$ (for which BEC phenomenon occurs), the reduced density matrix in open-trap limit will be infinite indicating that the leading-term of the asymptotic expansion in the large-opening regime of the trap diverges when performing the limit $\omega_{0}\downarrow 0$. To get more accurate information on the behavior of the reduced density matrix in $\omega_{0}$, a suitable $\kappa$-rescaling that takes into account its local nature, is needed. In view of the rewriting in \eqref{rdmscalepb}, we introduce:

\begin{definition}
\label{def6}
$\forall d \in \{1,2,3\}$, $\forall \kappa>0$, $\forall \beta>0$, $\forall \nu>0$ and $\forall (\bold{x},\bold{y}) \in \mathbb{R}^{2d}$, we define the rescaled reduced density matrix from \eqref{rdm} by setting:
\begin{align}
\label{rdmrescaled}
r_{\infty,\kappa}(\bold{x},\bold{y};\beta,\nu) :&= \kappa^{\frac{d}{2}} \rho_{\infty,\kappa}(\bold{x},\bold{y};\beta,\nu) \\
\label{rdmrescaled2}
&= \sum_{\bold{s} \in \mathbb{N}^{d}} \nu_{\infty,\kappa}(\beta,\overline{\mu}_{\infty,\kappa};\bold{s})
\Psi_{\infty,1}^{(\bold{s})}(\bold{x}\sqrt{\kappa}) \Psi_{\infty,1}^{(\bold{s})}(\bold{y}\sqrt{\kappa}),
\end{align}
where $\nu_{\infty,\kappa}(\beta,\overline{\mu}_{\infty,\kappa};\bold{s})$ is the rescaled average number of particles in the $\bold{s}$-state in \eqref{nustate}.\\
Also, we define the rescaled local density of particles at $\bold{x} \in \mathbb{R}^{d}$ from \eqref{rdmrescaled} by setting $\bold{y}=\bold{x}$.
\end{definition}

\subsection{The harmonically trapped Bose gas in \textit{open-trap limit}.}
\label{mainre}

Here, we investigate the global and local properties of the Bose gas in the open-trap limit.

\subsubsection{Statistical quantities in open-trap limit.}

In view of Definition \ref{defrescale}, we start by defining the \textit{open-trap rescaled average number of particles}:

\begin{definition}
\label{def2}
Consider a confined $d$-dimensional harmonically trapped Bose gas, $d \in \{1,2,3\}$ in the G-C ensemble. Provided that the limit exists (possibly infinite), $\forall \beta>0$ and $\forall \mu \in (-\infty,0)$, we define the open-trap rescaled average number of particles as:
\begin{equation}
\label{nuinf0'}
\nu_{\infty,0}(\beta,\mu) := \lim_{\kappa \downarrow 0} \nu_{\infty,\kappa}(\beta,\mu).
\end{equation}
\end{definition}

Motivated by Definition \ref{def2} and analogously to the critical density of particles for perfect Bose gas confined in boxes, we introduce the \textit{critical open-trap rescaled average number of particles} as:

\begin{definition}
\label{def22}
For any $\beta>0$ and $\mu<0$, assume that the limit in \eqref{nuinf0'} exists and is finite. Provided that the limit exists (possibly infinite), we define the critical open-trap rescaled average number of particles as:
\begin{equation}
\label{defnuc}
\nu_{c}(\beta):=\lim_{\mu\uparrow0}\nu_{\infty,0}(\beta,\mu) = \sup_{\mu<0} \nu_{\infty,0}(\beta,\mu).
\end{equation}
\end{definition}

In view of Definition \ref{def3}, we introduce similarly to Definition \ref{def2}, the \textit{open-trap rescaled average number of particles in the $\bold{s}$-state} as follows:

\begin{definition}
Consider a confined $d$-dimensional harmonically trapped Bose gas, $d \in \{1,2,3\}$ in the G-C ensemble. Provided that the limit exists (possibly infinite), $\forall \beta>0$, $\forall \nu>0$ and $\forall \bold{s} \in \mathbb{N}^{d}$, we define the open-trap rescaled average number of particles in the $\bold{s}$-state as:
\begin{equation}
\label{nu(s)}
\nu_{\infty,0}(\beta,\nu;\bold{s}) := \lim_{\kappa \downarrow 0} \nu_{\infty,\kappa}\left(\beta,\nu;\bold{s}\right) = \lim_{\kappa \downarrow 0} \nu_{\infty,\kappa}\left(\beta,\overline{\mu}_{\infty,\kappa};\bold{s}\right),
\end{equation}
where $\overline{\mu}_{\infty,\kappa} = \overline{\mu}_{\infty,\kappa}(\beta,\nu) \in (-\infty,E_{\infty,\kappa}^{(\bold{0})})$ satisfies \eqref{musolEq}.
\end{definition}

From Definitions \ref{def5} and \ref{def6}, we finally introduce the \textit{open-trap (rescaled) reduced density matrix and the open-trap local (rescaled) density function} as:

\begin{definition}
\label{defd}
Consider a confined $d$-dimensional harmonically trapped Bose gas, $d \in \{1,2,3\}$ in the G-C ensemble.\\
$\mathrm{(i)}$. Provided that the limit exists (possibly infinite), $\forall \beta>0$ and $\forall \nu>0$, we define the open-trap reduced density matrix as:
\begin{equation}
\label{opentrr}
\forall(\bold{x},\bold{y}) \in \mathbb{R}^{2d},\quad \rho_{\infty,0}(\bold{x},\bold{y};\beta,\nu):=
\lim_{\kappa\downarrow0} \rho_{\infty,\kappa}(\bold{x},\bold{y};\beta,\nu).
\end{equation}
The open-trap local density of particles at $\bold{x} \in \mathbb{R}^{d}$ is defined from \eqref{opentrr} by setting $\bold{y}=\bold{x}$.\\
$\mathrm{(ii)}$. Provided that the limit exists (possibly infinite), $\forall \beta>0$ and $\forall \nu>0$ we define the open-trap rescaled reduced density matrix as:
\begin{equation}
\label{opentrpr}
\forall(\bold{x},\bold{y}) \in \mathbb{R}^{2d},\quad r_{\infty,0}(\bold{x},\bold{y};\beta,\nu):=
\lim_{\kappa\downarrow0} r_{\infty,\kappa}(\bold{x},\bold{y};\beta,\nu).
\end{equation}
The open-trap rescaled local density of particles at $\bold{x} \in \mathbb{R}^{d}$ is defined from \eqref{opentrpr} by setting $\bold{y}=\bold{x}$.
\end{definition}

\subsubsection{The global properties--Open-trap BEC.}
\label{p2}

Here, we focus on the thermodynamics of the Bose gas in the open-trap limit. When dealing with global properties, the open-trap limit mimics either the regime of weak angular frequencies $\omega_{0}$ of the trap, or the  semiclassical regime corresponding to small values of $\hbar$ (seen as a parameter).\\
\indent We start by writing down an explicit expression for the open-trap rescaled average number of particles and by investigating its critical value. To do that, introduce for any real $\theta >0$:

\begin{equation}
\label{polylog}
\mathpzc{g}_{\theta}(\xi):=\sum_{n=1}^{\infty} \frac{\xi^{n}}{n^{\theta}},\qquad \forall \xi \in \mathbb{C}\,\,\, s.t.\,\,\,\left\{ \begin{array}{ll}
0 \leq \vert \xi \vert<1,\,&\textrm{if $0< \theta < 1$}, \\
0\leq \vert \xi \vert \leq 1,\,\xi \neq 1,\,&\textrm{if $\theta =1$},\\
0\leq \vert \xi \vert \leq 1,\, &\textrm{if $\theta >1$},
\end{array}\right.
\end{equation}
which is the analytic continuation of the polylogarithm initially defined on the open ball $\mathcal{B}(0,1)$.\\
\indent From Definitions \ref{def2}-\ref{def22}, we establish:

\begin{lema}
\label{Lemma0}
For any $d \in \{1,2,3\}$ and $\beta>0$:\\
$\mathrm{(i)}$. The open-trap rescaled average number of particles exists and reads as:
\begin{equation}
\label{nu(beta,mu)}
\forall \mu<0,\quad \nu_{\infty,0}(\beta,\mu)= \frac{1}{\Gamma(d)\left(\hbar \omega_{0}\right)^{d}} \int_{0}^{\infty} \mathrm{d}\tau\,
\frac{\tau^{d-1}}{\mathrm{e}^{\beta(\tau-\mu)}-1}= \frac{\mathpzc{g}_{d}\left(\mathrm{e}^{\beta\mu}\right)}{\left(\hbar \omega_{0} \beta\right)^d},
\end{equation}
where $\Gamma(\cdot\,)$ denotes the usual Euler Gamma function.\\
$\mathrm{(ii)}$. The critical open-trap rescaled average number of particles exists and satisfies:
\begin{subnumcases}
{\label{nucEq}\nu_{c}(\beta)=}
\infty,&\textrm{if $d=1$}, \label{nucEq1} \\
\frac{\mathpzc{g}_{d}(1)}{\left(\hbar \omega_{0} \beta\right)^d}<\infty,&\textrm{if $d=2,3$}. \label{nucEq2}
\end{subnumcases}
$\mathrm{(iii)}$. For any $0<\nu < \nu_{c}(\beta)$, there exists a unique $\overline{\mu}_{\infty,0}=\overline{\mu}_{\infty,0}(\beta,\nu) \in (-\infty,0)$ satisfying:
\begin{equation}
\label{muinf0}
\nu=\nu_{\infty,0}\left(\beta,\overline{\mu}_{\infty,0}(\beta,\nu)\right).
\end{equation}
\end{lema}

\begin{remark} From Lemma \ref{Lemma0} $\mathrm{(ii)}$, the equality $\nu=\nu_{c}(\beta)$ defines the critical temperature $T_{c}$:
\begin{equation*}
k_{B} T_{c} := \hbar \omega_{0} \left(\frac{\nu}{\mathpzc{g}_{d}(1)}\right)^{\frac{1}{d}},\quad d=2,3.
\end{equation*}
\end{remark}

Subsequently, we give a definition for the so-called \textit{open-trap BEC} which is analogous to the BEC concept (within the 'Einstein's formulation') for Bose gas confined in boxes, see e.g. \cite{vdBLP}:

\begin{definition}\textbf{Open-trap BEC criterion.}
\label{OpenTrapBECdef}
Consider a confined $d$-dimensional harmonically trapped Bose gas, $d \in \{1,2,3\}$ in the G-C ensemble. For any $\beta>0$, we say that the Bose gas manifests an open-trap BEC for a fixed rescaled number of particles $\nu>0$ if:\\
$\mathrm{(i)}$. The critical open-trap rescaled average number of particles is finite: $\nu_c(\beta) < \infty$, and moreover,\\
$\mathrm{(ii)}$. The open-trap rescaled average number of particles on the ground-state is strictly positive, i.e.,
\begin{equation*}
\nu_{\infty,0}(\beta,\nu;\bold{0})>0,\quad \nu>\nu_c(\beta).
\end{equation*}
\end{definition}

The main preliminary result states that the confined harmonically trapped Bose gas manifests an open-trap BEC in the sense of Definition \ref{OpenTrapBECdef} provided that $d>1$:

\begin{proposition}
\label{Prop1}
Consider a confined $d$-dimensional harmonically trapped Bose gas, $d \in \{1,2,3\}$ in the G-C ensemble. Then, for any $\beta>0$ and $\nu>0$:\\
$\mathrm{(i)}$. If $d=1$, the Bose gas manifests no open-trap BEC. Moreover, $\nu_{\infty,0}(\beta,\nu;s)=0$  $\forall s \in \mathbb{N}$.\\
$\mathrm{(ii)}$. If $d=2,3$, the Bose gas manifests an open-trap BEC. Furthermore, the open-trap rescaled average number of particles on the ground-state satisfies:
\begin{subnumcases}{\label{nu0} \nu_{\infty,0}(\beta,\nu;\bold{0})=}
0,&\textrm{when $\nu<\nu_c(\beta)$}, \label{nu01}\\
\nu-\nu_c(\beta),&\textrm{when $\nu\geq\nu_c(\beta)$}. \label{nu02}
\end{subnumcases}
Here, $\nu_{c}(\beta)$ is defined by \eqref{defnuc} and satisfies \eqref{nucEq2}. Moreover, $\nu_{\infty,0}(\beta,\nu;\bold{s})=0$ $\forall \bold{s} \in (\mathbb{N}^{*})^{d}$.\\
$\mathrm{(iii)}$. $\overline{\mu}_{\infty,\kappa} = \overline{\mu}_{\infty,\kappa}(\beta,\nu) \in (-\infty,E_{\infty,\kappa}^{(\bold{0})})$ satisfying \eqref{musolEq} admits the asymptotics in the limit $\kappa \downarrow 0$:
\begin{subnumcases}{\label{musol} \overline{\mu}_{\infty,\kappa} =}
E_{\infty,\kappa}^{(\bold{0})} + \overline{\mu}_{\infty,0} + o(1),
&\textrm{when $\nu<\nu_{c}(\beta)$ if $d=1,2,3$}, \label{musol1} \\
E_{\infty,\kappa}^{(\bold{0})} +  o(1), &\textrm{when $\nu=\nu_{c}(\beta)$ if $d=2,3$}, \label{musol2} \\
E_{\infty,\kappa}^{(\bold{0})} -\frac{\kappa^d}{\beta(\nu-\nu_{c}(\beta))}+o\left(\kappa^d\right),&\textrm{when $\nu>\nu_{c}(\beta)$ if $d=2,3$}. \label{musol3}
\end{subnumcases}
Here, $\overline{\mu}_{\infty,0}=\overline{\mu}_{\infty,0}(\beta,\nu) \in (-\infty,0)$ satisfies the equation \eqref{muinf0}.
\end{proposition}

\begin{remark} When $\nu>\nu_{c}(\beta)$ and $\nu/\nu_{c}(\beta) = \eta >1$, then \eqref{musol3} can be rewritten as:
\begin{equation*}
\beta \overline{\mu}_{\infty,\kappa} = \frac{d}{2} \hbar \omega_{0} \kappa \beta - \frac{(\hbar \omega_{0} \kappa \beta)^{d}}{\mathpzc{g}_{d}(1) \left(\eta - 1\right)} + o\left((\hbar \omega_{0} \kappa \beta)^{d}\right),\quad \textrm{when $\kappa \downarrow 0$},
\end{equation*}
and identifies with the asymptotic expansion in the semiclassical regime $\hbar \omega_{0}\beta \ll1$ by setting $\kappa = 1$.
\end{remark}

The results of Proposition \ref{Prop1} are based on the 'Einstein formulation' of the condensation in Definition \ref{OpenTrapBECdef}. However, there exists another kind of condensation named \textit{generalized BEC (g-BEC)}. The g-BEC concept was initially introduced in \cite{vdBLP} for perfect Bose gas in 'Dirichlet boxes'; for a review of definitions and classifications of g-BEC, see e.g. \cite{Beau}. Based on the \textit{open-trap limit} concept and analogously to the van den Berg-Lewis-Pul\'e formulation of the g-BEC in \cite{vdBLP}:

\begin{definition}\textbf{Open-trap g-BEC criterion.}
\label{OpenTrapgBECdef}
Consider a confined $d$-dimensional harmonically trapped Bose gas, $d \in \{1,2,3\}$ in the G-C ensemble.
For any $\beta>0$, we say that the Bose gas manifests an open-trap generalized BEC for a fixed rescaled number of particles $\nu>0$ if:
\begin{equation}
\label{gBECform}
\lim_{\varepsilon \downarrow 0} \lim_{\kappa \downarrow 0} \sum_{\bold{s} \in \mathbb{N}^{d}\,:\, 0<\sum_{j=1}^{d} \kappa s_{j} \leq \varepsilon} \nu_{\infty,\kappa}(\beta,\nu;\bold{s})>0.
\end{equation}
\end{definition}

We emphasize that, unlike the van den Berg-Lewis-Pul\'e formulation in \cite{vdBLP}, our definition of g-BEC \textit{excludes the ground-state} from the sum in \eqref{gBECform} (as mostly encountered in literature). Our definition distinguishes the 'usual' BEC (ground-state macroscopically occupied) from the 'non-usual' (states in a punctured neighborhood of the ground-state macroscopically occupied).\\
\indent The following proposition states that the confined $d$-dimensional harmonically trapped Bose gas does not manifests an open-trap g-BEC in the sense of Definition \ref{OpenTrapgBECdef}:

\begin{proposition}
\label{Prop1'}
Consider a confined $d$-dimensional harmonically trapped Bose gas, $d \in \{1,2,3\}$ in the G-C ensemble. Then for any $\beta>0$ and $\nu>0$:
\begin{equation}
\label{gBECform2}
\lim_{\varepsilon \downarrow 0} \lim_{\kappa \downarrow 0} \sum_{\bold{s} \in \mathbb{N}^{d}\,:\, 0<\sum_{j=1}^{d} \kappa s_{j} \leq \varepsilon} \nu_{\infty,\kappa}(\beta,\nu;\bold{s})=0.
\end{equation}
\end{proposition}

\subsubsection{The local properties (Part 1)--Equivalence of condensation criteria.}
\label{p3}

Involved in the Penrose-Onsager \textit{general criterion} of BEC in \cite{PO}, see also \cite{Yang}, the reduced density matrix allows to treat the Bose gases with interactions (whereas the Einstein criterion was originally formulated for the free Bose gas). Note that there is a huge amount of Physics literature dealing with this criterion for the Bose gas in boxes, see \cite{Beau} and references therein. From Definition \ref{def6} and analogously to the Penrose-Onsager criterion, we define the \textit{open-trap ODLRO criterion} as:

\begin{definition}\textbf{Open-trap ODLRO criterion.}
\label{OpenTrapODLROdef}
Consider a confined $d$-dimensional harmonically trapped Bose gas, $d\in \{1,2,3\}$ in the G-C ensemble. For any $\beta>0$ and $\nu>0$, assume that the limit in \eqref{opentrpr} exists and is finite. We say that the Bose-gas manifests an open-trap ODLRO for the fixed rescaled number of particles $\nu$ if the open-trap rescaled reduced density matrix satisfies:
\begin{equation}
\label{ODLRO}
r_{\infty,0}(\beta,\nu) := \lim_{\vert \bold{x}-\bold{y}\vert \uparrow \infty} r_{\infty,0}(\bold{x},\bold{y};\beta,\nu)>0.
\end{equation}
\end{definition}

For any $\beta>0$, introduce the thermal \textit{de Broglie} wavelength defined as:
\begin{equation}
\label{deBrog}
\lambda_{\beta} :=\sqrt{\frac{2\pi \hbar^{2} \beta}{m}} > 0.
\end{equation}

Here is our first main result focusing on the reduced density matrix in the open-trap limit. As emphasized below \eqref{rdmscalepb}, when dealing with the reduced density matrix, the open-trap limit mimics the large-opening regime of the trap (i.e., the regime of weak angular frequencies $\omega_{0}$).

\begin{theorem}
\label{Thm1}
Consider a confined $d$-dimensional harmonically trapped Bose gas, $d\in \{1,2,3\}$ in the G-C ensemble.
Then for any $\beta>0$, $\nu>0$ and $(\bold{x},\bold{y}) \in \mathbb{R}^{2d}$:\\
$\mathrm{(A)}$. The open-trap reduced density matrix exists and satisfies:
\begin{subnumcases}{\label{rdmd2} \rho_{\infty,0}(\bold{x},\bold{y};\beta,\nu) =} \frac{1}{\lambda_{\beta}^{d}}
\sum_{l=1}^{\infty} \frac{\mathrm{e}^{l \beta\overline{\mu}_{\infty,0}}}{l^{\frac{d}{2}}} \mathrm{e}^{- \frac{\pi}{\lambda_{\beta}^{2}} \frac{\vert \bold{x}-\bold{y}\vert^2}{l}},
&\textrm{when $\nu<\nu_c(\beta)$ if $d=1,2,3$}, \label{rdmd21}\\
\infty, &\textrm{when $\nu=\nu_{c}(\beta)$ if $d=2$},  \label{rdmd22} \\
\infty, &\textrm{when $\nu>\nu_c(\beta)$ if $d=2,3$}. \label{rdmd23}
\end{subnumcases}
$\mathrm{(B)}$. The open-trap rescaled reduced density matrix exists and satisfies:
\begin{equation}
\label{Thm1Eq2}
\begin{split}
r_{\infty,0}(\beta,\nu) &= \left(\frac{m\omega_{0}}{\pi \hbar}\right)^{\frac{d}{2}} \nu_{\infty,0}(\beta,\nu;\bold{0}) \\
&= 2^{\frac{d}{2}} \frac{\left(\hbar \omega_{0} \beta\right)^{\frac{d}{2}}}{\lambda_{\beta}^{d}} \times \left\{\begin{array}{ll}
0, &\textrm{when $\nu<\nu_{c}(\beta)$ if $d=1,2,3$}, \\
\left(\nu - \nu_{c}(\beta)\right), &\textrm{when $\nu \geq \nu_{c}(\beta)$ if $d=2,3$}.
\end{array}\right.
\end{split}
\end{equation}
As a result, the Bose gas manifests an open-trap ODLRO if $\nu>\nu_{c}(\beta)$ when $d=2,3$.\\
$\mathrm{(C)}$. In addition: the open-trap rescaled reduced density matrix satisfies:
\begin{equation}
\label{Thm1Eq1}
r_{\infty,0}(\bold{x},\bold{y};\beta,\nu) = \lim_{\kappa\downarrow0}
\nu_{\infty,\kappa}\left(\beta,\nu;\bold{0}\right) \Psi_{\infty,1}^{(\bold{0})}\left(\bold{x}\sqrt{\kappa}\right) \Psi_{\infty,1}^{(\bold{0})}\left(\bold{y}\sqrt{\kappa}\right),\quad \textrm{if $d=1,2,3$},
\end{equation}
as for the open-trap reduced density matrix without the ground-state:

\begin{subnumcases}{\label{Thm1Eq3}\lim_{\kappa\downarrow0}
\sum_{\bold{s}\in (\mathbb{N}^{*})^{d}}\frac{\Psi_{\infty,\kappa}^{(\bold{s})}(\bold{x}) \Psi_{\infty,\kappa}^{(\bold{s})}(\bold{y})}{\mathrm{e}^{\beta\left(E_{\infty,\kappa}^{(\bold{s})}- \overline{\mu}_{\infty,\kappa}\right)}-1}=}
\frac{1}{\lambda_{\beta}^{d}} \sum_{l=1}^{\infty} \frac{\mathrm{e}^{l \beta\overline{\mu}_{\infty,0}}}{l^{\frac{d}{2}}}
\mathrm{e}^{- \frac{\pi}{\lambda_{\beta}^{2}} \frac{\vert \bold{x}-\bold{y}\vert^2}{l}},
\!\!\!\!\!\!\!\!\!&\textrm{when $\nu<\nu_c(\beta)$ if $d\geq1$,\qquad} \label{Thm1Eq31}\\
\infty, &\textrm{when $\nu\geq\nu_{c}(\beta)$ if $d=2$},\label{Thm1Eq32}\\
\frac{1}{\lambda_{\beta}^{3}} \sum_{l=1}^{\infty} \frac{1}{l^{\frac{3}{2}}}
\mathrm{e}^{- \frac{\pi}{\lambda_{\beta}^{2}} \frac{\vert \bold{x}-\bold{y}\vert^2}{l}}, &\textrm{when $\nu>\nu_c(\beta)$ if $d=3$}.\label{Thm1Eq33}
\end{subnumcases}
\end{theorem}

\begin{remark}
\label{remthm1}
$\mathrm{(i)}$. In the case of $\nu<\nu_{c}(\beta)$ if $d=1,2,3$, then from \eqref{Thm1Eq2} along with \eqref{Thm1Eq31}, the contribution in \eqref{rdmd21} only comes from the reduced density matrix without the ground-state. We turn to the case of $\nu > \nu_{c}(\beta)$, $\nu/\nu_{c}= \eta > 1$ if $d=2,3$. By decomposing the sum involved in the reduced density matrix into two contributions (ground-state corresponding to the condensate gas plus the rest of the sum corresponding to the thermal gas), then from \eqref{Thm1Eq2}-\eqref{Thm1Eq1}:
\begin{equation*}
\lim_{\kappa \downarrow 0} \kappa^{\frac{d}{2}} \frac{\Psi_{\infty,\kappa}^{(\bold{0})}(\bold{x}) \Psi_{\infty,\kappa}^{(\bold{0})}(\bold{y})}{\mathrm{e}^{\beta\left(E_{\infty,\kappa}^{(\bold{0})} - \overline{\mu}_{\infty,\kappa}\right)} - 1} = \frac{2^{\frac{d}{2}}}{\lambda_{\beta}^{d}} \frac{\mathpzc{g}_{d}(1)}{\left(\hbar \omega_{0} \beta\right)^{\frac{d}{2}}} \left(\eta -1\right),
\end{equation*}
and in the case of $d=3$, from \eqref{Thm1Eq33}:
\begin{equation*}
\lim_{\kappa \downarrow 0} \sum_{\bold{s}\in (\mathbb{N}^{*})^{3}}\frac{\Psi_{\infty,\kappa}^{(\bold{s})}(\bold{x}) \Psi_{\infty,\kappa}^{(\bold{s})}(\bold{y})}{\mathrm{e}^{\beta\left(E_{\infty,\kappa}^{(\bold{s})}- \overline{\mu}_{\infty,\kappa}\right)}-1} = \frac{1}{\lambda_{\beta}^{3}} \sum_{l=1}^{\infty} \frac{1}{l^{\frac{3}{2}}}
\mathrm{e}^{- \frac{\pi}{\lambda_{\beta}^{2}} \frac{\vert \bold{x}-\bold{y}\vert^2}{l}}.
\end{equation*}
Therefore, the long range order is due to the condensate on the ground-state, the finite part of the reduced density matrix is due to the thermal gas. When $\nu \geq \nu_c(\beta)$ if $d=2$, the open-trap reduced density matrix diverges, even if the gas manifests no BEC when $\nu=\nu_{c}(\beta)$. This arises from the divergence of the non-condensate part of the open-trap reduced density matrix, see \eqref{Thm1Eq32}. In Annex \ref{AppendixC}, we investigate its behavior when $\kappa \downarrow 0$ and prove that:
\begin{equation*}
\sum_{\bold{s}\in (\mathbb{N}^{*})^{2}}\frac{\Psi_{\infty,\kappa}^{(\bold{s})}(\bold{x}) \Psi_{\infty,\kappa}^{(\bold{s})}(\bold{y})}{\mathrm{e}^{\beta\left(E_{\infty,\kappa}^{(\bold{s})}- \overline{\mu}_{\infty,\kappa}\right)}-1} \sim \frac{1}{\lambda_{\beta}^{2}} \ln\left(\frac{1}{\hbar \omega_{0} \kappa \beta}\right),\quad \textrm{when $\kappa \downarrow 0$}.
\end{equation*}
$\mathrm{(ii)}$. As a result of $\mathrm{(B)}$, there is equivalence between Definitions \ref{OpenTrapBECdef} and \ref{OpenTrapODLROdef}: the Bose gas manifests an open-trap BEC if and only if it manifests an open-trap ODLRO.
\end{remark}

\begin{remark}
\label{Remloop}
The sum-decomposition in the proof of Theorem \ref{Thm1} brings out that the condensate part comes from the macroscopic-loops, i.e., $l > \lfloor \kappa^{-\sigma} \rfloor$ with $1< \sigma <\frac{3}{2}$ and $\kappa < 1$, whereas the non-condensate part comes from the short-loops, i.e., $1 \leq l \leq \lfloor \kappa^{-\sigma} \rfloor$. Here $\lfloor \cdot\,\rfloor$ is the floor function. Note that in our sum-decomposition, the $\kappa$ plays the role of $\hbar \omega_{0} \beta$ in the loop-gas approach in \cite{Mullin1}. It is found in \cite{Mullin1} that the condensate-part comes from the large-loops, i.e., $l > (\hbar \omega \beta)^{-1}$. In fact, from our analysis the critical exponent turns out to be $\sigma = \frac{3}{2}$.
\end{remark}

\subsubsection{The local properties (Part 2)--Localization of the condensate/thermal gas.}
\label{p4}

Here, we focus on the diagonal part of the (rescaled) reduced density matrix,
interpreted as the (rescaled) local density of particles, in open-trap limit.
By introducing a scaling of the spatial variable, initially introduced by van den Berg \textit{et al.} in \cite{vdBLP} to derive the so-called \textit{barometric formula}, we state some results concerning the spatial localization of the condensate/thermal gas in open-trap limit. We can relate our statements with some well-known results in Physics literature concerning the shape of the condensate/thermal gas in the space. This is discussed in Sec. \ref{partie3}.

\begin{theorem}
\label{Thm2}
Consider a confined $d$-dimensional harmonically trapped Bose gas, $d \in \{1,2,3\}$ in the G-C ensemble. Let $\nu_{c}(\beta)$ be the critical density of particles in \eqref{defnuc} satisfying \eqref{nucEq}. Then $\forall d\in\{1,2,3\}$, $\forall \beta>0$, $\forall \nu>0$, $\forall \bold{x} \in (\mathbb{R}^{*})^{d}$ and
$\forall 0 \leq \delta \leq 1$ the two following limits exist:
\begin{equation*}
\rho_{\infty,0}^{(\delta)}(\bold{x};\beta,\nu) := \lim_{\kappa \downarrow 0} \rho_{\infty,\kappa}\left(\bold{x} \kappa^{-\delta},\bold{x} \kappa^{-\delta};\beta,\nu\right),\quad r_{\infty,0}^{(\delta)}(\bold{x};\beta,\nu) := \lim_{\kappa \downarrow 0} r_{\infty,\kappa}\left(\bold{x} \kappa^{-\delta},\bold{x} \kappa^{-\delta};\beta,\nu\right).
\end{equation*}
$\mathrm{(A)}$. With $\mathpzc{g}_{\theta}$ in \eqref{polylog}, $\lambda_{\beta}$ in \eqref{deBrog}  and $\overline{\mu}_{\infty,0}= \overline{\mu}_{\infty,0}(\beta,\nu)$ obeying \eqref{muinf0}, one has more precisely:
\begin{itemize}
\item $\forall d \in \{1,2,3\}$ and $\forall \nu<\nu_{c}(\beta)$:
\end{itemize}
\begin{subnumcases}{\label{dilat1}
\rho_{\infty,0}^{(\delta)}(\bold{x};\beta,\nu)= \frac{1}{\lambda_{\beta}^{d}}\times}
\mathpzc{g}_{\frac{d}{2}}\left(\mathrm{e}^{\beta \overline{\mu}_{\infty,0}}\right),&\textrm{if $0\leq \delta<1$}, \label{dilat11}\\
\mathpzc{g}_{\frac{d}{2}}\left(\mathrm{e}^{\beta \left(\overline{\mu}_{\infty,0}- \frac{1}{2} m \omega_{0}^{2} \vert \bold{x}\vert^{2}\right)}\right),&\textrm{if $\delta=1$}. \label{dilat12}
\end{subnumcases}
\begin{equation}
\label{dilat2}
r_{\infty,0}^{(\delta)}(\bold{x};\beta,\nu)= 0, \quad \textrm{if $\delta \geq 0$}.
\end{equation}
\begin{itemize}
\item $\forall d \in \{2,3\}$ and $\forall \nu \geq \nu_{c}(\beta)$:
\end{itemize}
\begin{subnumcases}{\label{dilat3}\textrm{If $d=2$},\quad \rho_{\infty,0}^{(\delta)}(\bold{x};\beta,\nu) = \frac{1}{\lambda_{\beta}^{2}}\times}
\infty,&\textrm{if\,\, $0\leq \delta < 1$,} \label{dilat31}\\
\mathpzc{g}_{1}\left(\mathrm{e}^{- \beta \frac{m\omega_{0}^{2}}{2} \vert \bold{x}\vert^{2}}\right),&\textrm{if\,\, $\delta=1$}.\label{dilat32}
\end{subnumcases}
\begin{subnumcases}{\label{dilat3'}\textrm{If $d=3$},\quad \rho_{\infty,0}^{(\delta)}(\bold{x};\beta,\nu) = \frac{1}{\lambda_{\beta}^{3}} \times}
\infty,&\textrm{if\,\, $0\leq \delta \leq \frac{1}{2}$ and $\nu \neq \nu_{c}(\beta)$,\qquad}\label{dilat3'1}\\
\mathpzc{g}_{\frac{3}{2}}(1),&\textrm{if\,\, $\frac{1}{2}<\delta < 1$},\label{dilat3'2}\\
\mathpzc{g}_{\frac{3}{2}}\left(\mathrm{e}^{- \beta \frac{m\omega_{0}^{2}}{2} \vert \bold{x}\vert^{2}}\right),\!\!\!\!\!\!\!\!\!&\textrm{if\,\, $\delta=1$}.\label{dilat3'3}
\end{subnumcases}
\begin{subnumcases}{\label{dilat4} r_{\infty,0}^{(\delta)}(\bold{x};\beta,\nu) = 2^{\frac{d}{2}} \frac{\left(\hbar \omega_{0} \beta\right)^{\frac{d}{2}}}{\lambda_{\beta}^{d}} \left(\nu-\nu_{c}(\beta)\right)\times}
1,&\textrm{if\,\,$0\leq \delta <\frac{1}{2}$},\label{dilat41}\\
\mathrm{e}^{- \frac{m \omega_{0}}{\hbar} \vert \bold{x}\vert^{2}},&\textrm{if\,\, $\delta =\frac{1}{2}$},\label{dilat42}\\
0, &\textrm{if\,\,$\frac{1}{2}<\delta \leq 1$}.\label{dilat43}
\end{subnumcases}
$\mathrm{(B)}$. In addition: the open-trap rescaled local density of particles in the condensate satisfies:
\begin{equation}
\label{Thm2BEq1}
\forall 0\leq\delta\leq 1,\quad r_{\infty,0}^{(\delta)}(\bold{x};\beta,\nu)= \lim_{\kappa\downarrow0}
\nu_{\infty,\kappa}(\beta,\nu;\bold{0}) \left\vert \Psi_{\infty,1}^{(\bold{0})}\left(\bold{x}\kappa^{-\delta}\sqrt{\kappa}\right)\right\vert^{2},\quad \textrm{if $d=1,2,3$},
\end{equation}
as for the open-trap local density of particles outside of the condensate:
\begin{equation*}
\lim_{\kappa\downarrow0}\sum_{\bold{s} \in (\mathbb{N}^{*})^{d}} \frac{\left\vert \Psi_{\infty,\kappa}^{(\bold{s})}\left(\bold{x}\kappa^{-\delta}\right)\right\vert^2}
{\mathrm{e}^{\beta\left(E_{\infty,\kappa}^{(\bold{s})}-\overline{\mu}_{\infty,\kappa}\right)}-1} =
\end{equation*}
\begin{subnumcases}{\label{Thm2BEq2}}
\rho_{\infty,0}^{(\delta)}(\bold{x};\beta,\nu),
&\textrm{when $\nu<\nu_c(\beta)$ if $0\leq \delta\leq1$ and $d=1,2,3$}, \label{Thm2BEq21}\\
\infty,
&\textrm{when $\nu\geq\nu_c(\beta)$ if $0\leq \delta\leq1$ and $d=2$}, \label{Thm2BEq22}\\
\frac{1}{\lambda_{\beta}^{3}} \mathpzc{g}_{\frac{3}{2}}(1),
&\textrm{when $\nu>\nu_c(\beta)$ if $0\leq \delta\leq \frac{1}{2}$ and $d=3$}, \label{Thm2BEq23}\\
\frac{1}{\lambda_{\beta}^{3}} \mathpzc{g}_{\frac{3}{2}}\left(\mathrm{e}^{- \beta \frac{m\omega_{0}^{2}}{2} \vert \bold{x}\vert^{2}}\right),
&\textrm{when $\nu\geq\nu_c(\beta)$ if $\frac{1}{2}<\delta\leq 1$ and $d=3$}. \label{Thm2BEq24}
\end{subnumcases}
\end{theorem}

\begin{remark}
We restricted to $\bold{x} \in (\mathbb{R}^{*})^{d}$ since the case of $\bold{x}=\bold{0}$
is covered by Theorem \ref{Thm1}.
\end{remark}

\begin{remark}
\label{remthm2}
The meaning of Theorem \ref{Thm2} is discussed in the following section. Let us mention that, in the particular case of $\delta=1$, in view of \eqref{dilat32}-\eqref{dilat3'3} and by setting $V(\bold{x}):= m\omega_{0}^{2} \vert \bold{x}\vert^{2}/2$, then when $\nu \geq \nu_{c}(\beta)$ if $d=2,3$, the identity in \eqref{nu(beta,mu)} leads to the rewriting:
\begin{equation}
\label{sc}
\lim_{\kappa \downarrow 0} \rho_{\infty,\kappa}^{(\delta=1)}(\bold{x};\beta,\nu)  = \int_{\mathbb{R}^{d}}\frac{\mathrm{d}\bold{p}}{(2\pi\hbar)^d} \frac{1}{\mathrm{e}^{\beta \left(\frac{\vert\bold{p}\vert^2}{2m}+V(\bold{x})\right)}-1}.
\end{equation}
\eqref{sc} is often referred to as the semi-classical formula for the local density, see e.g.
\cite[Eqs. (10.25)-(10.27)]{Phys3}.
In Sec. \ref{methoddis}, we show that the open-trap limit of the reduced density matrix with spatial arguments rescaled by $\kappa^{-1}$ identifies with the leading term of the asymptotic expansion of the reduced density matrix in the semiclassical limit $\hbar \downarrow 0$ (here, $\hbar$ has to be seen as a parameter).
\end{remark}

\subsubsection{Meaning of Theorems \ref{Thm1} and \ref{Thm2}--Rebuilding the density profile.}

Here, we make the connection between the results of Theorem \ref{Thm1}-\ref{Thm2} and those stated in literature.
We consider the situation corresponding to the occurrence of BEC phenomenon. When $\nu > \nu_{c}(\beta)$, $\nu/\nu_{c}(\beta)= \eta>1$ if $d=2,3$, then from \eqref{dilat4} along with \eqref{Thm2BEq1}:
\begin{equation}
\label{apinfo1}
\lim_{\kappa \downarrow 0} \kappa^{\frac{d}{2}} \frac{\left\vert \Psi_{\infty,\kappa}^{(\bold{0})}(\bold{x} \kappa^{-\delta})\right\vert^{2}}{\mathrm{e}^{\beta\left(E_{\infty,\kappa}^{(\bold{0})} - \overline{\mu}_{\infty,\kappa}\right)} - 1} = \frac{2^{\frac{d}{2}}}{\lambda_{\beta}^{d}} \frac{\mathpzc{g}_{d}(1)}{\left(\hbar \omega_{0} \beta\right)^{\frac{d}{2}}} \left(\eta -1\right)\times \left\{\begin{array}{ll}
1, &\textrm{if $0 \leq \delta < \frac{1}{2}$},\\
\mathrm{e}^{-2 \left(\hbar \omega_{0} \beta\right) \frac{\pi}{\lambda_{\beta}^{2}} \vert \bold{x}\vert^{2}}, &\textrm{if $\delta = \frac{1}{2}$},\\
0, &\textrm{if $\frac{1}{2}<\delta \leq 1$,} \end{array}\right.,
\end{equation}
and concerning the non-condensate part, from \eqref{dilat2}, \eqref{dilat3} and \eqref{Thm2BEq2}:
\begin{equation}
\label{apinfo2}
\lim_{\kappa \downarrow 0} \sum_{\bold{s}\in (\mathbb{N}^{*})^{d}}\frac{\left\vert \Psi_{\infty,\kappa}^{(\bold{s})}(\bold{x} \kappa^{-\delta})\right\vert^{2}}{\mathrm{e}^{\beta\left(E_{\infty,\kappa}^{(\bold{s})}- \overline{\mu}_{\infty,\kappa}\right)}-1} = \frac{1}{\lambda_{\beta}^{d}} \times \left\{\begin{array}{ll}
\mathpzc{g}_{\frac{d}{2}}\left(1\right), &\textrm{if $0 \leq \delta <1$ and $d\neq 2$},\\
\mathpzc{g}_{\frac{d}{2}}\left(\mathrm{e}^{-\beta \frac{m \omega_{0}^{2}}{2} \vert \bold{x}\vert^{2}}\right), &\textrm{if $\delta=1$,}
\end{array}\right..
\end{equation}
Note that Theorem \ref{Thm2} gives no indication on the behavior of the non-condensate part of the reduced density matrix when $d=2$ if $0<\delta < 1$ (the case of $\delta=0$ is discussed in Remark \ref{remthm1}).\\
In the light of \eqref{apinfo1}-\eqref{apinfo2}, the first term in the r.h.s. of \eqref{Intro1} corresponds to \eqref{apinfo1} with $\delta=\frac{1}{2}$, and the second term in the r.h.s. of \eqref{Intro1} corresponds to \eqref{apinfo2} with $\delta=1$, see Remark \ref{remthm2}.\\
\indent From \eqref{apinfo1}-\eqref{apinfo2}, one can infer a range of information on the localization of the condensate and non-condensate gas (thermal gas). Clearly, the localization ranges of the thermal gas and condensate are not the same. Indeed, one can see that the \textit{length scale} of the localization of the thermal gas is $L_\beta:=(m\omega_0^2\kappa^2\beta)^{-\frac{1}{2}}=\mathcal{O}(\kappa^{-1})$ (corresponding to $\delta=1$), whereas the length scale of the condensate is $L_\hbar:=(\hbar/m\omega_0\kappa)^{\frac{1}{2}}=\mathcal{O}(\kappa^{-\frac{1}{2}})$ (corresponding to $\delta=1/2$). This means that the condensate and the thermal gas do not coexist at the same scale of spatial distances, as it is stated in \cite[Eq. (10.28)]{Phys3}. The $\delta$-scaling can be interpreted as follows. When $\delta=1$, \eqref{apinfo2} gives the density profile of the thermal gas at large scale in the units of $L_\beta$ and \eqref{apinfo1} shows that there is a peak for $x=0$ corresponding to the condensate. 
When $\delta=1/2$,
\eqref{apinfo1} gives the density profile of the condensate in the units of $L_\hbar$ and \eqref{apinfo2} shows that there is a plateau corresponding to the thermal gas. This latter means that the thermal gas is viewed as a constant for $\delta<1$, i.e., at scales very much smaller than $L_\beta$. The intermediate cases $0\leq\delta<1/2$ and $1/2<\delta<1$ follow a similar interpretation. By the scaling $x\mapsto x\kappa^{-\delta}$ with $0\leq\delta\leq1$, we investigate in fact the density profile of the condensate and thermal gas at the length scale $L(\delta):=L_\hbar \left(L_\beta/L_\hbar\right)^{2\delta-1}=L_\beta \left(L_\hbar/L_\beta\right)^{2-2\delta}$. Since $L_\hbar/L_\beta=\hbar\beta\omega_0\kappa\ll1$ (the semiclassical regime), we conclude that $L(\delta)\ll L_\hbar$ if $\delta < 1/2$ and $L_\hbar\ll L(\delta)\ll L_\beta$ if $1/2 < \delta <1$. This explains why the condensate part vanishes for $\delta>1/2$, and also why the non-condensate part remains constant for $\delta<1$.
In addition, since $L_\hbar^2/L_\beta=\lambda_\beta/\sqrt{2\pi}=\sqrt{\hbar^2\beta/m}$ does not depend on $\kappa$, we have $L(\delta)=\lambda_\beta\times (2\pi)^{2\delta-1}(L_\hbar/\lambda_\beta)^{2\delta}$. The latter relation gives the $\kappa$-dependence of $L(\delta)$. Note that $L(0)=(2\pi)^{2\delta-1}\lambda_\beta\ll L_\hbar$.\\
\indent Turning to a more geometric interpretation, define $\forall d \in \{2,3\}$, $\forall \beta>0$ and $\forall \nu>\nu_{c}(\beta)$ the large scale (i.e., $\delta=1$) average square radius in the $j$-th direction of the open-trap reduced local density function as:
\begin{equation}
\label{avsquden}
\left\langle x_{j}^{2}\right\rangle_{\infty,0}^{(T)}(\beta,\nu):=\frac{\displaystyle{\int_{\mathbb{R}^{d}}\mathrm{d}\bold{x}\,
x_j^2\, \rho^{(\delta=1)}_{\infty,0}(\bold{x};\beta,\nu)}}
{\displaystyle{\int_{\mathbb{R}^d}\mathrm{d}\bold{x}\, \rho^{(\delta=1)}_{\infty,0}(\bold{x};\beta,\nu)}},\quad j=1,\ldots,d.
\end{equation}
Similarly, define under the same conditions the medium scale (i.e., $\delta=1/2$) average square radius in the $j$-th direction of the open-trap rescaled reduced local density function as:
\begin{equation}
\label{avsquden2}
\left\langle x_{j}^{2}\right\rangle_{\infty,0}^{(0)}(\beta,\nu)
:=\frac{\displaystyle{\int_{\mathbb{R}^d} \mathrm{d}\bold{x}\, x_{j}^{2}\ r^{(\delta=\frac{1}{2})}_{\infty,0}(\bold{x};\beta,\nu)}}
{\displaystyle{\int_{\mathbb{R}^{d}} \mathrm{d}\bold{x}\, r^{(\delta=\frac{1}{2})}_{\infty,0}(\bold{x};\beta,\nu)}},\quad j=1,\ldots,d.
\end{equation}
We mention that from \eqref{dilat32}-\eqref{dilat3'3} and \eqref{dilat42} the quantities in \eqref{avsquden}-\eqref{avsquden2} are well-defined. By a direct calculation, we get $\forall d \in \{2,3\}$, $\forall \beta>0$ and $\forall \nu>\nu_{c}(\beta)$ the following ratio:
\begin{equation}
\label{ratio}
\frac{\left\langle x_j^2\right\rangle_{\infty,0}^{(T)}(\beta,\nu)}
{\left\langle x_j^2\right\rangle_{\infty,0}^{(0)}(\beta,\nu)}= \frac{2}{\hbar \omega_{0} \beta} \frac{\mathpzc{g}_{d+1}(1)}{\beta \mathpzc{g}_{d}(1)}= \frac{2}{\hbar \omega_{0} \beta} \frac{\zeta(d+1)}{\beta\zeta(d)},\quad j=1,\ldots,d,
\end{equation}
The r.h.s. of \eqref{ratio} is $j$-independent since the trap is isotropic, and corresponds to
\cite[Eq. (10.28)]{Phys3} for $d=3$. The meaning of \eqref{ratio} is as follows: the density profile of the thermal gas is much more spread out than the density profile of the condensate (large vs medium scale).\\
\indent As a final remark, the local density of particles in the condensate is of the order of $\kappa^{-\frac{3}{2}}$ for $d=3$, see \eqref{apinfo1}. Ergo, it is infinite in open-trap limit, whereas the local density of particles in the thermal gas is finite in open-trap limit, see \eqref{apinfo2}. Hence, one can talk about a \textit{spatial Bose-Einstein condensation} since a very large number of particles is localized in a small region of the space compared with the region where the thermal gas is spread. Note that the first experimental demonstrations of the condensate is based on this latter feature since it is enough to 'take pictures' of the gas to bring out the spatial density of the particles distribution, see e.g. \cite{C-al,K-al}.

\section{Concluding remarks \& Extension to anisotropic traps.}
\label{partie3}

\subsection{Open-trap limit vs semiclassical regime.}
\label{methoddis}


\subsubsection*{The thermodynamics: open-trap limit vs semiclassical regime.}

The usual method to investigate the thermodynamic functions of the Bose gas from \eqref{grandpotentialTL} consists in approximating to the first-order the sum over the $\bold{s}$-index by an integral. Turning to the average number of particles, one has from \eqref{averagenumTL} along with \eqref{trace} and \eqref{nu(beta,mu)}, by setting $\kappa=1$:
\begin{equation*}
\overline{N}_{\infty,1}(\beta,\mu) =  \int_{0}^{\infty} \mathrm{d}\tau\, \frac{\tau^{d-1}}{\Gamma(d) \left(\hbar \omega_{0}\right)^{d}} \mathfrak{f}_{BE}(\beta,\mu;\tau) + \sum_{l=1}^{\infty} \mathrm{e}^{l\beta \mu} \left( \left(2 \sinh\left(\frac{\hbar \omega_{0} \beta}{2} l\right)\right)^{-d} - \left(\hbar \omega_{0} \beta l\right)^{-d}\right),
\end{equation*}
and the remainder behaves like $\mathcal{O}((\hbar \omega_{0} \beta)^{1-d}))$ when $(\hbar \omega_{0} \beta) \downarrow 0$. Its behavior is investigated further into details in \cite{2HR,KT1,KT2} in which the relevance of the semiclassical regime is discussed. In view of the leading term in the above expansion, then the density of states in the semiclassical regime is approximated to the first-order by its high-energy asymptotic. Such a result is recovered in our open-trap formulation mimicking the regime of weak angular frequencies $\omega_{0}$ of the trap. For completeness' sake, we mention that the leading term can also be derived in the 'true' semiclassical regime corresponding to small values of $\hbar$ (seen as a parameter). Indeed, consider the operator:
\begin{equation}
\label{tildHinfini}
\tilde{H}_{\infty,\kappa} := -\frac{(\hbar \kappa)^{2}}{2m} \Delta + \frac{m}{2} \omega_{0}^{2} \vert \bold{x}\vert^{2}, \quad \kappa>0.
\end{equation}
Under the unitary transformation on $L^{2}(\mathbb{R}^{d})$ defined as:
\begin{equation*}
\forall \kappa >0,\quad (\mathcal{U}(\kappa)\varphi)(\bold{x}) := \kappa^{-\frac{d}{2}} \varphi\left(\bold{x}\kappa^{-1}\right),\quad \bold{x} \in \mathbb{R}^{d},\, \varphi \in L^{2}(\mathbb{R}^{d}),
\end{equation*}
then $H_{\infty,\kappa}$ in \eqref{Hinfini} is unitary equivalent to $\tilde{H}_{\infty,\kappa}$ in \eqref{tildHinfini}, i.e.,
\begin{equation}
\label{unitequiv}
\mathcal{U}(\kappa) H_{\infty,\kappa} \mathcal{U}^{*}(\kappa) = \tilde{H}_{\infty,\kappa}.
\end{equation}
Denoting by $\{\tilde{G}_{\infty,\kappa}(t)\}_{t \geq 0}$ the strongly-continuous semigroup generated by $\tilde{H}_{\infty,\kappa}$, \eqref{unitequiv} leads to:
\begin{equation*}
\mathrm{Tr}_{L^{2}(\mathbb{R}^{d})}\left\{\tilde{G}_{\infty,\kappa}(\beta)\right\} = \mathrm{Tr}_{L^{2}(\mathbb{R}^{d})}\left\{G_{\infty,\kappa}(\beta)\right\},\quad \beta>0,\, \kappa>0.
\end{equation*}
In view of \eqref{lthaveragenum2}, then we obtain the same result than \eqref{nu(beta,mu)} when performing the limit $\kappa \downarrow 0$.

\subsubsection*{Reduced density matrix: Open-trap limit and semiclassical limit.}

The investigations essentially lean on the representation \eqref{otrep} of the reduced density matrix by the kernel of the Bose-Einstein function of the operator $H_{\infty,\kappa}$ in \eqref{fBE}, see \eqref{otrep}. From such a representation, investigating the behavior in open-trap limit of the (rescaled) reduced density matrix requires some sharp estimates on the kernel of the semigroup generated by $H_{\infty,\kappa}$ for small values of $\kappa$. Since this kernel is explicitly known (see \eqref{Mehler}-\eqref{multd}), then our approach turns out to be more robust than the one based on the representation in \eqref{rdm} involving the eigenfunctions of $H_{\infty,\kappa}$. Indeed, the control of the sum in \eqref{rdm} for small values of $\kappa$ is made difficult by the behavior of the Hermite polynomials which oscillate especially as the $\bold{s}$-index gets larger.\\
\indent In Theorem \ref{Thm1}, we investigate the reduced density matrix in open-trap limit. This allows us to derive the first-order approximation in the zero angular frequency limit. The semiclassical limit corresponding to small values of $\hbar$ (seen as a parameter) is investigated in Theorem \ref{Thm2} when using the scale $\delta=1$. From the definition in \eqref{fBE}, \eqref{unitequiv} leads in the kernels sense on $\mathbb{R}^{2d}$ to:
\begin{equation*}
\forall \kappa>0,\quad (\mathfrak{f}_{BE}(\beta,\overline{\mu}_{\infty,\kappa};\tilde{H}_{\infty,\kappa}))(\bold{x},\bold{y}) = \kappa^{-d} (\mathfrak{f}_{BE}(\beta,\overline{\mu}_{\infty,\kappa}; H_{\infty,\kappa}))(\bold{x}\kappa^{-1},\bold{y}\kappa^{-1}).
\end{equation*}
Then, it follows from the definition in \eqref{rdm} together with \eqref{sc}:
\begin{equation}
\label{tenddsto}
(\mathfrak{f}_{BE}(\beta,\overline{\mu}_{\infty,\kappa};\tilde{H}_{\infty,\kappa}))(\bold{x},\bold{y}) \sim \int_{\mathbb{R}^{d}}\frac{\mathrm{d}\bold{p}}{(2\pi \hbar \kappa)^d} \frac{1}{\mathrm{e}^{\beta(\frac{\vert\bold{p}\vert^2}{2m}+V(\bold{x}))}-1} \quad \textrm{when $\kappa \downarrow 0$}.
\end{equation}

\subsection{Homogeneous versus inhomogeneous systems.}

We stress the point that our results obtained in open-trap limit for the harmonically trapped Bose gas differ from the ones stated in \cite{LP,vdBL1} for the perfect Bose gas confined in 'Dirichlet boxes' (commonly referred to as 'homogeneous systems'), and from the ones in \cite{vdB, vdBL, Pule} stated for the free Bose gas in a weak harmonic trap model. The main difference concerns the critical density. From a rescaling of the average number of particles (see Definition \ref{defrescale}), we find that the open-trap critical rescaled average number of particles for the harmonically trapped Bose gas is finite if $d=2,3$ and proportional to $\mathpzc{g}_{d}(1)$, see Lemma \ref{Lemma0} $\mathrm{(ii)}$. This result contrasts with the case of homogeneous systems in which the bulk critical density of particles (by bulk, we mean in thermodynamic limit) is finite  if $d=3$ and proportional to $\mathpzc{g}_{\frac{d}{2}}(1)$, and also with the case of the weak harmonic trap model in which it is finite if $d=2,3$ and
proportional to $\int_0^1 du\, \mathpzc{g}_{\frac{d}{2}}(\mathrm{e}^{-\beta\frac{u^2}{2}})$.\\
\indent However, we mention that our expressions for the non-condensate part of the open-trap reduced density matrix for the harmonically trapped Bose gas in Theorem \ref{Thm1} $\mathrm{(C)}$ are exactly the same than the ones for the non-condensate part of the bulk reduced density matrix for homogeneous systems in any dimension, and thus it diverges if $d=2$ and converges if $d=3$. Even at the scales $0<\delta<1$, the non-condensate part of the open-trap scaled local density in Theorem \ref{Thm2} $\mathrm{(B)}$ is still equal to the non-condensate part of the bulk local density for homogeneous systems. This means that the non-condensate bosons do not feel the trap for those scales, and behave like free particles in the whole space $\mathbb{R}^d$. But at the scale $\delta=1$ and when $\nu \geq \nu_{c}(\beta)$ if $d=2,3$, the open-trap scaled local density has a gaussian decay whereas the bulk scaled local density (with the scaling $\bold{u}:=\bold{x}L^{-1}$, $\bold{u}\in[-1,1]^d$) for homogeneous systems is
constant on $(-1,1)^d$ and vanishes on the boundaries, see e.g.  \cite{vdBLL}. Concerning the condensate part of the open-trap rescaled density function for the harmonically trapped Bose gas, it is constant at the scales $0 \leq \delta<1/2$ and has a gaussian decay at the scale $\delta=1/2$ when $\nu \geq \nu_{c}(\beta)$ if $d=2,3$, see Theorem \ref{Thm2} $\mathrm{(A)}$, whereas the condensate part of the bulk scaled density function for homogenous systems oscillates on $[-1,1]^d$.

\subsection{Insight into BEC in some anisotropic harmonic traps.}

Here, we extend some of the results established for the isotropic harmonic trap to some models of three-dimensional anisotropic harmonic traps. In particular, we focus on two kind of anisotropic trap models: a \textit{quasi-1D} and \textit{quasi-2D} trap model. They are analogous to the anisotropic (van den Berg) boxes models for the homogeneous Bose gas investigated in \cite{vdBL1,vdB1,vdBLP,vdBLL}.\\
\indent The infinite-volume Hamiltonian in $L^{2}(\mathbb{R}^{3})$ which determines the dynamics of a single spin-0 particle trapped in a general three-dimensional anisotropic harmonic trap is given by:
\begin{equation}
\label{Hk3D}
H_{\infty,\underline{\kappa}}:=\frac{1}{2}\left(-i\nabla_{\bold{x}}\right)^2
+\frac{1}{2} \sum_{j=1}^{3} \left(\omega_{j} \kappa_{j}\right)^{2} x_{j}^{2},
\end{equation}
where $\omega_{j}>0$, $j=1,2,3$ are kept fixed in the following, and $\underline{\kappa}=(\kappa_1,\kappa_2,\kappa_3)$, $\kappa_{j}>0$. Below, we focus on some specific anisotropic traps. In particular, we consider the situation in which $\omega_{2}=\omega_{3} = \omega_{\perp}$ and $\omega_{1} \neq \omega_{\perp}$. Further, for any $\kappa>0$, we consider $\kappa_1=\kappa_1(\kappa)$ and $\kappa_2=\kappa_3=\kappa_{\perp}(\kappa)$ which satisfy $\kappa_1(\kappa)\neq\kappa_{\perp}(\kappa)$ $\forall \kappa>0$ along with $\kappa_{1}(\kappa),\kappa_{\perp}(\kappa)\downarrow0$ when $\kappa\downarrow0$. Since the $\kappa_{j}$'s are functions of $\kappa$, then in the following, for such models, we set $H_{\infty,\kappa}=H_{\infty,\underline{\kappa}}$. From \eqref{vapd} and \eqref{funcpd}, the eigenvalues and eigenfunctions are respectively given for any $\kappa>0$ by:
\begin{gather*}
E_{\infty,\kappa}^{(\bold{s})} = \hbar \omega_{1} \kappa_{1}\left(s_{1}+\frac{1}{2}\right) + \hbar \omega_{\perp} \kappa_{\perp} \sum_{j=2}^{3} \left( s_{j} + \frac{1}{2}\right),\quad \bold{s}=\{s_{j}\}_{j=1}^{3} \in \mathbb{N}^{3},\\
\Psi_{\infty,\kappa}^{(\bold{s})}(\bold{x}) = \psi_{\kappa_{1}}^{(s_{1})}(x_{1}) \prod_{j=2}^{3} \psi_{\infty,\kappa_{\perp}}^{(s_{j})}(x_{j}),\quad \bold{x} = \{x_{j}\}_{j=1}^{3} \in \mathbb{R}^{3}.
\end{gather*}
Next, $\forall \kappa>0$ denote $\vert\underline{\kappa}\vert:=(\kappa_1\kappa_2\kappa_3)^{1/3}= (\kappa_{1} \kappa_{\perp}^{2})^{\frac{1}{3}}$ and $\omega_{0} = (\omega_{1} \omega_{2}\omega_{3})^{\frac{1}{3}}=(\omega_{1} \omega_{\perp}^{2})^{\frac{1}{3}}$. Then, one has $\vert\underline{\kappa}\vert^{3/2}=\sqrt{\kappa_1}\kappa_\perp$ and $\vert\underline{\kappa}\vert^{3}=\kappa_1\kappa_\perp^{2}$. To take into account the anisotropy of the harmonic trap, the rescaling by $\kappa^{3}$ in Definitions \ref{defrescale}-\ref{def3} (resp. $\kappa^{\frac{3}{2}}$ in Definition \ref{def6}) when $d=3$ has to be replace with $\vert\underline{\kappa}\vert^{3}$ (resp.  $\vert\underline{\kappa}\vert^{3/2}$). Therefore,
similarly to \eqref{rdm}-\eqref{otrep} and \eqref{rdmrescaled}, the reduced density matrix and the rescaled reduced density matrix read $\forall \kappa>0$, $\forall \beta>0$, $\forall \nu>0$ and $\forall (\bold{x},\bold{y}) \in \mathbb{R}^{6}$ as:
\begin{gather}
\label{rhoAnisa}
\rho_{\infty,\kappa}(\bold{x},\bold{y};\beta,\nu)
=\sum_{l=1}^{\infty}\mathrm{e}^{l\beta\overline{\mu}_{\infty,\kappa}} G_{\infty,\kappa_1}^{(1)}(x_1,y_1;l\beta) G_{\infty,\kappa_{\perp}}^{(2)}(\bold{x}_\perp,\bold{y}_\perp;l\beta),\\
\label{rAnisa}
r_{\infty,\kappa}(\bold{x},\bold{y};\beta,\nu)
=\vert\underline{\kappa}\vert^{3/2} \rho_{\infty,\kappa}(\bold{x},\bold{y};\beta,\nu),
\end{gather}
where $\bold{x}_\perp:=(x_2,x_3)$, $\bold{y}_{\perp}:=(y_{2},y_{3})$ and
$\overline{\mu}_{\infty,\kappa}:=\overline{\mu}_{\infty,\kappa}(\beta,\nu) < E_{\infty,\kappa}^{(\bold{0})}$ satisfies similarly to \eqref{musolEq}:
\begin{equation}
\label{reuniq}
\nu=\nu_{\infty,\kappa}(\beta,\mu) = \vert \underline{\kappa}\vert^{3} \overline{N}_{\infty,\kappa}(\beta,\mu).
\end{equation}
When dealing with the above bulk quantities in open-trap limit, Definitions \ref{def2}-\ref{defd} still hold for the type of anisotropic harmonic traps we consider here (the $\kappa_{j}$'s are functions of $\kappa$). We stress the point that the results of Lemma \ref{Lemma0} still hold true.
In the rest of this section, our purpose consists in stating the counterpart of Propositions \ref{Prop1}-\ref{Prop1'} and Theorem \ref{Thm1} in the case of a quasi-1D and quasi-2D trap model (we do not consider the counterpart of Theorem \ref{Thm2}).

\subsubsection{A Quasi-1D trap model.}
\label{quasi1D}

A quasi-1D trap model was first introduced in \cite{BZ} adapting the exponentially anisotropic (van den Berg) boxes model for the homogeneous Bose gas studied in \cite{vdB}. It has been reviewed in \cite{Mullin}. It is a three-dimensional anisotropic trap model defined (in accordance with our formalism) as:
\begin{equation}
\label{kappaQ1Da}
\left\{ \begin{array}{ll}
\kappa_1&=\kappa \exp{\left(-\frac{\kappa_c^2}{\kappa^2}\right)},\quad \kappa_{c}>0, \\
\kappa_\perp&=\kappa\,(=\kappa_{2}=\kappa_{3}).
\end{array}\right.
\end{equation}
From \eqref{kappaQ1Da}, for small values of $\kappa$, the characteristic length $\sqrt{\hbar (m\omega_{1} \kappa_{1})^{-1}}$ along the $x_{1}$-direction is very large compared to the one along the $x_j$-directions, $j=2,3$ (hence the name of quasi-1D trap). For this kind of quasi 1-D trap model, Beau \textit{at al.} pointed out in \cite{BZ} (see also \cite{Mullin}) that the Bose gas can manifests both BEC and generalized-BEC in a suitable regime corresponding to a second kind of transition. Since such a model produces some very different results compared to the isotropic harmonic trap, this justifies its relevance. Let us go further into details.\\
\indent The counterpart of Propositions \ref{Prop1}-\ref{Prop1'} is contained in:

\begin{proposition}
\label{Prop1Q1D}
Consider a quasi-1D harmonically trapped Bose gas (the anisotropy is defined by \eqref{kappaQ1Da}), in the G-C ensemble. Then, for any $\beta>0$, $\nu>0$ and $\kappa_c>0$:\\
$\mathrm{(i)}$. The Bose gas manifests an open-trap BEC in the sense of Definition \ref{OpenTrapBECdef}. Furthermore, the open-trap rescaled average number of particles on the ground-state satisfies:
\begin{subnumcases}
{\label{nu00Q1D} \nu_{\infty,0}(\beta,\nu;\bold{0})=}
0,&\textrm{when $\nu<\nu_m(\beta)$}, \label{nu001Q1D}\\
\nu-\nu_m(\beta),&\textrm{when $\nu\geq\nu_m(\beta)$}, \label{nu002Q1D}
\end{subnumcases}
where $\nu_{m}(\beta)$ stands for a second critical open-trap rescaled average number of particles defined as:
\begin{equation}
\label{numaQ1D}
\nu_m(\beta) = \nu_{m}(\beta,\kappa_{c}) :=\nu_c(\beta)+\frac{\omega_{c}^{2}}{\hbar \beta \omega_{0}^{3}} > \nu_{c}(\beta),
\end{equation}
where $\omega_{0} = (\omega_{1} \omega_{\perp}^{2})^{\frac{1}{3}}$ and $\omega_{c} := \omega_{\perp} \kappa_{c}$. We recall that $\nu_{c}(\beta) = \mathpzc{g}_{3}(1) (\hbar \beta \omega_{0})^{-3}$ is the critical open-trap rescaled average number of particles in \eqref{defnuc} obeying \eqref{nucEq2}.\\
$\mathrm{(ii)}$. The Bose gas manifests an open-trap generalized-BEC in the sense of Definition \ref{OpenTrapgBECdef}. Moreover, the open-trap rescaled average number of particles nearby the ground-state satisfies:
\begin{subnumcases}{\label{nu0Ga}
\lim_{\varepsilon \downarrow 0} \lim_{\kappa \downarrow 0} \sum_{\bold{s} \in(\mathbb{N}^{*})^{3}\,:\, \sum_{j=1}^{3} \kappa_{j} s_{j} \leq \varepsilon} \nu_{\infty,\kappa}(\beta,\nu;\bold{s})=}
0,&\textrm{when $\nu<\nu_c(\beta)$}, \label{nu0G1a}\\
\nu-\nu_c(\beta), &\textrm{when $\nu_c(\beta)< \nu < \nu_m(\beta)$,\qquad} \label{nu0G2a}\\
\nu_m(\beta)-\nu_c(\beta), \!\!\!\!\!\!&\textrm{when $\nu \geq \nu_m(\beta)$.} \label{nu0G3a}
\end{subnumcases}
Moreover, $\nu_{\infty,0}(\beta,\nu;\bold{s})=0$ $\forall \bold{s} \in (\mathbb{N}^{*})^{3}$.\\
$\mathrm{(iii)}$. $\overline{\mu}_{\infty,\kappa} = \overline{\mu}_{\infty,\kappa}(\beta,\nu) \in (-\infty,E_{\infty,\kappa}^{(\bold{0})})$
satisfying \eqref{reuniq} admits the asymptotics in the limit $\kappa \downarrow 0$:
\begin{subnumcases}{\label{musolGa} \overline{\mu}_{\infty,\kappa} =}
E_{\infty,\kappa}^{(\bold{0})} + \overline{\mu}_{\infty,0} + o(1),
&\textrm{when $\nu<\nu_{c}(\beta)$}, \label{musolG1a} \\
E_{\infty,\kappa}^{(\bold{0})} -  \beta^{-1}\mathrm{e}^{-\frac{\hbar \omega_{1}\beta(\nu-\nu_c(\beta))}{\kappa^2}}+o \left(\mathrm{e}^{-\frac{\hbar \beta \omega_{1} (\nu-\nu_c(\beta))}{\kappa^2}}\right), \!\!\!\!\!\!&\textrm{when $\nu_c(\beta)<\nu \leq \nu_{m}(\beta)$,\qquad}\label{musolG2a} \\
E_{\infty,\kappa}^{(\bold{0})} -\frac{\kappa_1\kappa_\perp^2}{\beta \left(\nu-\nu_{m}(\beta)\right)}+o\left(\kappa_1\kappa_\perp^2\right),&\textrm{when $\nu>\nu_{m}(\beta)$}. \label{musolG3a}
\end{subnumcases}
Here, $E_{\infty,\kappa}^{(\bold{0})} = \frac{1}{2} \hbar \omega_{1} \kappa_{1} +  \hbar\omega_{\perp} \kappa_{\perp}$, and $\overline{\mu}_{\infty,0} = \overline{\mu}_{\infty,0}(\beta,\nu) \in (-\infty,0)$ satisfies \eqref{muinf0}.
\end{proposition}

The proof of Proposition \ref{Prop1Q1D} can be found in \cite{BZ}. Contrary to the case of the isotropic harmonic trap, the Bose gas manifests an open-trap g-BEC in the sense of \eqref{gBECform} when $\nu>\nu_c(\beta)$. Moreover, the Bose gas manifests an open-trap BEC in the sense of Definition \ref{OpenTrapBECdef} only when $\nu>\nu_m(\beta)> \nu_{c}(\beta)$. Therefore, the open-trap g-BEC and open-trap BEC coexist when $\nu>\nu_{m}(\beta)$. We mention that an article investigating the measurement of such a chemical potential \eqref{musolG2a} has been recently published \cite{MaBra}.

\begin{remark}
\label{RemQ1D}
Let us discuss the physical relevance of such an exponential-quasi-1D model. Preparing experimentally the system in a quasi-1D regime requires that the conditions $\hbar\omega_1\ll \hbar \omega_\perp$ along with
$\hbar\omega_\perp \beta \ll 1$ are fulfilled.
On the contrary, the condition $\hbar\omega_1 \beta \ll\  1 \ll\hbar\omega_\perp \beta$ implies that the Bose gas behaves like a one-dimensional system, see e.g. \cite{Ketterle2}. However, we need another condition on $\omega_c:=\omega_{\perp} \kappa_{c}$ since it has to be sufficiently large so that the difference $\nu_m-\nu_c$ is at least of the same order than $\nu_c$. Thus, by \eqref{numaQ1D}, $\omega_c$ has to be of the order of $\omega_\beta := (\hbar\beta)^{-1}$.
\end{remark}

Subsequently, we turn to the properties of the (rescaled) reduced density matrix in open-trap limit. As a counterpart of Theorem \ref{Thm1}, we establish:

\begin{corollary}
\label{Thm1Q1D}
Consider a quasi-1D harmonically trapped Bose gas (the anisotropy is defined by \eqref{kappaQ1Da}), in the G-C ensemble. Then for any $\beta>0$, $\nu>0$, $\kappa_{c}>0$ and $(\bold{x},\bold{y}) \in \mathbb{R}^{6}$:\\
$\mathrm{(A)}$. The open-trap reduced density matrix exists and satisfies:
\begin{subnumcases}
{\label{rdmd2Q1D}  \rho_{\infty,0}(\bold{x},\bold{y};\beta,\nu) =}
\frac{1}{\lambda_{\beta}^{3}} \sum_{l=1}^{\infty} \frac{\mathrm{e}^{l \beta\overline{\mu}_{\infty,0}}}{  l^{\frac{3}{2}}} \mathrm{e}^{- \frac{\pi}{\lambda_{\beta}^{2}}\frac{\vert \bold{x}-\bold{y}\vert^2}{l}},
&\textrm{when $\nu<\nu_c(\beta)$},  \label{rdmd21Q1D} \\
\infty, &\textrm{when $\nu>\nu_c(\beta)$}. \label{rdmd22Q1D}
\end{subnumcases}
$\mathrm{(B)}$. The open-trap rescaled reduced density matrix exists and satisfies:
\begin{equation}
\label{Thm1Eq2Q1D}
r_{\infty,0}(\beta,\nu)= \left(\frac{m \omega_{0}}{\pi \hbar}\right)^{\frac{3}{2}} \frac{\nu_{\infty,0}(\beta,\nu;\bold{0})}{\pi^{\frac{3}{2}}} = \left(\frac{m \omega_{0}}{\pi \hbar}\right)^{\frac{3}{2}} \times \left\{\begin{array}{ll}
0, &\textrm{when $\nu<\nu_{m}(\beta)$}, \\
\displaystyle{\frac{\nu - \nu_{m}(\beta)}{\pi^{\frac{3}{2}}}}, &\textrm{when $\nu \geq \nu_{m}(\beta)$}.
\end{array}\right.
\end{equation}
Here, $\nu_{m}(\beta)>\nu_{c}(\beta)$ is the second critical open-trap rescaled number of particles defined by \eqref{numaQ1D}. As a result of \eqref{Thm1Eq2Q1D}, the Bose gas manifests an open-trap ODLRO if $\nu>\nu_{m}(\beta)$.\\
$\mathrm{(C)}$. In addition: the open-trap rescaled reduced density matrix satisfies:
\begin{equation}
\label{Thm1Eq1Q1D}
r_{\infty,0}(\bold{x},\bold{y};\beta,\nu) = \lim_{\kappa\downarrow0}
\nu_{\infty,\kappa}\left(\beta,\nu;\bold{0}\right) \Psi_{\infty,1}^{(\bold{0})}\left(\bold{x}\sqrt{\kappa}\right) \Psi_{\infty,1}^{(\bold{0})}\left(\bold{y}\sqrt{\kappa}\right),
\end{equation}
as for the open-trap reduced density matrix without the ground-state:
\begin{subnumcases}{\label{Thm1Eq3Q1D}
\lim_{\kappa\downarrow0}
\sum_{\bold{s}\in (\mathbb{N}^{*})^{3}}\frac{\Psi_{\infty,\kappa}^{(\bold{s})}(\bold{x}) \Psi_{\infty,\kappa}^{(\bold{s})}(\bold{y})}{\mathrm{e}^{\beta\left(E_{\infty,\kappa}^{(\bold{s})}- \overline{\mu}_{\infty,\kappa}\right)}-1}=}
\frac{1}{\lambda_{\beta}^{3}} \sum_{l=1}^{\infty} \frac{\mathrm{e}^{l \beta\overline{\mu}_{\infty,0}}}{ l^{\frac{3}{2}}}
\mathrm{e}^{-\frac{\pi}{\lambda_{\beta}^{2}} \frac{\vert \bold{x}-\bold{y}\vert^2}{l}},
&\textrm{when $\nu<\nu_c(\beta)$}, \label{Thm1Eq31Q1D}\\
\infty, &\textrm{when $\nu>\nu_{c}(\beta)$}. \label{Thm1Eq32Q1D}
\end{subnumcases}
\end{corollary}

The proof of Corollary \ref{Thm1Q1D} is sketched in Sec. \ref{AppendixD0} and leans on the same methods than the ones used to prove Theorem \ref{Thm1}. Unlike the isotropic case, the Bose gas manifests an open-trap ODLRO if and only if $\nu>\nu_m(\beta)$, and even if the Bose gas manifests an open-trap g-BEC when $\nu_c(\beta)<\nu<\nu_m(\beta)$, see Proposition \ref{Prop1Q1D} $\mathrm{(ii)}$. We emphasize that, in the regime $\nu>\nu_{c}(\beta)$, the g-BEC has an impact on the reduced density matrix since its non-condensate part diverges.

\begin{remark}
\label{RemQ1D3}
The 'proof' of Corollary \ref{Thm1Q1D} in Sec. \ref{AppendixD0} allows to bring out that the divergence in the regime $\nu>\nu_{c}(\beta)$ of the non-condensate part of the reduced density matrix in \eqref{Thm1Eq32Q1D} arises from  the mesoscopic-loops contribution. With $\omega_{c}:= \omega_{\perp} \kappa_{c}$, we prove that when $\kappa \downarrow 0$:
\begin{multline}
\label{RemQ1D2Eq}
\sum_{l= N_{\kappa,\sigma}+1}^{M_{\kappa}} \left( \mathrm{e}^{l \beta\overline{\mu}_{\infty,\kappa}} G_{\infty,\kappa_{1}}^{(1)}\left(x_{1},y_{1};l\beta\right) G_{\infty,\kappa_{\perp}}^{(2)}\left(\bold{x}_{\perp},\bold{y}_{\perp};l\beta\right) - \mathrm{e}^{l \beta \left(\overline{\mu}_{\infty,\kappa} - E_{\infty,\kappa}^{(\bold{0})}\right)} \Psi_{\infty,\kappa}^{(\bold{0})}(\bold{x})\Psi_{\infty,\kappa}^{(\bold{0})}(\bold{y})\right) \\
\sim \frac{1}{\lambda_\beta}\frac{m\omega_\perp\kappa_\perp}{\sqrt{\pi} \hbar}\times\left\{\begin{array}{ll}
\displaystyle{\exp{\left(\frac{\mathpzc{g}_{3}(1)}{\left(\hbar \omega_\perp \kappa \beta\right)^2} \frac{\left(\eta-1\right)}{2}\right)}},\ &\textrm{when $1 < \frac{\nu}{\nu_{c}(\beta)} = \eta \leq \frac{\nu_{m}(\beta)}{\nu_{c}(\beta)}$},\\
\displaystyle{\exp{\left(\frac{\omega_c^2}{2\omega_\perp^2\kappa^2}\right)}},\ &\textrm{when $\nu>\nu_{m}(\beta)$},
\end{array}\right.,
\end{multline}
where $N_{\kappa,\sigma} := \lfloor \kappa^{-\sigma} \rfloor$, $\sigma >0$ and $M_{\kappa} = M_{\kappa,\kappa_{c}}:= \lfloor \mathrm{e}^{\frac{\kappa_{c}^{2}}{\kappa^{2}}}\rfloor$ with $\kappa<1$ (here $\lfloor \cdot\,\rfloor$ is the floor function). The 'proof' of Corollary \ref{Thm1Q1D}  brings also out that  the sum over the short-loops, i.e., loops in the range $1 \leq l \leq \lfloor\kappa^{-\sigma}\rfloor$ gives rises to the usual \textit{thermal gas} contribution given in \eqref{Thm1Eq33}, whereas the sum over the macroscopic-loops, i.e., $l > \lfloor \mathrm{e}^{\frac{\kappa_c^2}{\kappa^2}}\rfloor$ identically vanishes. Note that in our sum-decomposition, $\kappa$ plays the role of $\hbar \omega_{0} \beta$ in the loop-gas approach in \cite{Mullin1}. Due to the latter feature, the three-terms decomposition of the local density function as stated in \eqref{Intro2} is then justified. The first term corresponds to the usual BEC (macroscopic-loops contribution) and the third term to the thermal gas (short-
loops contribution). Moreover from \eqref{RemQ1D2Eq}, the additional term corresponds to the mesoscopic-loops contribution with loops of length in the range $\lfloor \kappa^{-\sigma} \rfloor < l \leq \lfloor \mathrm{e}^{\frac{\kappa_c^2}{\kappa^2}}\rfloor$.
\end{remark}

\subsubsection{A Quasi-2D trap model.}
\label{quasi2D}

In Theorem \ref{Thm1} $\mathrm{(C)}$, we stated that the non-condensate part of the open-trap reduced density matrix diverges when $\nu \geq \nu_{c}(\beta)$ if $d=2$. In Annex \ref{AppendixC}, we precise this result and prove that:
\begin{equation*}
\sum_{\bold{s} \in (\mathbb{N}^{*})^{2}} \frac{\Psi_{\infty,\kappa}^{(\bold{s})}(\bold{x}) \Psi_{\infty,\kappa}^{(\bold{s})}(\bold{y})}{\mathrm{e}^{\beta \left(E_{\infty,\kappa}^{(\bold{s})} - \overline{\mu}_{\infty,\kappa}\right)} - 1} \sim \frac{1}{\lambda_{\beta}^{2}} \ln\left(\frac{1}{\hbar \omega_{0} \kappa \beta}\right)\quad \textrm{when $\kappa \downarrow 0$},
\end{equation*}
uniformly in $(\bold{x},\bold{y}) \in \mathbb{R}^{4}$. We mention that a similar behavior has already  been pointed out in \cite{Mullin2}. In this paragraph, we introduce a three-dimensional anisotropic harmonic trap model mimicking the two-dimensional properties (for certain values of $\nu>\nu_c(\beta)$) while 'regularizing' the two-dimensional logarithmic divergence mentioned below. This trap model is defined as follows:
\begin{equation}
\label{kappaQ2Da}
\left\{ \begin{array}{ll}
 \kappa_1&=\kappa  \\
 \kappa_\perp&=\kappa \exp{\left(-\sqrt{\frac{\kappa_c}{\kappa}}\right)}\, (=\kappa_{2}=\kappa_{3}),\quad \kappa_{c}>0.
 \end{array}\right.
\end{equation}
From \eqref{kappaQ2Da}, for small values of $\kappa$, the characteristic length $\sqrt{\hbar (m\omega_{j}\kappa_{j})^{-1}}$ along the $x_{j}$-direction, $j=2,3$ is very large compared to the one along the $x_1$-direction (hence the name of quasi-2D trap). This anisotropic model is inspired by a model introduced in \cite{vdB} for homogeneous systems. Contrary to the quasi-1D model in Sec. \ref{quasi1D}, this quasi-2D model does not exhibit a second kind of transition. In fact, the counterpart of Propositions \ref{Prop1}-\ref{Prop1'} is similar to the isotropic case:

\begin{proposition}
\label{Prop1Q2Da}
Consider a quasi-2D harmonically trapped Bose gas (the anisotropy is defined by \eqref{kappaQ2Da}), in the G-C ensemble. Then, for any $\beta>0$, $\nu>0$ and $\kappa_0>0$:\\
$\mathrm{(i)}$. The Bose gas manifests an open-trap BEC in the sense of Definition \ref{OpenTrapBECdef}. Furthermore, the open-trap rescaled average number of particles on the ground-state satisfies:
\begin{subnumcases}{\label{nu00Q2D} \nu_{\infty,0}(\beta,\nu;\bold{0})=}
0,&\textrm{when $\nu<\nu_c(\beta)$}, \label{nu001Q2D}\\
\nu-\nu_c(\beta),&\textrm{when $\nu\geq\nu_c(\beta)$}, \label{nu002Q2D}
\end{subnumcases}
where $\nu_{c}(\beta)$ is the critical open-trap rescaled average number of particles in \eqref{defnuc} obeying \eqref{nucEq2}.\\
$\mathrm{(ii)}$. The Bose gas manifests no open-trap generalized-BEC in the sense of Definition \ref{OpenTrapgBECdef}:
\begin{equation}
\label{nu0Ga2}
\lim_{\varepsilon \downarrow 0} \lim_{\kappa \downarrow 0} \sum_{\bold{s} \in(\mathbb{N}^{*})^{3}\,:\, \sum_{j=1}^{3} \kappa_{j} s_{j} \leq \varepsilon} \nu_{\infty,\kappa}(\beta,\nu;\bold{s})=
0.
\end{equation}
Moreover, $\nu_{\infty,0}(\beta,\nu;\bold{s})=0$ $\forall \bold{s} \in (\mathbb{N}^{*})^{3}$.\\
$\mathrm{(iii)}$. $\overline{\mu}_{\infty,\kappa} = \overline{\mu}_{\infty,\kappa}(\beta,\nu) \in (-\infty,E_{\infty,\kappa}^{(\bold{0})})$
satisfying \eqref{reuniq} admits the asymptotics in the limit $\kappa \downarrow 0$:
\begin{subnumcases}{\label{musolG2D} \overline{\mu}_{\infty,\kappa} =}
E_{\infty,\kappa}^{(\bold{0})} + \overline{\mu}_{\infty,0} + o(1),
&\textrm{when $\nu<\nu_{c}(\beta)$}, \label{musolGG1a2D} \\
E_{\infty,\kappa}^{(\bold{0})} -\frac{\kappa_1\kappa_\perp^2}{\beta\left(\nu-\nu_{c}(\beta)\right)}+o\left(\kappa_1\kappa_\perp^2\right), &\textrm{when  $\nu>\nu_{c}(\beta)$}. \label{musolGG2a2D}
\end{subnumcases}
Here, $E_{\infty,\kappa}^{(\bold{0})} = \frac{1}{2} \hbar \omega_{1} \kappa_{1} +  \hbar \omega_{\perp} \kappa_{\perp}$, and $\overline{\mu}_{\infty,0} = \overline{\mu}_{\infty,0}(\beta,\nu) \in (-\infty,0)$ satisfies the equation \eqref{muinf0}.
\end{proposition}

Subsequently, we turn to the properties of the (rescaled) reduced density matrix in open-trap limit. As a counterpart of Theorem \ref{Thm1}, we establish:

\begin{corollary}
\label{Thm1Q2D}
Consider a quasi-2D harmonically trapped Bose gas (the anisotropy is defined by \eqref{kappaQ2Da}), in the G-C ensemble. Then for any $\beta>0$, $\nu>0$, $\kappa_{c}>0$ and $(\bold{x},\bold{y}) \in \mathbb{R}^{6}$:\\
$\mathrm{(A)}$. The open-trap reduced density matrix exists and satisfies:
\begin{subnumcases}
{\label{rdmd2Q2D}  \rho_{\infty,0}(\bold{x},\bold{y};\beta,\nu) =}
\frac{1}{\lambda_{\beta}^{3}} \sum_{l=1}^{\infty} \frac{\mathrm{e}^{l \beta\overline{\mu}_{\infty,0}}}{ l^{\frac{3}{2}}} \mathrm{e}^{- \frac{\pi}{\lambda_{\beta}^{2}} \frac{\vert \bold{x}-\bold{y}\vert^2}{l}},
&\textrm{when $\nu<\nu_c(\beta)$},  \label{rdmd21Q2D} \\
\infty, &\textrm{when $\nu>\nu_c(\beta)$}. \label{rdmd22Q2D}
\end{subnumcases}
$\mathrm{(B)}$. The open-trap rescaled reduced density matrix exists and satisfies:
\begin{equation}
\label{Thm1Eq2Q2D}
r_{\infty,0}(\beta,\nu)= \left(\frac{m \omega_{0}}{\pi \hbar}\right)^{\frac{3}{2}} \frac{\nu_{\infty,0}(\beta,\nu;\bold{0})}{\pi^{\frac{3}{2}}} = \left(\frac{m \omega_{0}}{\pi \hbar}\right)^{\frac{3}{2}} \times \left\{\begin{array}{ll}
0, &\textrm{when $\nu<\nu_{c}(\beta)$}, \\
\displaystyle{\frac{\nu - \nu_{c}(\beta)}{\pi^{\frac{3}{2}}}}, &\textrm{when $\nu \geq \nu_{c}(\beta)$}.
\end{array}\right.
\end{equation}
As a result of \eqref{Thm1Eq2Q2D}, the Bose gas manifests an open-trap ODLRO if $\nu>\nu_{c}(\beta)$.\\
$\mathrm{(C)}$. In addition: the open-trap rescaled reduced density matrix satisfies:
\begin{equation}
\label{Thm1Eq1Q2D}
r_{\infty,0}(\bold{x},\bold{y};\beta,\nu) = \lim_{\kappa\downarrow0}
\nu_{\infty,\kappa}\left(\beta,\nu;\bold{0}\right) \Psi_{\infty,1}^{(\bold{0})}\left(\bold{x}\sqrt{\kappa}\right) \Psi_{\infty,1}^{(\bold{0})}\left(\bold{y}\sqrt{\kappa}\right),
\end{equation}
as for the open-trap reduced density matrix without the ground-state:
\begin{subnumcases}{\label{Thm1Eq3Q2D}
\lim_{\kappa\downarrow0}
\sum_{\bold{s}\in (\mathbb{N}^{*})^{3}}\frac{\Psi_{\infty,\kappa}^{(\bold{s})}(\bold{x}) \Psi_{\infty,\kappa}^{(\bold{s})}(\bold{y})}{\mathrm{e}^{\beta\left(E_{\infty,\kappa}^{(\bold{s})}- \overline{\mu}_{\infty,\kappa}\right)}-1}=}
\frac{1}{\lambda_{\beta}^{3}} \sum_{l=1}^{\infty} \frac{\mathrm{e}^{l \beta\overline{\mu}_{\infty,0}}}{ l^{\frac{3}{2}}}
\mathrm{e}^{-\frac{\pi}{\lambda_{\beta}^{2}}\frac{\vert \bold{x}-\bold{y}\vert^2}{l}},
&\textrm{when $\nu<\nu_c(\beta)$}, \label{Thm1Eq31Q2D}\\
\sqrt{\frac{m \omega_{c}}{\pi \hbar}} \frac{2}{\lambda_{\beta}^{2}} + \frac{1}{\lambda_{\beta}^{3}}\sum_{l=1}^{\infty} \frac{1}{l^{\frac{3}{2}}}
\mathrm{e}^{-\frac{\pi}{\lambda_{\beta}^{2}} \frac{\vert \bold{x}-\bold{y}\vert^2}{l}}, \!\!\!\!\!\!\!\!\!&\textrm{when $\nu>\nu_{c}(\beta)$,\qquad} \label{Thm1Eq32Q2D}
\end{subnumcases}
where $\omega_{0} = (\omega_{1} \omega_{\perp}^{2})^{\frac{1}{3}}$ and $\omega_{c}:= \omega_{1} \kappa_{c}$.
\end{corollary}

The proof of Corollary \ref{Thm1Q2D} is sketched in Sec. \ref{AppendixD0}. The results of Corollary \ref{Thm1Q2D}  are similar to Theorem \ref{Thm1} but with the difference that the non-condensate part of the open-trap reduced density matrix has an additional term in the regime $\nu>\nu_{c}(\beta)$, see \eqref{Thm1Eq32Q2D} (compared to \eqref{Thm1Eq33}). Let us comment this result. In fact, we prove that when $\nu>\nu_c(\beta)$:
\begin{align*}
\sum_{\bold{s} \in (\mathbb{N}^{*})^{3}} \frac{\Psi_{\infty,\kappa}^{(\bold{s})}(\bold{x}) \Psi_{\infty,\kappa}^{(\bold{s})}(\bold{y})}{\mathrm{e}^{\beta\left(E_{\infty,\kappa}^{(\bold{s})} - \overline{\mu}_{\infty,\kappa}\right)} - 1}
&\sim \sqrt{\frac{m \omega_{1}}{\pi \hbar}} \frac{1}{\lambda_{\beta}^{2}}  \ln\left(\frac{1}{\kappa_\perp^{2}}\right) + \frac{1}{\lambda_{\beta}^{3}} \sum_{l=1}^{\infty} \frac{1}{ l^{\frac{3}{2}}}\mathrm{e}^{-\frac{\pi}{\lambda_{\beta}^{2}} \frac{\vert \bold{x}-\bold{y}\vert^2}{l}} \\
&=2 \sqrt{\frac{m \omega_{c}}{\pi \hbar}} \frac{1}{\lambda_{\beta}^{2}}  + \frac{1}{\lambda_{\beta}^{3}} \sum_{l=1}^{\infty} \frac{1}{ l^{\frac{3}{2}}}\mathrm{e}^{-\frac{\pi}{\lambda_{\beta}^{2}} \frac{\vert \bold{x}-\bold{y}\vert^2}{l}}.
\end{align*}
While mimicking the properties of the two-dimensional isotropic case, our quasi-2D trap model provides a non-condensate part of the open-trap reduced density matrix that is finite. The logarithmic divergence is then 'regularized' by the third dimension (multiplication by $\sqrt{\kappa_1}=\sqrt{\kappa}$). We point out that the additional contribution arises from the 'mesoscopic loops' that we can interpret as a \textit{local g-BEC}, see \eqref{filp0}-\eqref{filp3} in Sec. \ref{AppendixD0}. Note that this additional term still occurs in the open-trap local density (i.e., when $\bold{x}=\bold{y}$).

\begin{remark}
\label{RemQ2D1}
Let us discuss the physical relevance of such an exponential-quasi-2D model.
Preparing experimentally the system in a quasi-2D regime requires that the conditions $\hbar\omega_\perp \ll \hbar \omega_1$ and $\hbar\omega_1 \beta\ll 1$ are fulfilled. On the contrary, the condition $\hbar\omega_\perp \beta \ll 1 \ll\hbar\omega_1 \beta$ implies that the Bose gas behaves like a two-dimensional system, see e.g. \cite{Ketterle2}, and under the condition $\omega_\perp \ll\omega_1$ along with $\hbar\omega_1\beta \approx 1$, the Bose gas is in another quasi-2D regime, see \cite{Holzmann}. As in Remark \ref{RemQ1D}, we need a condition on $\omega_c:= \omega_{1} \kappa_{c}$ so that the first term of the local density (diagonal part of \eqref{Thm1Eq32Q2D}) is of the same order than the local density $\lambda_\beta^{-3} \mathpzc{g}_{\frac{3}{2}}(1)$. Thus, $\omega_c$ has to be of the order of $\omega_\beta = (\hbar\beta)^{-1}$.
\end{remark}

\begin{remark}
\label{RemQ2D2}
We mention that the result in \eqref{Thm1Eq32Q2D} is similar to the one derived for the perfect Bose gas when considering the analogous exponential-quasi-2D boxes in which one takes $L_1=L$ and $L_\perp = L \mathrm{e}^{\alpha L}$, where $\alpha>0$ is the exponential rate of the model. We refer to \cite{vdB,vdBLL}, and also to \cite{BZ}. If one compares the results from \cite{BZ} along with Corollary \ref{Thm1Q2D} versus the results stated for the isotropic case in Theorem \ref{Thm1}, one can interpret the additional term in the local density matrix as a g-BEC having the density of particles equals to $\rho_m(\bold{x})-\rho_c(\bold{x})=2\alpha \lambda_\beta^{-2}$. Here, $\alpha=\sqrt{m\omega_c/(\pi \hbar)}$ is analogous to the exponential rate appearing in the exponential-quasi-2D boxes, and $\rho_m(\bold{x})$ is analogous to the second critical density of particles ($\rho_c(\bold{x})$ is the usual critical density).
\end{remark}

\begin{remark}
\label{RemQ2D3}
In the proof of Corollary \ref{Thm1Q2D}, see Sec. \ref{AppendixD0}, the reduced density matrix is decomposed into three sums corresponding to different sizes of loops: the short-loops giving rise to the second term of the r.h.s. of \eqref{Thm1Eq32Q2D}, the mesoscopic-loops to the first term, plus the macroscopic-loops giving rise to \eqref{Thm1Eq1Q2D}. From \eqref{filp3}, mesoscopic-loops of different scales contribute to the first term of the r.h.s. of \eqref{Thm1Eq32Q2D}. For instance, the sum over the loops in the range $\lfloor \kappa^{-\sigma_{1}}\rfloor < l \leq \lfloor\kappa^{-\sigma_{2}} \kappa_{\perp}^{-1}\rfloor$ with $\sigma_{1}>0$, $\sigma_{2}\geq 0$ and $\kappa<1$ gives rise to half of the contribution; the sum in the range $\lfloor\kappa^{-\sigma_{2}} \kappa_{\perp}^{-1}\rfloor < l \leq \lfloor\kappa^{-\sigma_{2}} \kappa_{\perp}^{-2}\rfloor$ gives rise to the other half. Note that in our sum-decomposition, the $\kappa$ plays the role of $\hbar \omega_{0} \beta$ in the loop-gas approach in \cite{
Mullin1}. Therefore, if one uses numerical simulations with PIMC method (see e.g. \cite{Krauth1}) for investigations on the ideal Bose gas in exponential-quasi-2D harmonic traps, then one should observe a non-negligible mesoscopic-loop-length distribution.
\end{remark}

\section{Proofs of the main results.}

In all the proofs that we give in this section, we set $\hbar=m=\omega_{0}=1$ for the sake of simplicity.

\subsection{Proof of Theorem \ref{Thm1}.}

\textbf{Part $\mathrm{(A)}$.} Let $\beta,\nu>0$ be fixed. We start with the case of $\nu<\nu_{c}(\beta)$ if $d=1,2,3$. Consider the representation in \eqref{otrep}. From \eqref{musol1}, there exists a $\kappa_{0}>0$ s.t. $\forall 0<\kappa \leq \kappa_{0}$, $\overline{\mu}_{\infty,\kappa} \leq \frac{\overline{\mu}_{\infty,0}}{2} < 0$. This, together with the rough upper bound in the second inequality of \eqref{roughes}, lead to:
\begin{equation*}
\forall 0<\kappa \leq \kappa_{0},\quad \mathrm{e}^{l\beta \overline{\mu}_{\infty,\kappa}} G_{\infty,\kappa}^{(d)}(\bold{x},\bold{y};l\beta) \leq \mathrm{e}^{l\beta \frac{\overline{\mu}_{\infty,0}}{2}} G_{\infty,0}^{(d)}(\bold{x},\bold{y};l\beta) \leq \mathrm{e}^{l\beta \frac{\overline{\mu}_{\infty,0}}{2}}\left(2\pi l \beta\right)^{-\frac{d}{2}},
\end{equation*}
uniformly in $(\bold{x},\bold{y}) \in \mathbb{R}^{2d}$. From the above inequality, \eqref{rdmd21} follows by standard arguments.
We continue with the case of $\nu=\nu_{c}(\beta)$ if $d=2$. The strategy is to find a lower bound of the sum in \eqref{otrep} whose the limit $\kappa \downarrow 0$ diverges. Let us note that $\overline{\mu}_{\infty,\kappa} \geq 0$ for $\kappa>0$ small enough, see \eqref{musol2}. Then, from \eqref{Mehler}-\eqref{multd}, one has $\forall l \in \mathbb{N}^{*}$, $\forall(\bold{x},\bold{y}) \in \mathbb{R}^{4}$ and for $\kappa>0$ sufficiently small:
\begin{equation}
\label{compa1}
\mathrm{e}^{l\beta \overline{\mu}_{\infty,\kappa}} G_{\infty,\kappa}^{(d=2)}(\bold{x},\bold{y};l\beta) \geq \frac{\kappa}{2 \pi \sinh(\kappa l\beta)} \mathrm{e}^{- \frac{\kappa}{4} \vert \bold{x}+\bold{y}\vert^{2}} \mathrm{e}^{- \frac{1}{4}\left(\kappa + \frac{2}{\beta}\right)\vert \bold{x}-\bold{y}\vert^{2}},
\end{equation}
where we used the upper bounds in \eqref{Ek3}-\eqref{Ek4}. Then, under the conditions of \eqref{compa1}, one has:
\begin{equation}
\label{lowbo}
\rho_{\infty,\kappa}(\bold{x},\bold{y};\beta,\nu) \geq- \frac{1}{2\pi \beta} \mathrm{e}^{- \frac{\kappa}{4} \vert \bold{x}+\bold{y}\vert^{2}} \mathrm{e}^{- \frac{1}{4}\left(\kappa + \frac{2}{\beta}\right)\vert \bold{x}-\bold{y}\vert^{2}} \ln\left(\tanh\left(\frac{\beta \kappa}{2}\right)\right),
\end{equation}
and the above lower bound diverges in the limit $\kappa \downarrow 0$. To get \eqref{lowbo}, we used an integral comparison to minorize the sum, and then we performed explicitly the integral. Let us turn to the case of $\nu> \nu_{c}(\beta)$ if $d=2,3$. If $d=2$, it is enough to use a similar reasoning than the one leading to \eqref{lowbo}. If $d=3$, from \eqref{Mehler}-\eqref{multd}, one has $\forall l \in \mathbb{N}^{*}$, $\forall(\bold{x},\bold{y}) \in \mathbb{R}^{6}$ and for $\kappa>0$ sufficiently small:
\begin{equation*}
\mathrm{e}^{l\beta \overline{\mu}_{\infty,\kappa}} G_{\infty,\kappa}^{(d=3)}(\bold{x},\bold{y};l\beta) \geq  \pi^{-\frac{3}{2}} \kappa^{\frac{3}{2}} \mathrm{e}^{l\beta \left(\overline{\mu}_{\infty,\kappa}- E_{\infty,\kappa}^{(\bold{0})}\right)} \mathrm{e}^{- \frac{\kappa}{4} \vert \bold{x}+\bold{y}\vert^{2}} \mathrm{e}^{- \frac{1}{4}\left(\kappa + \frac{2}{\beta}\right)\vert \bold{x} - \bold{y}\vert^{2}},
\end{equation*}
where we used the upper bounds in \eqref{Ek2}-\eqref{Ek4}. Since $\overline{\mu}_{\infty,\kappa} < E_{\infty,\kappa}^{(\bold{0})}$, under the same conditions:
\begin{equation}
\label{lowbo1}
\rho_{\infty,\kappa}(\bold{x},\bold{y};\beta,\nu) \geq \frac{1}{\pi^{\frac{3}{2}}\beta} \mathrm{e}^{- \frac{\kappa}{4} \vert \bold{x}+\bold{y}\vert^{2}} \mathrm{e}^{- \frac{1}{4}\left(\kappa + \frac{2}{\beta}\right)\vert \bold{x} - \bold{y}\vert^{2}} \kappa^{\frac{3}{2}} \frac{\mathrm{e}^{-\beta\left(E_{\infty,\kappa}^{(\bold{0})} -\overline{\mu}_{\infty,\kappa}\right) }}{E_{\infty,\kappa}^{(\bold{0})}-\overline{\mu}_{\infty,\kappa}},
\end{equation}
where we used again an integral comparison, and then we performed explicitly the integral. From the asymptotic in \eqref{musol3}, the lower bound in \eqref{lowbo1} diverges in the limit $\kappa \downarrow 0$.\\

\textbf{Part $\mathrm{(B)}$.} Let $\beta,\nu>0$ be fixed. We start with the case of $\nu < \nu_{c}(\beta)$ if $d=1,2,3$. From \eqref{rdmrescaled} together with \eqref{otrep}, then \eqref{rdmd21} leads to $\lim_{\kappa \downarrow 0} r_{\infty,\kappa}(\bold{x},\bold{y};\beta,\nu) = 0$ uniformly on $\mathbb{R}^{2d}$.
Let us turn to the case of $\nu\geq\nu_{c}(\beta)$ if $d=2,3$. The key-idea consists in decomposing $\forall 0<\kappa < 1$ the quantity defined by \eqref{rdmrescaled} (knowing \eqref{otrep}) into two contributions:
\begin{equation}
\label{rdm4}
r_{\infty,\kappa}(\bold{x},\bold{y};\beta,\nu) = \kappa^{\frac{d}{2}} \sum_{l=1}^{N_{\kappa,\sigma}}
\mathrm{e}^{l\beta\overline{\mu}_{\infty,\kappa}}G_{\infty,\kappa}^{(d)}(\bold{x},\bold{y};l\beta)
+  \kappa^{\frac{d}{2}} \sum_{l=N_{\kappa,\sigma}+1}^{\infty}
\mathrm{e}^{l\beta\overline{\mu}_{\infty,\kappa}}G_{\infty,\kappa}^{(d)}(\bold{x},\bold{y};l\beta),
\end{equation}
where $N_{\kappa,\sigma} := \lfloor \kappa^{-\sigma} \rfloor$ with $0<\sigma<d$ for the moment (a limitation will appear when $\nu=\nu_{c}(\beta)$). Here, $\lfloor \cdot\, \rfloor$ denotes the floor function. Below, we prove that the contribution in \eqref{Thm1Eq2} when $\nu>\nu_{c}(\beta)$ only arises from the second quantity in the r.h.s. of \eqref{rdm4}. Let us investigate the first term in the r.h.s. of \eqref{rdm4}. From \eqref{morepre} followed by the lower bound in \eqref{Ink1}, one has $\forall (\bold{x},\bold{y}) \in \mathbb{R}^{2d}$:
\begin{equation}
\label{decomo2}
\mathrm{e}^{l\beta\overline{\mu}_{\infty,\kappa}}G_{\infty,\kappa}^{(d=2)}(\bold{x},\bold{y};l\beta)
\leq \frac{\kappa}{\pi} \mathrm{e}^{l\beta\left(\overline{\mu}_{\infty,\kappa} - E_{\infty,\kappa}^{(\bold{0})}\right)}  + \frac{\mathrm{e}^{l\beta\left(\overline{\mu}_{\infty,\kappa} - E_{\infty,\kappa}^{(\bold{0})}\right)}}{2 \pi l \beta},
\end{equation}
\begin{subnumcases}{\label{decomo2'}\mathrm{e}^{l\beta\overline{\mu}_{\infty,\kappa}}G_{\infty,\kappa}^{(d=3)} (\bold{x},\bold{y};l\beta)\leq \frac{\kappa^{\frac{3}{2}}}{\pi^{\frac{3}{2}}} \mathrm{e}^{l\beta\left(\overline{\mu}_{\infty,\kappa} - E_{\infty,\kappa}^{(\bold{0})}\right)} \times}
1 + \left( \frac{1}{l\beta \kappa} + \frac{1}{(2 l\beta \kappa)^{2}}\right) \label{decomo2'1},\\
1 + \left( \frac{3}{\sqrt{2l\beta \kappa}} + \frac{3}{2 l\beta \kappa} + \frac{1}{(2 l\beta \kappa)^{\frac{3}{2}}}\right). \label{decomo2'2}
\end{subnumcases}
In \eqref{decomo2'1} and \eqref{decomo2'2} we used that $\forall x >0$ $(1+ \frac{1}{x})^{\frac{3}{2}} \leq (1+ \frac{1}{x})^{2}$ and $(1+ \frac{1}{x})^{\frac{3}{2}} \leq (1+ \frac{1}{\sqrt{x}})^{3}$ respectively. Since $\overline{\mu}_{\infty,\kappa} - E_{\infty,\kappa}^{(\bold{0})} < 0$, one has for $\kappa<1$ sufficiently small the following upper bounds:
\begin{equation}
\label{ingamf}
\kappa^{\frac{d}{2}-m} \sum_{l=1}^{N_{\kappa,\sigma}} \frac{\mathrm{e}^{-l\beta\left(E_{\infty,\kappa}^{(\bold{0})} - \overline{\mu}_{\infty,\kappa}\right)}}{l^{m}} \leq \mathrm{e}^{-\beta\left(E_{\infty,\kappa}^{(\bold{0})} - \overline{\mu}_{\infty,\kappa}\right)} \times \left\{\begin{array}{ll}
\displaystyle{2\kappa^{\frac{d}{2}-m - \sigma(1-m)}},\,\,\, &\textrm{if $m\in\{0,\frac{1}{2}\}$}, \\
\displaystyle{\kappa^{\frac{d}{2}-1} \ln\left(\frac{\mathrm{e}}{\kappa^{\sigma}}\right)},\,\,\,&\textrm{if $m=1$}, \\
\displaystyle{3\kappa^{\frac{d}{2} -m}},\,\,\,&\textrm{if $m\in\{\frac{3}{2},2\}$}.
\end{array}\right.
\end{equation}
By  the squeeze theorem, when $\nu \geq \nu_{c}(\beta)$ if $d=2,3$ one has uniformly in $(\bold{x},\bold{y})\in\mathbb{R}^{2d}$:
\begin{equation}
\label{conclu01}
\forall 0 < \sigma < d,\quad \lim_{\kappa \downarrow 0} \kappa^{\frac{d}{2}} \sum_{l=1}^{N_{\kappa,\sigma}} \mathrm{e}^{l\beta\overline{\mu}_{\infty,\kappa}}G_{\infty,\kappa}^{(d)}(\bold{x},\bold{y};l\beta) =0.
\end{equation}
Subsequently, let us turn to the second term in the r.h.s. of \eqref{rdm4}. Since $\overline{\mu}_{\infty,\kappa} - E_{\infty,\kappa}^{(\bold{0})} < 0$, then one has $\forall d \in \{2,3\}$, $\forall 0<\sigma < d$, $\forall (\bold{x},\bold{y}) \in \mathbb{R}^{2d}$ and for $\kappa<1$ sufficiently small:
\begin{equation}
\label{fondlowb}
\kappa^{\frac{d}{2}}\sum_{l=N_{\kappa,\sigma}+1}^{\infty}
\mathrm{e}^{l\beta \overline{\mu}_{\infty,\kappa}}G_{\infty,\kappa}^{(d)}(\bold{x},\bold{y};l\beta)
\geq \frac{\kappa^{d}}{\pi^{\frac{d}{2}}} \mathrm{e}^{- \frac{\kappa}{4}\vert \bold{x}+\bold{y}\vert^{2}} \mathrm{e}^{-\frac{\kappa}{4} \vert \bold{x}-\bold{y}\vert^{2} \coth\left(\frac{\beta}{2} \kappa^{1-\sigma}\right)} \frac{\mathrm{e}^{- N_{\kappa,\sigma} \beta\left(E_{\infty,\kappa}^{(\bold{0})} - \overline{\mu}_{\infty,\kappa}\right)}}{\mathrm{e}^{\beta\left(E_{\infty,\kappa}^{(\bold{0})} - \overline{\mu}_{\infty,\kappa}\right)} -1}.
\end{equation}
Here, we used the upper bounds in \eqref{Ek2}-\eqref{Ek3} together with a formula to express the remainder of the geometric series. Now, we distinguish the case of $\nu >\nu_{c}(\beta)$ from the case of $\nu=\nu_{c}(\beta)$. When $\nu >\nu_{c}(\beta)$, we get from the asymptotic in \eqref{musol3} the existence of a $K_{\beta}>0$ s.t.
\begin{equation}
\label{majptio}
\forall 0< \kappa \leq K_{\beta},\quad 3\kappa^{d} \geq 2\beta\left(\nu - \nu_{c}(\beta)\right) \left(E_{\infty,\kappa}^{(\bold{0})} - \overline{\mu}_{\infty,\kappa}\right) \geq \kappa^{d} >0.
\end{equation}
By the upper bound in \eqref{majptio}, one has $\forall 0<\sigma<d$, $\forall(\bold{x},\bold{y})\in\mathbb{R}^{2d}$ and $\forall 0<\kappa < \min\{1,K_{\beta}\}$:
\begin{equation}
\label{fondlowb2}
\kappa^{\frac{d}{2}}\sum_{l=N_{\kappa,\sigma}+1}^{\infty}
\mathrm{e}^{l\beta \overline{\mu}_{\infty,\kappa}}G_{\infty,\kappa}^{(d)}(\bold{x},\bold{y};l\beta)
\geq \frac{\nu_{\infty,\kappa}(\beta,\nu;\bold{0})}{\pi^{\frac{d}{2}}} \mathrm{e}^{- \frac{\kappa}{4} \vert \bold{x}+\bold{y}\vert^{2}} \mathrm{e}^{-\frac{\kappa}{4} \vert \bold{x}-\bold{y}\vert^{2} \coth\left(\frac{\beta}{2} \kappa^{1-\sigma}\right)} \mathrm{e}^{- \frac{3  \kappa^{d-\sigma}}{2\left(\nu - \nu_{c}(\beta)\right)}}.
\end{equation}
Next, let us find an upper bound for the l.h.s. of \eqref{fondlowb2}. If $d=2$, from the upper bound in \eqref{decomo2} then $\forall 0<\sigma<2$, $\forall (\bold{x},\bold{y}) \in \mathbb{R}^{4}$ and  for $\kappa<\min\{1,K_{\beta}\}$ sufficiently small:
\begin{align}
\kappa \sum_{l=N_{\kappa,\sigma}+1}^{\infty}
\mathrm{e}^{l\beta \overline{\mu}_{\infty,\kappa}}G_{\infty,\kappa}^{(d=2)}(\bold{x},\bold{y};l\beta)
&\leq \frac{\kappa^{2}}{\pi} \sum_{l=1}^{\infty} \mathrm{e}^{-l\beta\left(E_{\infty,\kappa}^{(\bold{0})} - \overline{\mu}_{\infty,\kappa}\right)}
+ \frac{\kappa}{2 \pi \beta} \Gamma_{0}\left(\beta N_{\kappa,\sigma} \left(E_{\infty,\kappa}^{(\bold{0})} - \overline{\mu}_{\infty,\kappa}\right)\right)\nonumber\\
\label{majd=2}
&\leq \frac{\nu_{\infty,\kappa}(\beta,\nu;\bold{0})}{\pi} + \frac{\kappa}{2\pi \beta} \Gamma_{0}\left(\frac{\kappa^{2}\left(\kappa^{-\sigma} -1\right)}{2\left(\nu - \nu_{c}(\beta)\right)}\right),
\end{align}
where $\Gamma_{0}(\cdot\,)$ denotes the incomplete Gamma function (below $\gamma$ stands for the Euler constant):
\begin{equation}
\label{Euler}
\forall x>0,\quad \Gamma_{0}(x) := \int_{x}^{\infty} \mathrm{d}t\, \frac{\mathrm{e}^{-t}}{t} =  - \gamma - \ln(x) - \sum_{k=1}^{\infty} (-1)^{k} \frac{x^{k}}{k (k!)}.
\end{equation}
If $d=3$, from the upper bound in \eqref{decomo2'1}, $\forall \sigma \in (0,3)$, $\forall (\bold{x},\bold{y}) \in \mathbb{R}^{6}$ and for $\kappa<1$ small enough:
\begin{multline}
\label{majd=3}
\kappa^{\frac{3}{2}} \sum_{l=N_{\kappa,\sigma}+1}^{\infty} \mathrm{e}^{l\beta \overline{\mu}_{\infty,\kappa}} G_{\infty,\kappa}^{(d=3)}(\bold{x},\bold{y};l\beta) \\
\leq  \frac{\nu_{\infty,\kappa}(\beta,\nu;\bold{0})}{\pi^{\frac{3}{2}}} + \frac{\kappa^{2}}{\beta \pi^{\frac{3}{2}}} \Gamma_{0}\left(\beta N_{\kappa,\sigma} \left(E_{\infty,\kappa}^{(\bold{0})} - \overline{\mu}_{\infty,\kappa}\right)\right) + \frac{1}{(2\beta)^{2}\pi^{\frac{3}{2}}} \frac{\kappa}{N_{\kappa,\sigma}},
\end{multline}
where we used some integral comparisons. Since the r.h.s. of \eqref{fondlowb2} and \eqref{majd=2}-\eqref{majd=3} converge $\forall 0<\sigma<d$ and uniformly in $(\bold{x},\bold{y}) \in \mathbb{R}^{2d}$ to the same value when $\kappa \downarrow 0$, then $\lim_{\kappa \downarrow 0} r_{\infty,\kappa}(\bold{x},\bold{y};\beta,\nu)$ exists  by the squeeze theorem and equals \eqref{Thm1Eq2}. We emphasize that the result strongly relies on the asymptotic form of the chemical potential in \eqref{musol3}. When $\nu=\nu_{c}(\beta)$, we use a similar method but the upper bounds in \eqref{majd=2} and \eqref{majd=3} have to be replaced with some independent of the difference $E_{\infty,\kappa}^{(\bold{0})} - \overline{\mu}_{\infty,\kappa}$. Indeed, the asymptotic in \eqref{musol2} does not allow us to conclude from the bounds in \eqref{majd=2} and \eqref{majd=3} because of the presence of the $\ln$ in \eqref{Euler}. In the case of $d=2$, from \eqref{fondlowb} and \eqref{decomo2}, then one has $\forall 0<\sigma<2$, $\
\forall (\bold{x},\bold{y}) \in \mathbb{R}^{4}$ and for $\kappa<1$ sufficiently small:
\begin{multline}
\label{othmet}
\frac{\nu_{\infty,\kappa}(\beta,\nu;\bold{0})}{\pi} \mathrm{e}^{- \frac{\kappa}{4}\vert \bold{x} + \bold{y}\vert^{2}} \mathrm{e}^{-\frac{\kappa}{4} \vert \bold{x}-\bold{y}\vert^{2} \coth\left(\frac{\beta}{2} \kappa^{1-\sigma}\right)}  \mathrm{e}^{- \frac{\beta}{2} \kappa^{1-\sigma}}\\
\leq \kappa \sum_{l=N_{\kappa,\sigma}+1}^{\infty}
\mathrm{e}^{l\beta \overline{\mu}_{\infty,\kappa}}G_{\infty,\kappa}^{(d=2)}(\bold{x},\bold{y};l\beta) \leq \frac{\nu_{\infty,\kappa}(\beta,\nu;\bold{0})}{\pi} \left( 1+ \frac{1}{2\beta} \frac{1}{\kappa N_{\kappa,\sigma}}\right).
\end{multline}
To derive the upper bound, we minorized the $l$ in the denominator of the second term in the r.h.s. of \eqref{decomo2} before extending the sum up to $l=1$. In order to apply the squeeze theorem (remind that $\lim_{\kappa \downarrow 0} \nu_{\infty,\kappa}(\beta,\nu_{c}(\beta);\bold{0})=0$, see \eqref{nu02}), the limiting condition $d>\sigma>1$ is required. From \eqref{decomo2'} and by using similar arguments, if $d=3$ the same limitation is required as well.\\

\textbf{Part $\mathrm{(C)}$.} Let $\beta,\nu>0$ be fixed. We start with \eqref{Thm1Eq1}. From \eqref{funcpd}-\eqref{funcp1} with \eqref{nu(s)}-\eqref{nu0}:
\begin{equation*}
\lim_{\kappa\downarrow 0}\nu_{\infty,\kappa}(\beta,\nu;\bold{0})
\Psi_{\infty,1}^{(\bold{0})}(\bold{x}\sqrt{\kappa}) \Psi_{\infty,1}^{(\bold{0})}(\bold{y}\sqrt{\kappa})
=\lim_{\kappa\downarrow 0} \nu_{\infty,\kappa}(\beta,\nu;\bold{0}) \frac{\mathrm{e}^{-\frac{\kappa}{2}\left(\vert\bold{x}\vert^2+\vert\bold{y}\vert^2\right)}}{\pi^{\frac{d}{2}}}
=\frac{\nu_{\infty,0}(\beta,\nu;\bold{0})}{\pi^{\frac{d}{2}}}.
\end{equation*}
From \eqref{Thm1Eq2}, the r.h.s. is nothing but $r_{\infty,0}(\bold{x},\bold{y};\beta,\nu)$ which is independent of $\bold{x},\bold{y} \in \mathbb{R}^{2d}$. \\
We continue with \eqref{Thm1Eq3}. Let us mention that the reduced density matrix can be rewritten as:
\begin{equation}
\label{corr1}
\forall(\bold{x},\bold{y}) \in \mathbb{R}^{2d},\quad \rho_{\infty,\kappa}(\bold{x},\bold{y};\beta,\nu)
= \frac{\Psi_{\infty,\kappa}^{(\bold{0})}(\bold{x}) \Psi_{\infty,\kappa}^{(\bold{0})}(\bold{y})}{\mathrm{e}^{\beta\left(E_{\infty,\kappa}^{(\bold{0})} - \overline{\mu}_{\infty,\kappa}\right)} - 1}
+\sum_{\bold{s} \in (\mathbb{N}^{*})^{d}} \frac{\Psi_{\infty,\kappa}^{(\bold{s})}(\bold{x}) \Psi_{\infty,\kappa}^{(\bold{s})}(\bold{y})}{\mathrm{e}^{\beta\left(E_{\infty,\kappa}^{(\bold{s})} - \overline{\mu}_{\infty,\kappa}\right)} - 1}.
\end{equation}
When $\nu < \nu_{c}(\beta)$ if $d=1,2,3$, the first quantity in the r.h.s. of \eqref{corr1} vanishes in the limit $\kappa \downarrow 0$. Then \eqref{Thm1Eq31} follows from \eqref{rdmd21}. We turn to the cases of $\nu \geq \nu_{c}(\beta)$ if $d=2$, $\nu>\nu_{c}(\beta)$ if $d=3$.  Similarly to \eqref{rdm4}, we decompose $\forall 0<\kappa<1$ the reduced density matrix into two contributions:
\begin{equation}
\label{corr2}
\rho_{\infty,\kappa}(\bold{x},\bold{y};\beta,\nu)=
\sum_{l=1}^{N_{\kappa,\sigma}} \mathrm{e}^{l \beta \overline{\mu}_{\infty,\kappa}} G_{\infty,\kappa}^{(d)}(\bold{x},\bold{y};l\beta) +
\sum_{l=N_{\kappa,\sigma}+1}^{\infty} \mathrm{e}^{l \beta \overline{\mu}_{\infty,\kappa}} G_{\infty,\kappa}^{(d)}(\bold{x},\bold{y};l\beta),
\end{equation}
where $N_{\kappa,\sigma}=\lfloor \kappa^{-\sigma}\rfloor$ with $0<\sigma<d$ for the moment. When $\nu \geq \nu_{c}(\beta)$ if $d=2$, the strategy consists in finding a lower bound for the l.h.s. of \eqref{Thm1Eq3} involving the first term in the r.h.s. of \eqref{corr2}. $\forall (\bold{x},\bold{y})\in\mathbb{R}^{4}$, $\forall 0<\sigma<2$ and for $\kappa<1$ sufficiently small, one has from \eqref{rdm}-\eqref{otrep}:
\begin{multline}
\label{sum}
\sum_{l=1}^{N_{\kappa,\sigma}}\mathrm{e}^{l\beta\overline{\mu}_{\infty,\kappa}} G_{\infty,\kappa}^{(d=2)}(\bold{x},\bold{y};l\beta)
=\sum_{l=1}^{N_{\kappa,\sigma}} \sum_{\bold{s} \in \mathbb{N}^{2}} \mathrm{e}^{l\beta\left(\overline{\mu}_{\infty,\kappa}-E_{\infty,\kappa}^{(\bold{s})}\right)}
\Psi_{\infty,\kappa}^{(\bold{s})}(\bold{x})\Psi_{\infty,\kappa}^{(\bold{s})}(\bold{y})\\
\leq \frac{\kappa}{\pi} \frac{1-\mathrm{e}^{-\beta N_{\kappa,\sigma}\left(E_{\infty,\kappa}^{(\bold{0})}-\overline{\mu}_{\infty,\kappa}\right)}}
{\mathrm{e}^{\beta\left(E_{\infty,\kappa}^{(\bold{0})}-\overline{\mu}_{\infty,\kappa}\right)}-1} \mathrm{e}^{-\frac{\kappa}{2}\left(\vert\bold{x}\vert^{2}+\vert\bold{y}\vert^{2}\right)}
+\sum_{\bold{s} \in (\mathbb{N}^{*})^{2}}\frac{\Psi_{\infty,\kappa}^{(\bold{s})}(\bold{x}) \Psi_{\infty,\kappa}^{(\bold{s})}(\bold{y})}
{\mathrm{e}^{\beta\left(E_{\infty,\kappa}^{(\bold{s})}-\overline{\mu}_{\infty,\kappa}\right)}-1},
\end{multline}
where we separated the case $\bold{s}=\bold{0}$ from the sum over $\bold{s}$ before extending to $\infty$ the sum over $l$ in the second term of \eqref{sum}. Under the same conditions, and since $\overline{\mu}_{\infty,\kappa} \geq 0$ for $\kappa<1$ small enough:
\begin{multline}
\label{sum2}
\sum_{\bold{s}\neq\bold{0}}\frac{\Psi_{\infty,\kappa}^{(\bold{s})}(\bold{x}) \Psi_{\infty,\kappa}^{(\bold{s})}(\bold{y})}
{\mathrm{e}^{\beta\left(E_{\infty,\kappa}^{(\bold{s})}-\overline{\mu}_{\infty,\kappa}\right)}-1} \geq \sum_{l=1}^{N_{\kappa,\sigma}} \mathrm{e}^{l\beta\overline{\mu}_{\infty,\kappa}} G_{\infty,\kappa}^{(d=2)}(\bold{x},\bold{y};l\beta) - \frac{\kappa}{\pi} \frac{1-\mathrm{e}^{-\beta N_{\kappa,\sigma}\left(E_{\infty,\kappa}^{(\bold{0})}-\overline{\mu}_{\infty,\kappa}\right)}}
{\mathrm{e}^{\beta\left(E_{\infty,\kappa}^{(\bold{0})}-\overline{\mu}_{\infty,\kappa}\right)}-1} \mathrm{e}^{-\frac{\kappa}{2}\left(\vert\bold{x}\vert^{2}+\vert\bold{y}\vert^{2}\right)}\\
\geq \frac{1}{2\beta \pi} \left(\mathrm{e}^{-\frac{\kappa}{4} \vert\bold{x}+\bold{y}\vert^{2}} \mathrm{e}^{-\frac{1}{4}\left(\kappa+\frac{2}{\beta}\right)\vert\bold{x}-\bold{y}\vert^{2}} \ln{\left(\frac{\kappa^{-\sigma} - 1}{1+\beta\kappa^{1-\sigma}}\right)}
- 2\beta \kappa^{1-\sigma} \mathrm{e}^{-\frac{\kappa}{2}\left(\vert\bold{x}\vert^{2}+\vert\bold{y}\vert^{2}\right)}\right),
\end{multline}
and the above lower bound diverges when $\kappa \downarrow 0$ $\forall 0<\sigma<1$ and $\forall(\bold{x},\bold{y}) \in \mathbb{R}^{4}$. In the l.h.s. of the second inequality, we majorized the second term by the lower and upper bound in \eqref{Ink2} and \eqref{Ink1} respectively, then we minorized the sum by an integral (as we did in \eqref{lowbo} from \eqref{compa1}) and used \eqref{Ek4}. Next, we treat the case of $\nu > \nu_{c}(\beta)$ if $d=3$. The strategy consists in showing that the l.h.s. of \eqref{Thm1Eq3} equals the limit $\kappa \downarrow 0$ of the first term in the r.h.s. of \eqref{corr2} for some suitable $\sigma$. $\forall (\bold{x},\bold{y})\in\mathbb{R}^{6}$, $\forall 0<\sigma<3$ and $\forall 0<\kappa<\min\{1,K_{\beta}\}$ (see \eqref{majptio}), one has:
\begin{equation}
\label{estLB}
\begin{split}
&\sum_{l=N_{\kappa,\sigma}+1}^{\infty} \mathrm{e}^{l \beta \overline{\mu}_{\infty,\kappa}}
G_{\infty,\kappa}^{(d=3)}(\bold{x},\bold{y};l\beta)
-\frac{\Psi_{\infty,\kappa}^{(\bold{0})}(\bold{x}) \Psi_{\infty,\kappa}^{(\bold{0})}(\bold{y})}
{\mathrm{e}^{\beta\left(E_{\infty,\kappa}^{(\bold{0})} - \overline{\mu}_{\infty,\kappa}\right)} - 1} \\
&\geq \frac{\kappa^{\frac{3}{2}}}{\mathrm{e}^{\beta\left(E_{\infty,\kappa}^{(\bold{0})}- \overline{\mu}_{\infty,\kappa}\right)}-1} \frac{\mathrm{e}^{-\frac{\kappa}{2}\left(\vert\bold{x}\vert^2+ \vert \bold{y}\vert^2\right)}}{\pi^{\frac{3}{2}}}
\left(\mathrm{e}^{-\frac{\kappa}{4}\left[\coth\left(\frac{\beta}{2} \kappa^{1-\sigma}\right)-1\right]\vert \bold{x}-\bold{y}\vert^{2}}\mathrm{e}^{-\beta N_{\kappa,\sigma} \left(E_{\infty,\kappa}^{(\bold{0})}-\overline{\mu}_{\infty,\kappa}\right)}-1\right)\\
&\geq  - \frac{1}{2} \frac{\nu_{\infty,\kappa}(\beta,\nu;\bold{0})}{\kappa^{\frac{3}{2}}} \frac{\mathrm{e}^{-\frac{\kappa}{2}\left(\vert\bold{x}\vert^2+ \vert \bold{y}\vert^2\right)}}{\pi^{\frac{3}{2}}}
\left(\kappa \frac{\mathrm{e}^{- \beta \kappa^{1-\sigma}}}{1-\mathrm{e}^{- \beta \kappa^{1-\sigma}}}\vert \bold{x}-\bold{y}\vert^{2}
+ 3 \frac{\kappa^{3-\sigma}}{\left(\nu-\nu_c(\beta)\right)}\right),
\end{split}
\end{equation}
and the above lower bound vanishes when $\kappa \downarrow 0$ $\forall 1<\sigma< \frac{3}{2}$ and $\forall (\bold{x},\bold{y}) \in \mathbb{R}^{6}$. To get the r.h.s. of the second inequality from the l.h.s., we used the lower bound in \eqref{Ink2}. Here, the exponential decay in $\kappa^{1-\sigma}$ arises from the difference $\coth(\frac{\beta}{2} \kappa^{1-\sigma}) - 1$. Under the same conditions than \eqref{estLB}:
\begin{multline}
\label{estUB}
\sum_{l=N_{\kappa,\sigma}+1}^{\infty} \mathrm{e}^{l \beta \overline{\mu}_{\infty,\kappa}} G_{\infty,\kappa}^{(d=3)}(\bold{x},\bold{y};l\beta) -\frac{\Psi_{\infty,\kappa}^{(\bold{0})}(\bold{x}) \Psi_{\infty,\kappa}^{(\bold{0})}(\bold{y})}
{\mathrm{e}^{\beta\left(E_{\infty,\kappa}^{(\bold{0})} - \overline{\mu}_{\infty,\kappa}\right)} - 1} \leq
\frac{\kappa^{\frac{3}{2}}}{\mathrm{e}^{\beta(E_{\infty,\kappa}^{(\bold{0})}-\overline{\mu}_{\infty,\kappa})}-1} \frac{\mathrm{e}^{-\frac{\kappa}{2} \left(\vert\bold{x}\vert^2+\vert\bold{y}\vert^2\right)}}
{\pi^{\frac{3}{2}}} \\
\times
\left(\mathrm{e}^{\frac{\kappa}{4}\left(1-\tanh\left(\frac{\kappa}{2}\beta \kappa^{1-\sigma}\right)\right)\vert\bold{x}+\bold{y}\vert^{2}}-1\right)
 +\frac{\kappa^{\frac{3}{2}}}{\pi^{\frac{3}{2}}} \sum_{l=N_{\kappa,\sigma}+1}^{\infty}
\left(\frac{1}{\kappa l\beta}+\frac{1}{(2l\beta\kappa)^{2}}\right)
\mathrm{e}^{-l\beta\left(E_{\infty,\kappa}^{(\bold{0})}-\overline{\mu}_{\infty,\kappa}\right)},
\end{multline}
where we used \eqref{decomo2'1}. Afterwards, by using the upper bound in \eqref{Ink2} and the argument which lead to the estimate in \eqref{majd=3}, then under the conditions of \eqref{estLB}, the r.h.s. of \eqref{estUB} is less than:
\begin{multline}
\label{estUB2}
\frac{\nu_{\infty,\kappa}(\beta,\nu;\bold{0})}{\kappa^{\frac{3}{2}}} \frac{\mathrm{e}^{-\frac{\kappa}{2} \left(\vert\bold{x}\vert^2+\vert\bold{y}\vert^2\right)}}{\pi^{\frac{3}{2}}} \frac{\kappa}{2} \frac{\mathrm{e}^{-\beta \kappa^{1-\sigma}}}{1 + \mathrm{e}^{-\beta \kappa^{1 -\sigma}}} \vert\bold{x}+\bold{y}\vert^{2} \mathrm{e}^{\frac{\kappa}{2} \mathrm{e}^{-\beta \kappa^{1 -\sigma}} \vert\bold{x}+\bold{y}\vert^{2}} + \\+ \frac{\sqrt{\kappa}}{\beta \pi^{\frac{3}{2}}} \Gamma_{0}\left(\frac{\kappa^{3}\left(\kappa^{-\sigma}-1\right)}{2\left(\nu-\nu_{c}(\beta)\right)}\right) + \frac{1}{(2\beta)^{2} \pi^{\frac{3}{2}}}\frac{\kappa^{\sigma - \frac{1}{2}}}{1 - \kappa^{\sigma}}.
\end{multline}
Here, the exponential decay in $\kappa^{1-\sigma}$ arises from the difference $1 - \tanh(\frac{\beta}{2} \kappa^{1-\sigma}) - 1$. Since \eqref{estUB2} vanishes when $\kappa \downarrow 0$ $\forall \sigma \in (1,3)$ and $\forall (\bold{x},\bold{y}) \in \mathbb{R}^{6}$, we conclude from \eqref{estLB} by the squeeze theorem:
\begin{equation}
\label{efedesez}
\forall 1<\sigma<\frac{3}{2},\quad \lim_{\kappa\downarrow 0}\left(\sum_{l=N_{\kappa,\sigma}+1}^{\infty} \mathrm{e}^{l \beta \overline{\mu}_{\infty,\kappa}} G_{\infty,\kappa}^{(d=3)}(\bold{x},\bold{y};l\beta)
-\frac{\Psi_{\infty,\kappa}^{(\bold{0})}(\bold{x}) \Psi_{\infty,\kappa}^{(\bold{0})}(\bold{y})}{\mathrm{e}^{\beta\left(E_{\infty,\kappa}^{(\bold{0})} - \overline{\mu}_{\infty,\kappa}\right)} - 1}\right)
=0.
\end{equation}
In view of \eqref{corr1}, \eqref{corr2} and the foregoing, to prove \eqref{Thm1Eq33} it remains to show that:
\begin{equation}
\label{corr4}
\forall 1<\sigma<\frac{3}{2},\,\forall (\bold{x},\bold{y}) \in \mathbb{R}^{6},\quad \lim_{\kappa\downarrow0}
\sum_{l=1}^{N_{\kappa,\sigma}} \mathrm{e}^{l \beta \overline{\mu}_{\infty,\kappa}} G_{\infty,\kappa}^{(d=3)}(\bold{x},\bold{y};l\beta) = \sum_{l=1}^{\infty} \frac{1}{(2\pi l\beta)^{\frac{3}{2}}}
\mathrm{e}^{-\frac{\vert \bold{x}-\bold{y}\vert^2}{2l\beta}}.
\end{equation}
On the one hand, $\forall (\bold{x},\bold{y})\in\mathbb{R}^{6}$, $\forall 0<\sigma<3$ and $\forall 0<\kappa<\min\{1,K_{\beta}\}$ (see \eqref{majptio}), one has:
\begin{equation}
\label{jhg1}
\sum_{l=1}^{N_{\kappa,\sigma}} \mathrm{e}^{l \beta \overline{\mu}_{\infty,\kappa}} G_{\infty,\kappa}^{(d=3)}(\bold{x},\bold{y};l\beta) \geq \mathrm{e}^{-\frac{\kappa}{4} \vert \bold{x}+\bold{y}\vert^{2}} \mathrm{e}^{-\frac{\kappa}{4} \vert \bold{x}-\bold{y}\vert^{2}}\mathrm{e}^{-\frac{3}{2} \frac{\kappa^{3-\sigma}}{\nu-\nu_{c}(\beta)}} \sum_{l=1}^{N_{\kappa,\sigma}} \frac{\mathrm{e}^{-\frac{\vert \bold{x} - \bold{y}\vert^{2}}{2\beta l}}}{(2\pi \beta l)^{\frac{3}{2}}},
\end{equation}
where we used the upper bounds in \eqref{Ink1} and the ones in \eqref{Ek3}-\eqref{Ek4}. On the other hand, from \eqref{decomo2'2} together with \eqref{ingamf}, one has under the same conditions than \eqref{jhg1}:
\begin{equation*}
\sum_{l=1}^{N_{\kappa,\sigma}} \mathrm{e}^{l \beta \overline{\mu}_{\infty,\kappa}} G_{\infty,\kappa}^{(d=3)}(\bold{x},\bold{y};l\beta)
\leq \sum_{l=1}^{N_{\kappa,\sigma}} \frac{\mathrm{e}^{-\frac{\vert \bold{x} - \bold{y}\vert^{2}}{2\beta l}}}{(2\pi l\beta)^{\frac{3}{2}}} + 2 \frac{\mathrm{e}^{-\frac{\kappa^{3}}{2\left(\nu - \nu_{c}(\beta)\right)}}}{\pi^{\frac{3}{2}}} \left(\kappa^{\frac{3}{2}-\sigma} + \frac{3}{\sqrt{2\beta}} \kappa^{1 - \frac{\sigma}{2}} + \frac{3}{4\beta} \sqrt{\kappa} \ln\left(\frac{\mathrm{e}}{\kappa^{\sigma}}\right)\right).
\end{equation*}
From \eqref{jhg1} along with the above upper bound, \eqref{corr4} follows from the squeeze theorem.\qed

\subsection{Proof of Theorem \ref{Thm2}.}

\indent \textbf{Part $\mathrm{(A)}$}. Let $\beta,\nu>0$ be fixed.
\begin{itemize}
\item Case of $\nu< \nu_{c}(\beta)$ if $d=1,2,3$ - Proof of \eqref{dilat1}-\eqref{dilat2}.
\end{itemize}
At first, let us note that from \eqref{musol1} and \eqref{Mehler}-\eqref{multd}, one has $\forall l \in \mathbb{N}^{*}$ and $\forall \bold{x} \in \mathbb{R}^{d}$:
\begin{equation}
\label{pointlim}
\lim_{\kappa \downarrow 0} \mathrm{e}^{l\beta \overline{\mu}_{\infty,\kappa}} G_{\infty,\kappa}^{(d)}\left(\bold{x}\kappa^{-\delta},\bold{x} \kappa^{-\delta};l\beta\right) = \frac{\mathrm{e}^{l\beta \overline{\mu}_{\infty,0}}}{(2\pi l \beta)^{\frac{d}{2}}} \times \left\{\begin{array}{ll}
1,\,\,\,&\textrm{if $1>\delta \geq 0$}, \\
\mathrm{e}^{-\frac{1}{2} l\beta \vert \bold{x}\vert^{2}},\,\,&\textrm{if $\delta=1$}.
\end{array}\right.
\end{equation}
Here, we used the following:
\begin{equation*}
\forall l \in \mathbb{N}^{*},\quad \lim_{\kappa \downarrow 0} \left(\frac{\kappa}{2\pi \sinh(\kappa l\beta)}\right)^{\frac{d}{2}}= \frac{1}{(2\pi l\beta)^{\frac{d}{2}}},\quad \lim_{\kappa \downarrow 0} \kappa^{1 -2\delta} \tanh\left(\frac{\kappa}{2} l\beta\right) = \left\{\begin{array}{ll}
0,\,\,\,&\textrm{if $1- 2\delta > -1$},\\
\frac{l\beta}{2},\,\,\,&\textrm{if $1-2\delta = -1$}.
\end{array}\right.
\end{equation*}
Subsequently, from \eqref{musol1} again, there exists a $\kappa_{0}>0$ s.t. $\forall 0<\kappa \leq \kappa_{0}$, $\overline{\mu}_{\infty,\kappa} \leq \frac{\overline{\mu}_{\infty,0}}{2} < 0$. By using the rough upper bound in \eqref{roughes}, then one has $\forall 0 \leq \delta \leq 1$ and uniformly on $\mathbb{R}^{d}$:
\begin{equation*}
\forall 0 < \kappa \leq \kappa_{0}, \quad  \mathrm{e}^{l\beta \overline{\mu}_{\infty,\kappa}} G_{\infty,\kappa}^{(d)}\left(\bold{x}\kappa^{-\delta},\bold{x} \kappa^{-\delta};l\beta\right) \leq (2\pi l\beta)^{-\frac{d}{2}} \mathrm{e}^{l\beta \frac{\overline{\mu}_{\infty,0}}{2}}.
\end{equation*}
By standard arguments, it follows:
\begin{equation*}
\lim_{\kappa \downarrow 0} \sum_{l=1}^{\infty} \mathrm{e}^{l\beta \overline{\mu}_{\infty,\kappa}} G_{\infty,\kappa}^{(d)}\left(\bold{x}\kappa^{-\delta},\bold{x} \kappa^{-\delta};l\beta\right) = \sum_{l=1}^{\infty} \lim_{\kappa \downarrow 0}  \mathrm{e}^{l\beta \overline{\mu}_{\infty,\kappa}} G_{\infty,\kappa}^{(d)}\left(\bold{x}\kappa^{-\delta},\bold{x} \kappa^{-\delta};l\beta\right),
\end{equation*}
what proves \eqref{dilat1} because of \eqref{pointlim}. From \eqref{rdmrescaled}, \eqref{dilat2} results directly from \eqref{dilat1}.
\begin{itemize}
\item Case of $\nu \geq \nu_{c}(\beta)$ if $d=2,3$ - Proof of \eqref{dilat3}-\eqref{dilat3'}.
\end{itemize}
Let us start with \eqref{dilat31}-\eqref{dilat3'1}. We look for a lower bound of the sum in \eqref{otrep} whose the limit $\kappa \downarrow 0$ diverges. If $d=2$, from \eqref{Mehler}-\eqref{multd}, then $\forall l \in \mathbb{N}^{*}$, $\forall \bold{x} \in \mathbb{R}^{2}$ and for $\kappa>0$ small enough:
\begin{equation*}
\mathrm{e}^{l\beta \overline{\mu}_{\infty,\kappa}} G_{\infty,\kappa}^{(d=2)}\left(\bold{x} \kappa^{-\delta} ,\bold{x} \kappa^{-\delta};l\beta\right) \geq \frac{1}{l\beta} \frac{\kappa l\beta}{2\pi \sinh(\kappa l \beta)} \mathrm{e}^{- \frac{1}{2} \kappa^{2-2\delta} \vert\bold{x}\vert^{2} l\beta} \geq \frac{1}{2\pi} \frac{1}{l\beta} \mathrm{e}^{-\kappa l\beta} \mathrm{e}^{- \frac{1}{2} \kappa^{2-2\delta} \vert\bold{x}\vert^{2} l\beta}.
\end{equation*}
In the l.h.s. of the second inequality, we used \eqref{Ek4} and $\overline{\mu}_{\infty,\kappa} \geq 0$ for $\kappa>0$ small enough, see \eqref{musol2}-\eqref{musol3}. To get the r.h.s., we used the expansion in power series of the $\sinh$ which yields:
\begin{equation}
\label{expsri}
\frac{\kappa l\beta}{\sinh( \kappa l\beta)} = \frac{\kappa l\beta}{\sum_{m=0}^{\infty} \frac{(\kappa l\beta)^{2m+1}}{(2m+1)!}} = \left(\sum_{m=0}^{\infty} \frac{(\kappa l \beta)^{2m}}{(2m+1)!}\right)^{-1} \geq \left(\cosh(\kappa l\beta)\right)^{-1} \geq \mathrm{e}^{-\kappa l\beta}.
\end{equation}
Then, one gets for $\kappa>0$ sufficiently small and $\forall \bold{x} \in \mathbb{R}^{2}$:
\begin{equation*}
\rho_{\infty,\kappa}\left(\bold{x}\kappa^{-\delta},\bold{x}\kappa^{-\delta};\beta,\nu\right) \geq \frac{1}{2\pi \beta} \int_{1}^{\infty} \mathrm{d}t\, \frac{\mathrm{e}^{-\beta \left(\kappa + \frac{1}{2} \kappa^{2-2\delta} \vert\bold{x}\vert^{2}\right)t}}{t} = \Gamma_{0}\left(\beta\left(\kappa + \frac{1}{2}\kappa^{2-2\delta} \vert\bold{x}\vert^{2}\right)\right),
\end{equation*}
where $\Gamma_{0}$ is the incomplete Gamma function in \eqref{Euler}. In the limit $\kappa \downarrow 0$, the above lower bound diverges $\forall 0\leq \delta \leq 1$  if $\bold{x}=\bold{0}$, $\forall 0\leq \delta<1$ otherwise. If $d=3$, from \eqref{Mehler}-\eqref{multd} and by mimicking the arguments leading to \eqref{lowbo1}, then $\forall l \in \mathbb{N}^{*}$, $\forall \bold{x} \in \mathbb{R}^{3}$ and for $\kappa>0$ small enough:
\begin{equation*}
\rho_{\infty,\kappa}\left(\bold{x}\kappa^{-\delta},\bold{x}\kappa^{-\delta};\beta,\nu\right) \geq \frac{1}{(2\pi)^{\frac{3}{2}}\beta} \mathrm{e}^{-\kappa^{1-2\delta} \vert \bold{x}\vert^{2}} \frac{\kappa^{\frac{3}{2}}}{E_{\infty,\kappa}^{(\bold{0})}-\overline{\mu}_{\infty,\kappa}} \mathrm{e}^{-\beta\left(E_{\infty,\kappa}^{(\bold{0})} -\overline{\mu}_{\infty,\kappa}\right)}.
\end{equation*}
By \eqref{musol3}, the above lower bound diverges in the limit $\kappa \downarrow 0$ $\forall 0 \leq \delta \leq 1$ if $\bold{x}=\bold{0}$, $\forall 0\leq \delta\leq\frac{1}{2}$ otherwise.  Next, let us prove \eqref{dilat32} and \eqref{dilat3'2}-\eqref{dilat3'3}. To do so, we first give a lower bound for the quantity $\rho_{\infty,\kappa}(\bold{x}\kappa^{-\delta},\bold{x}\kappa^{-\delta};\beta,\nu)$ when $\delta=1$ if $d=2$ and when $\frac{1}{2} < \delta \leq 1$ if $d=3$.\\
If $d=2$ and $\delta=1$, for any $\bold{x} \in (\mathbb{R}^{*})^{2}$ and for $\kappa>0$ small enough:
\begin{equation}
\label{negf1}
\rho_{\infty,\kappa}\left(\bold{x}\kappa^{-1},\bold{x}\kappa^{-1};\beta,\nu\right) \geq \frac{1}{2\pi \beta} \sum_{l=1}^{\infty} \frac{\mathrm{e}^{-l\beta\left[\left(E_{\infty,\kappa}^{(\bold{0})} - \overline{\mu}_{\infty,\kappa}\right)+ \frac{1}{2}\vert \bold{x}\vert^{2}\right]}}{l} = \frac{\mathpzc{g}_{1}\left(\mathrm{e}^{-\beta\left[\left(E_{\infty,\kappa}^{(\bold{0})} - \overline{\mu}_{\infty,\kappa}\right)+ \frac{1}{2}\vert \bold{x}\vert^{2}\right]}\right)}{2\pi \beta},
\end{equation}
where we used \eqref{expsri}. If $d=3$ and $1 \geq \delta > \frac{1}{2}$, $\forall \bold{x} \in (\mathbb{R}^{*})^{3}$ and for $\kappa>0$ small enough:
\begin{equation}
\label{negf2}
\rho_{\infty,\kappa}\left(\bold{x}\kappa^{-\delta},\bold{x}\kappa^{-\delta};\beta,\nu\right) \geq \frac{\mathpzc{g}_{\frac{3}{2}}\left(\mathrm{e}^{-\beta\left[\left(E_{\infty,\kappa}^{(\bold{0})} - \overline{\mu}_{\infty,\kappa}\right)+ \frac{1}{2}\vert \bold{x}\vert^{2} \kappa^{2-2\delta}\right]}\right)}{(2\pi \beta)^{\frac{3}{2}}}.
\end{equation}
In the limit $\kappa \downarrow 0$, the lower bounds in \eqref{negf1} and \eqref{negf2} converge to \eqref{dilat32} and \eqref{dilat3'2}-\eqref{dilat3'3}  respectively  following the values of $\delta$. Secondly, we give an upper bound for $\rho_{\infty,\kappa}(\bold{x}\kappa^{-\delta},\bold{x}\kappa^{-\delta};\beta,\nu)$ when $\delta=1$ if $d=2$ and when $\frac{1}{2} < \delta \leq 1$ if $d=3$ whose the limit $\kappa \downarrow 0$ reduced to the announced results. Under these conditions, introduce $\forall 0<\kappa < 1$ and $\forall \bold{x} \in (\mathbb{R}^{*})^{d}$ the decomposition:
\begin{equation}
\label{newdecom}
\rho_{\infty,\kappa}\left(\bold{x}\kappa^{-\delta},\bold{x}\kappa^{-\delta};\beta,\nu\right) =  \left\{\sum_{l=1}^{N_{\kappa,\sigma}}
\mathrm{e}^{l\beta\overline{\mu}_{\infty,\kappa}} + \sum_{l=N_{\kappa,\sigma}+1}^{\infty}
\mathrm{e}^{l\beta\overline{\mu}_{\infty,\kappa}}\right\} G_{\infty,\kappa}^{(d)} \left(\bold{x}\kappa^{-\delta},\bold{x}\kappa^{-\delta};l\beta\right),
\end{equation}
where $N_{\kappa,\sigma}:=\lfloor \kappa^{-\sigma}\rfloor$ with $0<\sigma <d$ for the moment. Let us give an upper bound for the second term in the r.h.s. of \eqref{newdecom}. From \eqref{decomo2}-\eqref{decomo2'1} and by mimicking the arguments leading to \eqref{majd=2} and \eqref{majd=3}, one has $\forall d \in \{2,3\}$, $\forall \bold{x} \in (\mathbb{R}^{*})^{d}$, $\forall 0<\sigma <d$ and for $\kappa<1$ sufficiently small:
\begin{multline}
\label{vghu01}
\sum_{l=N_{\kappa,\sigma}+1}^{\infty}
\mathrm{e}^{l\beta\overline{\mu}_{\infty,\kappa}} G_{\infty,\kappa}^{(d)} \left(\bold{x}\kappa^{-\delta},\bold{x}\kappa^{-\delta};l\beta\right)
\leq \mathrm{e}^{-\frac{1}{2}\beta \vert \bold{x}\vert^{2} \frac{\kappa^{2-2\delta - \sigma}}{1 + \beta \kappa^{1-\sigma}}}  \\
\times  \frac{1}{\beta \pi^{\frac{d}{2}}} \left\{ \kappa^{\frac{d}{2}} \frac{\mathrm{e}^{-\beta\left(E_{\infty,\kappa}^{(\bold{0})}-\overline{\mu}_{\infty,\kappa}\right)}}{ E_{\infty,\kappa}^{(\bold{0})}-\overline{\mu}_{\infty,\kappa}} + \frac{\kappa^{\frac{d}{2} - 1}}{2^{3-d}} \Gamma_{0}\left(N_{\kappa,\sigma} \beta\left(E_{\infty,\kappa}^{(\bold{0})}-\overline{\mu}_{\infty,\kappa}\right)\right) + \frac{d-2}{4 \beta} \frac{1}{\sqrt{\kappa} N_{\kappa,\sigma}} \right\},
\end{multline}
where we used that $\forall l \geq N_{\kappa,\sigma}+1$, $\tanh(\frac{\kappa \beta}{2} l) \geq \tanh(\frac{\beta}{2}\kappa^{1-\sigma})$ followed by the upper bound in \eqref{Ek4}. From \eqref{musol3}, the upper bound in \eqref{vghu01} vanishes in the limit $\kappa \downarrow 0$ $\forall 2-2\delta <\sigma < d$ if $d=2,3$ and $\forall \bold{x} \in (\mathbb{R}^{*})^{d}$. Let us give an upper bound for the first term in the r.h.s. of \eqref{newdecom}. From \eqref{decomo2}, \eqref{decomo2'2} and \eqref{ingamf},  $\forall d \in \{2,3\}$, $\forall \bold{x} \in (\mathbb{R}^{*})^{d}$, $\forall 0<\sigma <d$ and for $\kappa<1$ small enough:
\begin{multline}
\label{vghu02}
\sum_{l=1}^{N_{\kappa,\sigma}}
\mathrm{e}^{l\beta\overline{\mu}_{\infty,\kappa}} G_{\infty,\kappa}^{(d)} \left(\bold{x}\kappa^{-\delta},\bold{x}\kappa^{-\delta};l\beta\right) \\
\leq \sum_{l=1}^{\infty} \frac{\mathrm{e}^{-l\beta\left(E_{\infty,\kappa}^{(\bold{0})}-\overline{\mu}_{\infty,\kappa}\right)}}{(2\pi l \beta)^{\frac{d}{2}}} \mathrm{e}^{- \frac{1}{2} l\beta \vert \bold{x}\vert^{2}  \frac{\kappa^{2 - 2\delta}}{1 + \beta \kappa^{1-\sigma}}}   - \sum_{l=N_{\kappa,\sigma}+1}^{\infty} \frac{\mathrm{e}^{-l\beta\left[\left(E_{\infty,\kappa}^{(\bold{0})}-\overline{\mu}_{\infty,\kappa}\right) + \frac{\vert \bold{x}\vert^{2}}{2} \frac{\kappa^{2-2\delta}}{1 + \beta \kappa^{1-\sigma}}\right]}}{(2\pi l \beta)^{\frac{d}{2}}}   +  \\+ 2 \frac{\mathrm{e}^{-\beta\left(E_{\infty,\kappa}^{(\bold{0})}- \overline{\mu}_{\infty,\kappa}\right)}}{\pi^{\frac{d}{2}}} \left(\kappa^{\frac{d}{2}-\sigma} + \frac{3(d-2)}{\sqrt{2\beta}} \kappa^{1 - \frac{\sigma}{2}} + \frac{3(d-2)}{4\beta} \sqrt{\kappa} \ln\left(\frac{\mathrm{e}}{\kappa^{\sigma}}\right)\right).
\end{multline}
From the asymptotic in \eqref{musol3}, the last term in the r.h.s. of \eqref{vghu02} vanishes in the limit $\kappa \downarrow 0$ $\forall 0<\sigma<\frac{d}{2}$ if $d=2,3$. By an integral comparison, one can prove that the second term vanishes in the limit $\kappa \downarrow 0$ $\forall 2-2\delta<\sigma<d$ if $d=2,3$ and $\forall \bold{x} \in (\mathbb{R}^{*})^{d}$. Finally by standard arguments, one has $\forall 0<\sigma<1$ if $d=2$, $\forall 0<\sigma<d$ if $d=3$ and $\forall \bold{x} \in (\mathbb{R}^{*})^{d}$:
\begin{equation*}
\lim_{\kappa \downarrow 0}\sum_{l=1}^{\infty} \frac{\mathrm{e}^{-l\beta\left(E_{\infty,\kappa}^{(\bold{0})}-\overline{\mu}_{\infty,\kappa}\right)}}{(2\pi l \beta)^{\frac{d}{2}}} \mathrm{e}^{- \frac{1}{2} l\beta \vert \bold{x}\vert^{2} \frac{\kappa^{2 - 2\delta }}{1 + \beta \kappa^{1-\sigma}}} = \sum_{l=1}^{\infty} \lim_{\kappa \downarrow 0} \frac{\mathrm{e}^{-l\beta\left(E_{\infty,\kappa}^{(\bold{0})}-\overline{\mu}_{\infty,\kappa}\right)}}{(2\pi l \beta)^{\frac{d}{2}}} \mathrm{e}^{- \frac{1}{2} l\beta \vert \bold{x}\vert^{2} \frac{\kappa^{2 - 2\delta }}{1 + \beta \kappa^{1-\sigma}}}.
\end{equation*}
By adding the r.h.s. of \eqref{vghu01} and \eqref{vghu02}, we got an upper bound for the l.h.s. of \eqref{newdecom} converging when $\delta=1$ to \eqref{dilat32}-\eqref{dilat3'3} $\forall 0<\sigma<\frac{d}{2}$ if $d=2,3$, when $\frac{1}{2}<\delta<1$ to \eqref{dilat3'2} $\forall 1<\sigma<\frac{3}{2}$ if $d=3$. In view of \eqref{negf1}-\eqref{negf2}, \eqref{dilat32} and \eqref{dilat3'2}-\eqref{dilat3'3} follow from the squeeze theorem.
\begin{itemize}
\item Case of $\nu \geq \nu_{c}(\beta)$ if $d=2,3$ - Proof of \eqref{dilat4}.
\end{itemize}
Similarly to \eqref{rdm4}, the starting-point is a decomposition of the quantity defined by \eqref{otrep} $\forall 0 \leq \delta \leq 1$, $\forall d\in\{2,3\}$ and $\forall 0<\kappa < 1$ into two contributions:
\begin{equation}
\label{rdm5}
r_{\infty,\kappa}\left(\bold{x}\kappa^{-\delta},\bold{x}\kappa^{-\delta};\beta,\nu\right) = \kappa^{\frac{d}{2}} \left\{ \sum_{l=1}^{N_{\kappa,\sigma}}
\mathrm{e}^{l\beta\overline{\mu}_{\infty,\kappa}}  + \sum_{l=N_{\kappa,\sigma}+1}^{\infty}
\mathrm{e}^{l\beta\overline{\mu}_{\infty,\kappa}}\right\} G_{\infty,\kappa}^{(d)}\left(\bold{x}\kappa^{-\delta}, \bold{x}\kappa^{-\delta};l\beta\right),
\end{equation}
where $N_{\kappa,\sigma}:= \lfloor \kappa^{-\sigma} \rfloor$ with $0<\sigma<d$ for the moment (a limitation will appear if $\delta = \frac{1}{2}$ when $\nu>\nu_{c}(\beta)$ or when $\nu=\nu_{c}(\beta)$). From \eqref{decomo2}-\eqref{ingamf}, one has $\forall 0 \leq \delta \leq 1$, $\forall d \in \{2,3\}$ and $\forall \bold{x} \in \mathbb{R}^{d}$:
\begin{equation*}
\forall 0<\sigma<d,\quad \lim_{\kappa \downarrow 0} \kappa^{\frac{d}{2}} \sum_{l=1}^{N_{\kappa,\sigma}} \mathrm{e}^{l\beta \overline{\mu}_{\infty,\kappa}} G_{\infty,\kappa}^{(d)}\left(\bold{x}\kappa^{-\delta},\bold{x} \kappa^{-\delta};l\beta\right) = 0.
\end{equation*}
Let us investigate the second term in the r.h.s. of \eqref{rdm5}. We distinguish the case of $\nu>\nu_{c}(\beta)$ from the case of $\nu=\nu_{c}(\beta)$. When $\nu>\nu_{c}(\beta)$, by using the same arguments leading to \eqref{fondlowb2} and \eqref{majd=2}-\eqref{majd=3}, one has $\forall 0 \leq \delta \leq 1$, $\forall d \in \{2,3\}$, $\forall \bold{x} \in (\mathbb{R}^{*})^{d}$, $\forall 0<\sigma<d$  and $\forall 0<\kappa<\min\{1,K_{\beta}\}$:
\begin{multline*}
\frac{\nu_{\infty,\kappa}(\beta,\nu;\bold{0})}{\pi^{\frac{d}{2}}} \mathrm{e}^{- \kappa^{1-2\delta} \vert \bold{x}\vert^{2}} \mathrm{e}^{- \frac{3 \kappa^{d-\sigma}}{2(\nu - \nu_{c}(\beta))}} \leq \kappa^{\frac{d}{2}} \sum_{l=N_{\kappa,\sigma}+1}^{\infty} \mathrm{e}^{l\beta \overline{\mu}_{\infty,\kappa}} G_{\infty,\kappa}^{(d)}\left(\bold{x}\kappa^{-\delta},\bold{x} \kappa^{-\delta};l\beta\right) \\
\leq \mathrm{e}^{-\kappa^{1-2\delta} \vert \bold{x}\vert^{2} \tanh\left(\frac{\beta}{2}\frac{1}{\kappa^{\sigma-1}}\right)} \times \left\{\begin{array}{ll}
\textrm{r.h.s. of \eqref{majd=2}},\,\,\, &\textrm{if $d=2$}, \\
\textrm{r.h.s. of \eqref{majd=3}},\,\,\, &\textrm{if $d=3$}.
\end{array}\right.
\end{multline*}
In the limit $\kappa \downarrow 0$, the above l.h.s. and r.h.s. tend to \eqref{dilat41} $\forall 0<\sigma <d$ if $0\leq \delta <\frac{1}{2}$, to \eqref{dilat42} $\forall 1< \sigma<d$ if $\delta =\frac{1}{2}$, and to $0$ $\forall 0<\sigma <d$ if $\frac{1}{2}<\delta \leq 1$. When $\nu > \nu_{c}(\beta)$, \eqref{dilat4} follows from the squeeze theorem. The case of $\nu=\nu_{c}(\beta)$ can be treated by the same arguments than the ones used at the end of the proof of Theorem \ref{Thm1}, see \eqref{othmet}. The limiting condition $d>\sigma>1$ is required.\\

\textbf{Part $\mathrm{(B)}$}. Let $\beta,\nu>0$ be fixed. In view of \eqref{dilat2}-\eqref{dilat4},
\eqref{Thm2BEq1} follows by direct calculations from \eqref{nu0} along with \eqref{funcpd}-\eqref{funcp1}.
Let us turn to $\mathrm{(ii)}$. By setting $\bold{x}=\bold{y}$ in \eqref{corr1} after dilating the spatial variables by $\kappa^{-\delta}$, then from \eqref{funcpd}-\eqref{funcp1} the first term of the r.h.s. can be rewritten as:
\begin{gather}
\label{reec1}
\forall x \in \mathbb{R}^{*}, \quad \frac{\left\vert\psi_{\infty,\kappa}^{(0)}\left(x\kappa^{-\delta}\right)\right\vert^{2}}{\mathrm{e}^{\beta \left(E_{\infty,\kappa}^{(0)}-\overline{\mu}_{\infty,\kappa}\right)}-1} = \frac{\sqrt{\kappa}}{\mathrm{e}^{\beta \left(E_{\infty,\kappa}^{(0)}-\overline{\mu}_{\infty,\kappa}\right)}-1} \frac{\mathrm{e}^{-x^{2} \kappa^{1-2\delta}}}{\sqrt{\pi}},\\
\label{reecd}
\forall d \in \{2,3\},\,\forall \bold{x} \in (\mathbb{R}^{*})^{d},\quad \frac{\left\vert\Psi_{\infty,\kappa}^{(\bold{0})}\left(\bold{x}\kappa^{-\delta}\right) \right\vert^{2}}{\mathrm{e}^{\beta \left(E_{\infty,\kappa}^{(\bold{0})}-\overline{\mu}_{\infty,\kappa} \right)}-1} =
\nu_{\infty,\kappa}(\beta,\nu;\bold{0}) \frac{\mathrm{e}^{-\vert \bold{x}\vert^{2} \kappa^{1-2\delta}}}{(\kappa\pi)^{\frac{d}{2}}}.
\end{gather}
From \eqref{musol1}, the r.h.s. of \eqref{reec1} vanishes in the limit $\kappa \downarrow 0$ $\forall 0\leq \delta \leq 1$. This leads to \eqref{Thm2BEq21}. From \eqref{nu0}, the r.h.s. of \eqref{reecd} vanishes in the limit $\kappa \downarrow 0$ $\forall \frac{1}{2} < \delta \leq 1$ and $\forall d\in \{2,3\}$. This proves \eqref{Thm2BEq24}. Let us turn to the case of $0 \leq \delta \leq \frac{1}{2}$ if $d=2,3$. To do that, it is enough to mimic the arguments used in the proof of Theorem \ref{Thm1}.
By setting $\bold{x}=\bold{y}$ in \eqref{corr1}-\eqref{corr2} after dilating the spatial variable
by $\kappa^{-\delta}$, then since $1-2\delta\geq0$ $\forall 0\leq \delta < \frac{1}{2}$,
the conclusions obtained from \eqref{sum}-\eqref{sum2} and from \eqref{estLB}-\eqref{estUB2} still hold true $\forall \bold{x}\in (\mathbb{R}^{*})^{2}$ and $\forall \bold{x}\in (\mathbb{R}^{*})^{3}$ respectively. \qed

\subsection{Proof of Lemma \ref{Lemma0} and Propositions \ref{Prop1}-\ref{Prop1'}.}

\noindent \textbf{Proof of Lemma \ref{Lemma0}.} Let $d\in \{1,2,3\}$, $\beta>0$ and $\mu<0$ kept fixed. From \eqref{nukappamu}, one has:
\begin{equation}
\label{RiemannSumLimit}
\lim_{\kappa\downarrow0} \sum_{s_1,\ldots,s_d\in \mathbb{N}^{d}}\frac{\kappa^d}{\mathrm{e}^{\beta \left(\kappa (s_1+\dotsb+s_d)+\kappa \frac{d}{2}-\mu\right)}-1} = \int_{0}^{\infty} \mathrm{d}\tau_1 \dotsb \int_{0}^{\infty} \mathrm{d}\tau_d \frac{1}{\mathrm{e}^{\beta(\tau_1+\dotsb+\tau_d-\mu)}-1},
\end{equation}
where the integrals over $\tau_j,\ j=1,\ldots,d$ are obtained by taking the limit $\kappa\downarrow0$
of the Darboux-Riemann sum in the l.h.s of \eqref{RiemannSumLimit}. Therefore, $\lim_{\kappa \downarrow 0} \nu_{\infty,\kappa}(\beta,\mu)$ exists, and by simple calculations:
\begin{equation}
\label{babar}
\nu_{\infty,0}(\beta,\mu) = \frac{1}{\Gamma(d)} \int_{0}^{\infty} \mathrm{d}\tau\,  \frac{\tau^{d-1}}{\mathrm{e}^{\beta(\tau-\mu)}-1}.
\end{equation}
Afterwards, by expanding $(\mathrm{e}^{\beta(\epsilon-\mu)}-1)^{-1}$ in power series and by using the Fubini theorem:
\begin{equation*}
\nu_{\infty,0}(\beta,\mu) = \sum_{l=1}^{\infty} \mathrm{e}^{l\beta\mu} \frac{1}{\Gamma(d)} \int_{0}^{\infty} \mathrm{d}\tau\, \tau^{d-1}\mathrm{e}^{-l\beta \tau} = \sum_{l=1}^{\infty} \frac{\mathrm{e}^{l\beta\mu}}{(l\beta)^d}=\frac{\mathpzc{g}_{d}\left(\mathrm{e}^{\beta\mu}\right)}{\beta^d}.
\end{equation*}
The proof of $\mathrm{(i)}$ is done. $\mathrm{(ii)}$ follows from the definition \eqref{defnuc} together with \eqref{nu(beta,mu)}. Finally, $\mathrm{(iii)}$ results from the fact that $\mu \mapsto \nu_{\infty,0}(\beta,\mu)$ is continuous and strictly increasing on $(-\infty,0)$. \qed \\

\noindent \textbf{Proof of Proposition \ref{Prop1}.} Let $\beta, \nu>0$ kept fixed. We start with the case of $d=1$. From Definition \ref{OpenTrapBECdef} along with \eqref{nucEq1}, then the Bose gas does not manifest an open-trap BEC. Besides,  $\overline{\mu}_{\infty,0}=\overline{\mu}_{\infty,0}(\beta,\nu)$ satisfies $\nu = \beta^{-1} \mathpzc{g}_{1}(\mathrm{e}^{\beta \overline{\mu}_{\infty,0}})$ from \eqref{nu(beta,mu)}. Ergo,  $\overline{\mu}_{\infty,\kappa}=\overline{\mu}_{\infty,\kappa}(\beta,\nu)$ has to obey the asymptotic in \eqref{musol1} since similarly to the calculus performed in the proof of Lemma \ref{Lemma0}:
\begin{equation*}
\label{overlinemu}
\lim_{\kappa\downarrow0}\sum_{s\in \mathbb{N}} \nu_{\infty,\kappa}\left(\beta,\overline{\mu}_{\infty,\kappa};s\right)
=\lim_{\kappa\downarrow0}\sum_{s\in \mathbb{N}}\frac{\kappa}{\mathrm{e}^{\beta\left(\kappa s-\overline{\mu}_{\infty,0} +o(1)\right)}-1}
=\frac{\mathpzc{g}_{1}\left(\mathrm{e}^{\beta\overline{\mu}_{\infty,0}}\right)}{\beta} =\nu.
\end{equation*}
Moreover, the definition in \eqref{nustate} and the asymptotic in \eqref{musol1} lead to  $\nu_{\infty,0}(\beta,\nu;s)=0$ $\forall s \in \mathbb{N}$.
Let us turn to the cases of $d=2,3$. When $\nu<\nu_{c}(\beta)$ and $\nu=\nu_{c}(\beta)$, the $\overline{\mu}_{\infty,\kappa}=\overline{\mu}_{\infty,\kappa}(\beta,\nu)$ has to obey the asymptotic in \eqref{musol1} and \eqref{musol2} respectively by the same arguments than the ones used to treat the case $d=1$. Hence, \eqref{nustate} and \eqref{musol1}-\eqref{musol2} together lead to $\nu_{\infty,0}(\beta,\nu;\bold{s})=0$ $\forall \bold{s} \in \mathbb{N}^{d}$. When $\nu>\nu_{c}(\beta)$, one has to investigate the open-trap
rescaled average number of particles on the ground-state to conclude the existence of an open-trap BEC, see Definition \ref{OpenTrapBECdef}. For such $\nu$'s, assume that the  $\overline{\mu}_{\infty,\kappa}=\overline{\mu}_{\infty,\kappa}(\beta,\nu)$ has the following asymptotic in the limit $\kappa \downarrow 0$:
\begin{equation}\label{musolC}
\overline{\mu}_{\infty,\kappa}=E_{\infty,\kappa}^{(\bold{0})} - C \kappa^d + o\left(\kappa^d\right),
\end{equation}
for some constant $C>0$. Set $\widetilde{\mu}_{\infty,\kappa}:=\overline{\mu}_{\infty,\kappa} - E_{\infty,\kappa}^{(\bold{0})}$. We decompose $\nu_{\infty,\kappa}$ into two contributions:
\begin{equation}
\label{nudecomp}
\forall \kappa>0,\quad \nu=\nu_{\infty,\kappa}\left(\beta,\overline{\mu}_{\infty,\kappa}\right)= \nu_{\infty,\kappa}\left(\beta,\overline{\mu}_{\infty,\kappa};\bold{0}\right) +  \sum_{\bold{s}\in (\mathbb{N}^{*})^d} \nu_{\infty,\kappa}\left(\beta,\overline{\mu}_{\infty,\kappa};\bold{s}\right).
\end{equation}
By mimicking the arguments leading to \eqref{RiemannSumLimit}-\eqref{babar}, the second term of the r.h.s. of \eqref{nudecomp} obeys:
\begin{equation}
\label{nudecomp1}
\lim_{\kappa\downarrow0} \sum_{\bold{s}\in \mathbb{N}^d\,:\,\sum_{j=1}^{d} s_{j} > 0} \frac{\kappa^{d}}{\mathrm{e}^{\beta \left(\kappa \sum_{j=1}^{d} s_{j} - \widetilde{\mu}_{\infty,\kappa}\right)} - 1}
= \frac{1}{\Gamma(d)} \int_{0}^{\infty} \mathrm{d}\tau\, \frac{\tau^{d-1}}{\mathrm{e}^{\beta\tau}-1}=\frac{\mathpzc{g}_{d}(1)}{\beta^d}=\nu_c(\beta).
\end{equation}
This means that the first term in the r.h.s. in \eqref{nudecomp} satisfies, see the definition in \eqref{nu(s)}:
\begin{equation}
\label{musolC2}
\nu_{\infty,0}\left(\beta,\nu;\bold{0}\right) := \lim_{\kappa \downarrow 0} \nu_{\infty,\kappa}\left(\beta,\overline{\mu}_{\infty,\kappa};\bold{0}\right) = \nu -\nu_{c}(\beta)>0.
\end{equation}
Finally, from \eqref{musolC} one has for $\bold{s}=\bold{0}$:
\begin{equation}
\label{musolC3}
\nu_{\infty,\kappa}\left(\beta,\overline{\mu}_{\infty,\kappa};\bold{0}\right)= \frac{\kappa^d}{\mathrm{e}^{\beta\left(C\kappa^d+o\left(\kappa^d\right)\right)}-1}=\frac{1}{\beta C}+o(1) \quad \textrm{when $\kappa \downarrow 0$}.
\end{equation}
Gathering \eqref{musolC3}, \eqref{musolC2} and \eqref{musolC} together, the asymptotic in \eqref{musol3} follows. We emphasize that the asymptotic form in \eqref{musolC} is determined by the limits $\kappa \downarrow 0$ in \eqref{nudecomp1}-\eqref{musolC2}.
Finally, \eqref{nu02} follows from the foregoing. Therefore, the Bose gas manifests an open-trap BEC for $d=2,3$. \qed \\

\noindent \textbf{Proof of Proposition \ref{Prop1'}.} Let $\beta>0$ and $\nu>0$ be fixed. From the results of Proposition \ref{Prop1}, the l.h.s. of \eqref{gBECform2} is identically zero if $d=1$, and if $d=2,3$ when $\nu \leq \nu_{c}(\beta)$. When $\nu>\nu_{c}(\beta)$ if $d=2,3$, the method consists in decomposing the sum in the r.h.s. of \eqref{nudecomp} into two contributions:
\begin{equation}
\label{redecpm}
\forall 0< \varepsilon \leq 1,\quad \sum_{\bold{s} \in \mathbb{N}^{d}\,:\, 0<\sum_{j=1}^{d} \kappa s_{j} \leq \varepsilon} \nu_{\infty,\kappa}(\beta,\overline{\mu}_{\infty,\kappa};\bold{s}) + \sum_{\bold{s} \in \mathbb{N}^{d}\,:\, \sum_{j=1}^{d} \kappa s_{j} > \varepsilon} \nu_{\infty,\kappa}(\beta,\overline{\mu}_{\infty,\kappa};\bold{s}),
\end{equation}
and then in investigating successively the limits $\kappa \downarrow 0$ and $\varepsilon \downarrow 0$. By a direct calculation, one gets:
\begin{gather*}
\lim_{\varepsilon \downarrow 0} \lim_{\kappa \downarrow 0} \sum_{\bold{s} \in \mathbb{N}^{d}\,:\, \sum_{j=1}^{d} \kappa s_{j} > \varepsilon} \nu_{\infty,\kappa}(\beta,\overline{\mu}_{\infty,\kappa};\bold{s}) = \lim_{\varepsilon \downarrow 0} \frac{1}{\Gamma(d)} \int_{\varepsilon}^{\infty} \mathrm{d}\tau\,  \frac{\tau^{d-1}}{\mathrm{e}^{\beta \tau}-1} = \frac{\mathpzc{g}_{d}(1)}{\beta^{d}} =\nu_{c}(\beta).
\end{gather*}
Since the limit $\kappa \downarrow 0$ of the first sum in \eqref{redecpm} exists, then \eqref{gBECform2} follows by Proposition \ref{Prop1}. \qed

\section{Annexes.}

\subsection{Annex 1 - An additional result for the two-dimensional case.}
\label{AppendixC}

We stated in \eqref{Thm1Eq32} that the open-trap limit of the non-condensate part of the reduced density matrix diverges when $\nu\geq\nu_{c}(\beta)$ if $d=2$. We can prove that the divergence is logarithmic in $\kappa^{-1}$:
\begin{proposition}
\label{divlogprop}
Let $d=2$. For any $\beta>0$ and $\nu\geq\nu_{c}(\beta)$, one has uniformly in $(\bold{x},\bold{y}) \in \mathbb{R}^{4}$:
\begin{equation}
\label{divlog}
\sum_{\bold{s} \in (\mathbb{N}^{*})^{2}} \frac{\Psi_{\infty,\kappa}^{(\bold{s})}(\bold{x}) \Psi_{\infty,\kappa}^{(\bold{s})}(\bold{y})}{\mathrm{e}^{\beta(E_{\infty,\kappa}^{(\bold{s})} - \overline{\mu}_{\infty,\kappa})} - 1} \sim \frac{1}{\lambda_{\beta}^{2}} \ln\left(\frac{1}{\hbar \omega_{0} \kappa \beta}\right)\quad \textrm{when $\kappa \downarrow 0$}.
\end{equation}
\end{proposition}

\noindent \textbf{Proof.} For the sake of simplicity, we set $\hbar=m=\omega_{0}=1$ in the following. On the one hand, for $\kappa < 1$ small enough, one has the following lower bound:
\begin{multline}
\label{koff2'}
\forall(\bold{x},\bold{y}) \in \mathbb{R}^{4},\quad \sum_{l=1}^{\infty} \mathrm{e}^{l \beta \overline{\mu}_{\infty,\kappa}} G_{\infty,\kappa}^{(d=2)}(\bold{x},\bold{y};l\beta) - \frac{\Psi_{\infty,\kappa}^{(\bold{0})}(\bold{x}) \Psi_{\infty,\kappa}^{(\bold{0})}(\bold{y})}{\mathrm{e}^{\beta(E_{\infty,\kappa}^{(\bold{0})} - \overline{\mu}_{\infty,\kappa})} - 1} \\ \geq
\frac{\kappa}{\pi} \mathrm{e}^{-\frac{\kappa}{2}\left(\vert \bold{x}\vert^{2} + \vert \bold{y}\vert^{2}\right)} \sum_{l=1}^{\infty} \mathrm{e}^{-l \beta\left(E_{\infty,\kappa}^{(\bold{0})} - \overline{\mu}_{\infty,\kappa}\right)} \left\{-\left(\frac{1 - \mathrm{e}^{-\frac{\kappa}{4}\left[ \coth\left(\frac{\beta}{2} \kappa l\right)-1\right]\vert \bold{x}-\bold{y}\vert^{2}}}{1 - \mathrm{e}^{-2\kappa l \beta}}\right) + \frac{\mathrm{e}^{-2\kappa l \beta}}{1 - \mathrm{e}^{-2\kappa l\beta}}\right\},
\end{multline}
and from the arguments leading to \eqref{estLB}, the r.h.s. of \eqref{koff2'} is greater than:
\begin{multline*}
- \frac{\kappa}{\pi} \mathrm{e}^{-\frac{\kappa}{2}\left(\vert \bold{x}\vert^{2} + \vert \bold{y}\vert^{2}\right)} \sum_{l=1}^{\infty} \mathrm{e}^{-l \beta\left(E_{\infty,\kappa}^{(\bold{0})} - \overline{\mu}_{\infty,\kappa}\right)} \left(\frac{\kappa}{2} \mathrm{e}^{-\kappa \beta l}\vert \bold{x}-\bold{y}\vert^{2}\right)\left(1 + \frac{1}{2 \kappa l \beta}\right)\left(1+\frac{1}{\kappa l \beta}\right) + \\
+  \frac{1}{2\pi \beta} \mathrm{e}^{-\frac{\kappa}{2}\left(\vert \bold{x}\vert^{2} + \vert \bold{y}\vert^{2}\right)} \sum_{l=1}^{\infty} \mathrm{e}^{-l \beta\left(E_{\infty,\kappa}^{(\bold{0})} - \overline{\mu}_{\infty,\kappa}\right)} \frac{\mathrm{e}^{- 2 \kappa l \beta}}{l}.
\end{multline*}
Here, we used the upper and lower bounds in \eqref{Ink1} to majorize the first term inside the braces in the r.h.s. of \eqref{koff2'}. By using some integral comparison, the above quantity is bigger than:
\begin{multline}
\label{koff2}
- \frac{\nu_{\infty,\kappa}(\beta,\nu;\bold{0})}{2\pi} \vert \bold{x}-\bold{y}\vert^{2}  -\frac{3 \kappa}{4 \beta \pi} \vert \bold{x} - \bold{y}\vert^{2} \left(\mathrm{e}^{-\beta \kappa} + \int_{1}^{\infty} \mathrm{d}t\, \frac{\mathrm{e}^{-\kappa \beta t}}{t}\right) + \\
- \frac{1}{4 \beta^{2} \pi} \vert \bold{x} - \bold{y}\vert^{2} \left(\mathrm{e}^{-\beta \kappa} + \int_{1}^{\infty} \mathrm{d}t\, \frac{1}{t^{2}}\right)
+ \frac{1}{2\pi\beta} \mathrm{e}^{-\frac{\kappa}{2} (\vert \bold{x}\vert^{2} + \vert \bold{y}\vert^{2})} \int_{1}^{\infty} \mathrm{d}t\, \frac{\mathrm{e}^{-3\kappa \beta t}}{t}.
\end{multline}
\eqref{koff2} is made independent of $E_{\infty,\kappa}^{(\bold{0})} - \overline{\mu}_{\infty,\kappa}$ by using that $\mathrm{e}^{-l \beta(E_{\infty,\kappa}^{(\bold{0})} - \overline{\mu}_{\infty,\kappa})} \leq 1$ in the second/third term, and $\mathrm{e}^{l \beta \overline{\mu}_{\infty,\kappa}} \geq 1$ in the fourth term. On the other hand, for $\kappa<1$ one has the upper bound:
\begin{multline}
\label{luo}
\forall(\bold{x},\bold{y}) \in \mathbb{R}^{4},\quad \sum_{l=1}^{\infty} \mathrm{e}^{l \beta \overline{\mu}_{\infty,\kappa}} G_{\infty,\kappa}^{(d=2)}(\bold{x},\bold{y};l\beta) - \frac{\Psi_{\infty,\kappa}^{(\bold{0})}(\bold{x}) \Psi_{\infty,\kappa}^{(\bold{0})}(\bold{y})}{\mathrm{e}^{\beta(E_{\infty,\kappa}^{(\bold{0})} - \overline{\mu}_{\infty,\kappa})} - 1} \\ \leq
\frac{\kappa}{\pi} \mathrm{e}^{-\frac{\kappa}{2}\left(\vert \bold{x}\vert^{2} + \vert \bold{y}\vert^{2}\right)} \sum_{l=1}^{\infty} \mathrm{e}^{-l \beta\left(E_{\infty,\kappa}^{(\bold{0})} - \overline{\mu}_{\infty,\kappa}\right)} \left(\frac{\mathrm{e}^{\frac{\kappa}{4}\left[1 - \tanh\left(\frac{\beta}{2} \kappa l\right)\right]\vert \bold{x}+\bold{y}\vert^{2}} -1}{1 - \mathrm{e}^{-2\kappa l \beta}} + \frac{\mathrm{e}^{-2\kappa l \beta}}{1 - \mathrm{e}^{-2\kappa l\beta}}\right).
\end{multline}
From the arguments leading to \eqref{estUB} along with the above method, the r.h.s. of \eqref{luo} is less than:
\begin{multline}
\label{koff3}
\frac{\kappa}{\pi} \sum_{l=1}^{\infty} \mathrm{e}^{-l \beta\left(E_{\infty,\kappa}^{(\bold{0})} - \overline{\mu}_{\infty,\kappa}\right)} \left(\frac{\kappa}{2} \frac{\mathrm{e}^{-\kappa \beta l}}{1 + \mathrm{e}^{-\kappa \beta l}} \vert \bold{x}+\bold{y}\vert^{2} \mathrm{e}^{\frac{\kappa}{2} \frac{\mathrm{e}^{-\kappa \beta l}}{1 + \mathrm{e}^{-\kappa \beta l}} \vert \bold{x}+\bold{y}\vert^{2}} + \mathrm{e}^{-2 \kappa \beta l}\right)\left(1 + \frac{1}{2\kappa \beta l}\right) \\
\leq \frac{\nu_{\infty,\kappa}(\beta,\nu;\bold{0})}{2\pi} \vert \bold{x}+\bold{y}\vert^{2} \mathrm{e}^{\frac{\kappa}{2} \vert \bold{x} + \bold{y}\vert^{2}} + \frac{\kappa}{4\pi \beta} \vert \bold{x}+\bold{y}\vert^{2} \mathrm{e}^{\frac{\kappa}{2} \vert \bold{x} + \bold{y}\vert^{2}}  \int_{0}^{\infty} \mathrm{d}t\, \mathrm{e}^{- \kappa \beta t} + \\
+ \frac{\kappa}{\pi} \int_{0}^{\infty} \mathrm{d}t\, \mathrm{e}^{-2\kappa \beta t} +  \frac{1}{2\pi \beta} \mathrm{e}^{-2\kappa \beta} + \frac{1}{2\pi \beta}  \int_{1}^{\infty} \mathrm{d}t\, \frac{\mathrm{e}^{-2 \kappa \beta t}}{t}.
\end{multline}
The equivalent in \eqref{divlog} arises from the last term of the r.h.s. of \eqref{koff3} and \eqref{koff2} via \eqref{Euler}.\qed

\subsection{Annex 2 - Sketch of the proof of Corollaries \ref{Thm1Q1D} and \ref{Thm1Q2D}.}
\label{AppendixD0}

In this section, we set $\hbar=m=\omega_{0}=1$ for the sake of simplicity.\\

\noindent \textbf{Sketch of the proof of Corollary \ref{Thm1Q1D}}.\\
\indent \textbf{Part $\mathrm{(A)}$.} The proof of \eqref{rdmd21Q1D} is similar to the proof of \eqref{rdmd21}. We turn to \eqref{rdmd22Q1D}. In view of \eqref{rhoAnisa}, then from \eqref{Mehler}, one has $\forall l \in \mathbb{N}^{*}$, $\forall (\bold{x},\bold{y}) \in \mathbb{R}^{6}$ and for $\kappa>0$ sufficiently small:
\begin{multline}
\label{llowb}
\mathrm{e}^{l \beta \overline{\mu}_{\infty,\kappa}} G_{\infty,\kappa_{1}}^{(1)}(x_{1},y_{1};l\beta) G_{\infty,\kappa_{\perp}}^{(2)}(\bold{x}_{\perp},\bold{y}_{\perp};l\beta) \geq
\pi^{-\frac{3}{2}} \sqrt{\kappa_{1}} \kappa_{\perp}  \mathrm{e}^{-\frac{\kappa_{1}}{4}(x_{1}+y_{1})^{2}}  \mathrm{e}^{-\frac{1}{4}(x_{1}-y_{1})^{2}\left(\kappa_{1} + \frac{2}{\beta}\right)} \\ \times \mathrm{e}^{-\frac{\kappa_{\perp}}{4}\vert \bold{x}_{\perp}+ \bold{y}_{\perp}\vert^{2}} \mathrm{e}^{-\frac{1}{4}\vert \bold{x}_{\perp} - \bold{y}_{\perp}\vert^{2}\left(\kappa_{\perp} + \frac{2}{\beta}\right)} \times \left\{\begin{array}{ll}
\displaystyle{\frac{\mathrm{e}^{-l\beta \left(E_{\infty,\kappa}^{(\bold{0})} - \overline{\mu}_{\infty,\kappa}\right)}}{\sqrt{2 \kappa_{1} l \beta}}}, &\textrm{when $\nu_{c}(\beta) < \nu \leq \nu_{m}(\beta)$},\\
\mathrm{e}^{-l\beta \left(E_{\infty,\kappa}^{(\bold{0})} - \overline{\mu}_{\infty,\kappa}\right)}, &\textrm{when $\nu > \nu_{m}(\beta)$},
\end{array}\right.
\end{multline}
where we used that $(1-\mathrm{e}^{-2\kappa_{1} l\beta})^{\frac{1}{2}} \leq \sqrt{2 \kappa_{1} l\beta}$ if $\nu_{c}(\beta) < \nu \leq \nu_{m}(\beta)$ ($\leq 1$ otherwise). From \eqref{rhoAnisa}, it remains to use an integral comparison as in \eqref{lowbo1} and the asymptotics in Proposition \ref{Prop1Q1D} $\mathrm{(iii)}$.\\
\indent \textbf{Part $\mathrm{(B)}$.} When $\nu<\nu_{c}(\beta)$, the result is obvious. When $\nu > \nu_{c}(\beta)$,  the key-idea consists, as in the proof of \eqref{Thm1Eq2},  in decomposing $\forall 0<\kappa<1$ the sum involved in \eqref{rhoAnisa} into two contributions:
\begin{equation}
\label{sumdecopoopp}
r_{\infty,\kappa}(\bold{x},\bold{y};\beta,\nu) = \vert \underline{\kappa}\vert^{\frac{3}{2}} \left\{ \sum_{l=1}^{N_{\kappa,\sigma}} + \sum_{l= N_{\kappa,\sigma}+1}^{\infty} \right\} \mathrm{e}^{l\beta \overline{\mu}_{\infty,\kappa}} G_{\infty,\kappa_{1}}^{(1)}(x_{1},y_{1};l\beta) G_{\infty,\kappa_{\perp}}^{(2)}(\bold{x}_{\perp},\bold{y}_{\perp};l\beta),
\end{equation}
where $N_{\kappa,\sigma} := \lfloor \kappa^{-\sigma} \rfloor$ with $\sigma>0$. For the following, one has the upper bounds:
\begin{equation*}
\forall(\bold{x},\bold{y}) \in \mathbb{R}^{6},\quad \mathrm{e}^{l\beta \overline{\mu}_{\infty,\kappa}} G_{\infty,\kappa_{1}}^{(1)}(x_{1},y_{1};l\beta) G_{\infty,\kappa_{\perp}}^{(2)}(\bold{x}_{\perp},\bold{y}_{\perp};l\beta)
\end{equation*}
\begin{subnumcases}
{\label{uuuper}
 \leq
\frac{\vert \underline{\kappa}\vert^{\frac{3}{2}}}{\pi^{\frac{3}{2}}} \mathrm{e}^{-l\beta \left(E_{\infty,\kappa}^{(\bold{0})} - \overline{\mu}_{\infty,\kappa}\right)} \times} \left(1 + \frac{1}{\sqrt{2 \kappa_{1} l\beta}} + \frac{1}{2 \kappa_{\perp} l\beta} + \frac{1}{(2 \vert \underline{\kappa} \vert l \beta)^{\frac{3}{2}}}\right), \label{uuuper1} \\
\left(1 + \frac{1}{4 \kappa_{1} l\beta} + \frac{1}{2 \kappa_{\perp} l\beta} + \frac{1}{8 \kappa_{1} \kappa_{\perp} (l \beta)^{2}}\right) \label{uuuper2},
\end{subnumcases}
where we used that $\forall x>0$ $\frac{1}{\sqrt{1 - \mathrm{e}^{- x}}} \leq (1 + \frac{1}{\sqrt{x}})$ and $\frac{1}{\sqrt{1 - \mathrm{e}^{- x}}} \leq (1 + \frac{1}{2x})$ respectively. From \eqref{uuuper1}:
\begin{equation*}
\forall \sigma>0,\quad \lim_{\kappa \downarrow 0} \vert \underline{\kappa}\vert^{\frac{3}{2}} \sum_{l=1}^{N_{\kappa,\sigma}} \mathrm{e}^{l\beta \overline{\mu}_{\infty,\kappa}} G_{\infty,\kappa_{1}}^{(1)}(x_{1},y_{1};l\beta) G_{\infty,\kappa_{\perp}}^{(2)}(\bold{x}_{\perp},\bold{y}_{\perp};l\beta) = 0,
\end{equation*}
where we mimicked the arguments leading to \eqref{conclu01}. Let us turn to the second sum in the r.h.s. of \eqref{sumdecopoopp}. By mimicking the arguments leading to \eqref{fondlowb2}, one has on $\mathbb{R}^{6}$ for $\kappa<1$ sufficiently small:
\begin{multline}
\label{parie}
\vert \underline{\kappa} \vert^{\frac{3}{2}} \sum_{l=N_{\kappa,\sigma}+1}^{\infty} \mathrm{e}^{l\beta \overline{\mu}_{\infty,\kappa}} G_{\infty,\kappa_{1}}^{(1)}(x_{1},y_{1};l\beta) G_{\infty,\kappa_{\perp}}^{(2)}(\bold{x}_{\perp},\bold{y}_{\perp};l\beta) \geq   \frac{\nu_{\infty,\kappa}(\beta,\nu;\bold{0})}{\pi^{\frac{3}{2}}} \mathrm{e}^{-\frac{\kappa_{1}}{4} (x_{1}+y_{1})^{2}} \\ \times \mathrm{e}^{-\frac{\kappa_{\perp}}{4}\vert \bold{x}_{\perp}+ \bold{y}_{\perp}\vert^{2}}\mathrm{e}^{-\frac{\kappa_{1}}{4}(x_{1}-y_{1})^{2}\coth\left(\frac{\beta}{2} \kappa_{1} N_{\kappa,\sigma}\right)}\mathrm{e}^{-\frac{\kappa_{\perp}}{4}\vert \bold{x}_{\perp} - \bold{y}_{\perp}\vert^{2}\coth\left(\frac{\beta}{2} \kappa_{\perp} N_{\kappa,\sigma}\right)} \mathrm{e}^{-N_{\kappa,\sigma} \beta\left(E_{\infty,\kappa}^{(\bold{0})} - \overline{\mu}_{\infty,\kappa}\right)}.
\end{multline}
From  \eqref{uuuper} along with the arguments leading to \eqref{majd=3}, one has on $\mathbb{R}^{6}$ for $\kappa<1$ small enough:
\begin{multline*}
\vert \underline{\kappa} \vert^{\frac{3}{2}} \sum_{l=N_{\kappa,\sigma}+1}^{\infty} \mathrm{e}^{l\beta \overline{\mu}_{\infty,\kappa}} G_{\infty,\kappa_{1}}^{(1)}(x_{1},y_{1};l\beta) G_{\infty,\kappa_{\perp}}^{(2)}(\bold{x}_{\perp},\bold{y}_{\perp};l\beta)
\leq   \frac{\nu_{\infty,\kappa}(\beta,\nu;\bold{0})}{\pi^{\frac{3}{2}}} \\
 + \frac{\sqrt{\kappa_{1}} \kappa_{\perp}^{2}}{ \sqrt{2} \pi^{\frac{3}{2}} \sqrt{\beta}} \int_{N_{\kappa,\sigma}}^{\infty} \mathrm{d}t\, \frac{\mathrm{e}^{-\beta\left(E_{\infty,\kappa}^{(\bold{0})} - \overline{\mu}_{\infty,\kappa}\right)}}{\sqrt{t}} + \frac{\kappa_{1} \kappa_{\perp}}{2 \pi^{\frac{3}{2}} \beta} \int_{N_{\kappa,\sigma}}^{\infty} \mathrm{d}t\, \frac{\mathrm{e}^{-\beta\left(E_{\infty,\kappa}^{(\bold{0})} - \overline{\mu}_{\infty,\kappa}\right)}}{t} + \frac{\sqrt{\kappa_{1}} \kappa_{\perp}}{(2 \pi \beta)^{\frac{3}{2}}} \int_{N_{\kappa,\sigma}}^{\infty} \mathrm{d}t\, \frac{1}{t^{\frac{3}{2}}}.
\end{multline*}
Here, we used the upper bound in \eqref{uuuper1}. Under the same conditions, the above r.h.s. is less than:
\begin{equation}
\label{pari2}
\frac{\nu_{\infty,\kappa}(\beta,\nu;\bold{0})}{\pi^{\frac{3}{2}}} + \frac{\sqrt{\kappa_{1}} \kappa_{\perp}^{2}}{\sqrt{2} \pi \beta} \frac{1}{\sqrt{E_{\infty,\kappa}^{(\bold{0})} - \overline{\mu}_{\infty,\kappa}}} + \frac{\kappa_{1} \kappa_{\perp}}{2 \pi^{\frac{3}{2}} \beta} \Gamma_{0}\left( N_{\kappa,\sigma} \beta \left(E_{\infty,\kappa}^{(\bold{0})} - \overline{\mu}_{\infty,\kappa}\right)\right) + \frac{\sqrt{\kappa_{1}} \kappa_{\perp}}{\sqrt{2}( \pi \beta)^{\frac{3}{2}}} \frac{1}{\sqrt{N_{\kappa,\sigma}}}.
\end{equation}
In view of Proposition \ref{Prop1Q1D} $\mathrm{(iii)}$, the lower bound in \eqref{parie} and the upper bound in \eqref{pari2} converge $\forall \sigma>0$ and uniformly in $(\bold{x},\bold{y}) \in \mathbb{R}^{6}$ to the same value when $\kappa \downarrow 0$. The squeeze theorem leads to \eqref{Thm1Eq2Q1D} when $\nu > \nu_{c}(\beta)$. Turning to the case of $\nu=\nu_{c}(\beta)$, one has to use the decomposition:
\begin{equation}
\label{sumdecopoopp2}
r_{\infty,\kappa}(\bold{x},\bold{y};\beta,\nu) = \vert \underline{\kappa}\vert^{\frac{3}{2}} \left\{ \sum_{l=1}^{M_{\kappa,\sigma}} + \sum_{l= M_{\kappa,\sigma}+1}^{\infty} \right\} \mathrm{e}^{l\beta \overline{\mu}_{\infty,\kappa}} G_{\infty,\kappa_{1}}^{(1)}(x_{1},y_{1};l\beta) G_{\infty,\kappa_{\perp}}^{(2)}(\bold{x}_{\perp},\bold{y}_{\perp};l\beta),
\end{equation}
where $M_{\kappa,\sigma} = M_{\kappa,\sigma,\kappa_{c}} := \lfloor \kappa^{-\sigma} \mathrm{e}^{\frac{\kappa_{c}^{2}}{\kappa^{2}}}\rfloor$, with $\sigma>0$ for the moment. Firstly, the open-trap limit of the first sum in the r.h.s. of \eqref{sumdecopoopp2} vanishes $\forall 0<\sigma<3$. Subsequently, one has the inequality:
\begin{multline}
\label{kkkil}
\frac{\nu_{\infty,\kappa}(\beta,\nu;\bold{0})}{\pi^{\frac{3}{2}}} \mathrm{e}^{-\frac{\kappa_{1}}{4} (x_{1}+y_{1})^{2}} \mathrm{e}^{-\frac{\kappa_{\perp}}{4}\vert \bold{x}_{\perp}+ \bold{y}_{\perp}\vert^{2}}\mathrm{e}^{-\frac{\kappa_{1}}{4}(x_{1}-y_{1})^{2}\coth\left(\frac{\beta}{2} \kappa_{1} M_{\kappa,\sigma}\right)}\mathrm{e}^{-\frac{\kappa_{\perp}}{4}\vert \bold{x}_{\perp} - \bold{y}_{\perp}\vert^{2}\coth\left(\frac{\beta}{2} \kappa_{\perp} M_{\kappa,\sigma}\right)} \\
\times \mathrm{e}^{-\beta M_{\kappa,\sigma}\left[\kappa_{1} + 2 \kappa_{\perp}\right]}\leq \vert \underline{\kappa} \vert^{\frac{3}{2}} \sum_{l=M_{\kappa,\sigma}+1}^{\infty} \mathrm{e}^{l\beta \overline{\mu}_{\infty,\kappa}} G_{\infty,\kappa_{1}}^{(1)}(x_{1},y_{1};l\beta) G_{\infty,\kappa_{\perp}}^{(2)}(\bold{x}_{\perp},\bold{y}_{\perp};l\beta) \\ \leq \frac{\nu_{\infty,\kappa}(\beta,\nu;\bold{0})}{\pi^{\frac{3}{2}}} \left( 1
 + \frac{1}{\sqrt{2 \beta} \pi^{\frac{3}{2}}} \frac{1}{\sqrt{\kappa_{1} M_{\kappa,\sigma}}} + \frac{1}{2 \pi^{\frac{3}{2}} \beta} \frac{1}{\kappa_{\perp} M_{\kappa,\sigma}} \right) + \frac{\sqrt{\kappa_{1}} \kappa_{\perp}}{(2 \pi \beta)^{\frac{3}{2}}} \int_{M_{\kappa,\sigma}}^{\infty} \mathrm{d}t\, \frac{1}{t^{\frac{3}{2}}}.
\end{multline}
The r.h.s. of \eqref{kkkil} converges to $\pi^{-\frac{3}{2}}(\nu_{c}(\beta)-\nu)=0$ $\forall \sigma>1$. Therefore, \eqref{Thm1Eq2Q1D} when $\nu=\nu_{c}(\beta)$ follows from the foregoing provided that $1<\sigma<3$ in \eqref{sumdecopoopp2}.\\
\indent \textbf{Part $\mathrm{(C)}$.} The proof of \eqref{Thm1Eq31Q1D} is similar to the proof of \eqref{Thm1Eq31}. We turn to the proof of \eqref{Thm1Eq32Q1D}. The strategy consists in decomposing $\forall 0 < \kappa < 1$ the non condensate part of the reduced density matrix into 3 contributions:
\begin{multline*}
\sum_{\bold{s} \in (\mathbb{N}^{*})^{3}} \frac{\Psi_{\infty,\kappa}^{(\bold{s})}(\bold{x}) \Psi_{\infty,\kappa}^{(\bold{s})}(\bold{y})}{\mathrm{e}^{\beta \left(E_{\infty,\kappa}^{(\bold{s})} - \overline{\mu}_{\infty,\kappa}\right)} - 1} = \left\{\sum_{l= 1}^{N_{\kappa,\sigma}} + \sum_{l= 1+N_{\kappa,\sigma}}^{M_{\kappa}} + \sum_{l= 1+M_{\kappa}}^{\infty}\right\} \\
\times \left( \mathrm{e}^{l \beta\overline{\mu}_{\infty,\kappa}} G_{\infty,\kappa_{1}}^{(1)}\left(x_{1},y_{1};l\beta\right) G_{\infty,\kappa_{\perp}}^{(2)}\left(\bold{x}_{\perp},\bold{y}_{\perp};l\beta\right)- \mathrm{e}^{l \beta \left(\overline{\mu}_{\infty,\kappa} - E_{\infty,\kappa}^{(\bold{0})}\right)} \Psi_{\infty,\kappa}^{(\bold{0})}(\bold{x})\Psi_{\infty,\kappa}^{(\bold{0})}(\bold{y})\right),
\end{multline*}
where $N_{\kappa,\sigma}:=\lfloor \kappa^{-\sigma} \rfloor$ with $\sigma>0$ and $M_{\kappa} := \lfloor \mathrm{e}^{\frac{\kappa_{c}^{2}}{\kappa^{2}}}\rfloor$. From Proposition \ref{Prop1Q1D} and by using accurate estimates as we did in the proof of Part $\mathrm{(B)}$, one can prove that for any $\nu>\nu_{c}(\beta)$:
\begin{multline*}
\lim_{\kappa \downarrow 0}
\sum_{l= a_{\kappa}}^{b_{\kappa}} \left( \mathrm{e}^{l \beta\overline{\mu}_{\infty,\kappa}} G_{\infty,\kappa_{1}}^{(1)}\left(x_{1},y_{1};l\beta\right) G_{\infty,\kappa_{\perp}}^{(2)}\left(\bold{x}_{\perp},\bold{y}_{\perp};l\beta\right) - \mathrm{e}^{l \beta \left(\overline{\mu}_{\infty,\kappa} - E_{\infty,\kappa}^{(\bold{0})}\right)} \Psi_{\infty,\kappa}^{(\bold{0})}(\bold{x})\Psi_{\infty,\kappa}^{(\bold{0})}(\bold{y})\right) \\
= \left\{\begin{array}{ll}
0, &\textrm{when $a_{\kappa} = 1+M_{\kappa}$, $b_{\kappa}=\infty$},\\
\displaystyle{\sum_{l=1}^{\infty} \frac{1}{\left(2\pi l \beta\right)^{\frac{3}{2}}} \mathrm{e}^{-\frac{\vert \bold{x}-\bold{y}\vert^{2}}{2l \beta}}}, &\textrm{when $a_{\kappa} = 1$, $b_{\kappa} = N_{\kappa,\sigma}$},
\end{array}\right.,\\
\sum_{l= N_{\kappa,\sigma}+1}^{M_{\kappa}} \left( \mathrm{e}^{l \beta\overline{\mu}_{\infty,\kappa}} G_{\infty,\kappa_{1}}^{(1)}\left(x_{1},y_{1};l\beta\right) G_{\infty,\kappa_{\perp}}^{(2)}\left(\bold{x}_{\perp},\bold{y}_{\perp};l\beta\right) - \mathrm{e}^{l \beta \left(\overline{\mu}_{\infty,\kappa} - E_{\infty,\kappa}^{(\bold{0})}\right)} \Psi_{\infty,\kappa}^{(\bold{0})}(\bold{x})\Psi_{\infty,\kappa}^{(\bold{0})}(\bold{y})\right)\\
\sim \frac{\kappa_\perp}{\sqrt{2\pi^2 \beta}}  \times \left\{\begin{array}{ll}
\mathrm{e}^{\beta \frac{\nu - \nu_{c}(\beta)}{2 \kappa^2}},\ &\textrm{when $\nu_{c}(\beta) < \nu \leq \nu_{m}(\beta)$},\\
\mathrm{e}^{\frac{\kappa_c^2}{2 \kappa^2}},\ &\textrm{when $\nu>\nu_{m}(\beta)$}, \end{array}\right.\quad \textrm{when $\kappa \downarrow 0$}.
\end{multline*}

\noindent \textbf{Sketch of the proof of Corollary \ref{Thm1Q2D}.} \\
\indent \textbf{Part $\mathrm{(A)}$.} The proof of \eqref{rdmd21Q2D} is similar to the proof of \eqref{rdmd21}. The proof of \eqref{rdmd22Q2D} follows from the lower bound in \eqref{llowb} in the case of $\nu> \nu_{m}(\beta)$ along with the asymptotic in \eqref{musolGG2a2D}.\\
\indent \textbf{Part $\mathrm{(B)}$.} When $\nu < \nu_{c}(\beta)$ the proof is obvious. When $\nu>\nu_{c}(\beta)$, it is enough to use the decomposition in \eqref{sumdecopoopp}. The lower bound in \eqref{parie} and the upper bound in \eqref{pari2} still hold true, and from the asymptotic in \eqref{musolGG2a2D}, then \eqref{Thm1Eq2Q2D} follows by the squeeze theorem. When $\nu = \nu_{c}(\beta)$, one has to use the decomposition as in \eqref{sumdecopoopp2} but with $\tilde{M}_{\kappa,\sigma} = \tilde{M}_{\kappa,\sigma,\kappa_{c}} := \lfloor \kappa^{-\sigma} \mathrm{e}^{\sqrt{\frac{\kappa_{c}}{\kappa}}} \rfloor$, $\sigma>1$.\\
\indent \textbf{Part $\mathrm{(C)}$.} The proof of \eqref{Thm1Eq31Q2D} is similar to the proof of \eqref{Thm1Eq31}. We turn to the proof of \eqref{Thm1Eq32Q2D}. The strategy consists in decomposing $\forall 0<\kappa<1$ the reduced density matrix into 3 contributions:
\begin{equation}
\label{filp0}
\rho_{\infty,\kappa}(\bold{x},\bold{y};\beta,\nu) = \left\{\sum_{l=1}^{N_{\kappa,\sigma_{1}}} + \sum_{l=1 + N_{\kappa,\sigma_{1}}}^{\tilde{M}_{\kappa,\sigma_{2},\chi}} + \sum_{l= 1+ \tilde{M}_{\kappa,\sigma_{2},\chi}}^{\infty} \right\} \mathrm{e}^{l\beta \overline{\mu}_{\infty,\kappa}} G_{\infty,\kappa_{1}}^{(1)}(x_{1},y_{1};l\beta) G_{\infty,\kappa_{\perp}}^{(2)}(\bold{x}_{\perp},\bold{y}_{\perp};l\beta),
\end{equation}
where $N_{\kappa,\sigma_{1}} = \lfloor \kappa^{-\sigma_{1}} \rfloor$ with $\sigma_{1}>0$ and $\tilde{M}_{\kappa,\sigma_{2},\chi} = \tilde{M}_{\kappa,\sigma_{2},\chi,\kappa_{c}} := \lfloor \kappa^{-\sigma_{2}} \mathrm{e}^{\chi \sqrt{\frac{\kappa_{c}}{\kappa}}} \rfloor$ with $\sigma_{2}\geq0$, $\chi>0$ for the moment. By mimicking the arguments leading to \eqref{efedesez}, $\forall \sigma_{2}\geq 0$ and for $\chi=2$ on $\mathbb{R}^{6}$:
\begin{equation}
\label{filp1}
\lim_{\kappa \downarrow 0} \left(\sum_{l= 1+ \tilde{M}_{\kappa,\sigma_{2},2}}^{\infty} \mathrm{e}^{l\beta \overline{\mu}_{\infty,\kappa}} G_{\infty,\kappa_{1}}^{(1)}(x_{1},y_{1};l\beta) G_{\infty,\kappa_{\perp}}^{(2)}(\bold{x}_{\perp},\bold{y}_{\perp};l\beta) - \frac{\Psi_{\infty,\kappa}^{(\bold{0})}(\bold{x}) \Psi_{\infty,\kappa}^{(\bold{0})}(\bold{y})}{\mathrm{e}^{\beta(E_{\infty,\kappa}^{(\bold{0})} - \overline{\mu}_{\infty,\kappa})} - 1} \right)=0.
\end{equation}
To derive \eqref{filp1}, the upper bound in \eqref{uuuper2} turns out to be more convenient. Next, by a similar method than the one leading to \eqref{corr4}, one has  $\forall \sigma_{1}$, $\forall \sigma_{2}\geq 0$ and $\forall 0<\chi < 2$ on $\mathbb{R}^{6}$:
\begin{gather}
\label{filp2}
\lim_{\kappa \downarrow 0} \sum_{l=1}^{N_{\kappa,\sigma_{1}}} \mathrm{e}^{l\beta \overline{\mu}_{\infty,\kappa}} G_{\infty,\kappa_{1}}^{(1)}(x_{1},y_{1};l\beta) G_{\infty,\kappa_{\perp}}^{(2)}(\bold{x}_{\perp},\bold{y}_{\perp};l\beta) = \sum_{l=1}^{\infty} \frac{1}{(2\pi l \beta)^{\frac{3}{2}}} \mathrm{e}^{- \frac{\vert \bold{x} - \bold{y}\vert^{2}}{2\beta l}},\\
\label{filp3}
\lim_{\kappa \downarrow 0} \left\{\sum_{l=1+ N_{\kappa,\sigma_{1}}}^{\tilde{M}_{\kappa,\sigma_{2},\chi}} + \sum_{1+ \tilde{M}_{\kappa,\sigma_{2},\chi}}^{\tilde{M}_{\kappa,\sigma_{2},2}} \right\} \mathrm{e}^{l\beta \overline{\mu}_{\infty,\kappa}} G_{\infty,\kappa_{1}}^{(1)}(x_{1},y_{1};l\beta) G_{\infty,\kappa_{\perp}}^{(2)}(\bold{x}_{\perp},\bold{y}_{\perp};l\beta) = \left\{\chi + \left(2-\chi\right)\right\} \frac{\sqrt{\kappa_{c}}}{2 \pi^{\frac{3}{2}} \beta}.
\end{gather}

\section{Acknowledgments.}

B.S. was partially supported by the Lundbeck Foundation, and the European Research Council under the European Community's Seventh Framework Program (FP7/2007--2013)/ERC grant agreement 202859. A part of this work was done while the second author was visiting DIAS-STP, B.S. is grateful for invitation and financial support. Both authors thank Tony C. Dorlas for helpful and stimulating discussions.

\appendix

\renewcommand{\thedefinition}{\Alph{section}.\arabic{definition}}

\section{Appendices.}

\subsection{Appendix 1 - The large volume behavior.}
\label{AppendixA}

In this section, we prove the thermodynamic limit of the grand-canonical potential and average number of particles associated to the harmonically trapped Bose gas in the G-C situation.\\

\indent We start by introducing the one-parameter semigroup generated by $H_{L,\kappa}$ in \eqref{HL}. It is defined $\forall L>0$ and $\forall \kappa>0$ by $\{G_{L,\kappa}(t) := \mathrm{e}^{- t H_{L,\kappa}} : L^{2}(\Lambda_{L}^{d}) \rightarrow L^{2}(\Lambda_{L}^{d})\}_{t \geq 0}$. It is strongly continuous, and it is a self-adjoint and positive operator by the spectral theorem and the functional calculus. By standard arguments, $\{G_{L,\kappa}(t)\}_{t>0}$ is an integral operator whose the integral kernel, denoted by $G_{L,\kappa}^{(d)}(\cdot\,,\cdot\,;t)$, is jointly continuous in $(\bold{x},\bold{y},t) \in \overline{\Lambda_{L}^{d}}\times\overline{\Lambda_{L}^{d}} \times (0,\infty)$ and vanishes if $\bold{x} \in \partial \Lambda_{L}^{d}$ or $\bold{y} \in \partial \Lambda_{L}^{d}$. Moreover, the mapping $L \mapsto G_{L,\kappa}^{(d)}(\bold{x},\bold{y};t)$ is positive and monotone increasing, see \cite[Coro. 6.3.13]{BR2}. This leads to the following pointwise inequality which holds $\forall\kappa>0$
and $\forall L>0$:
\begin{equation}
\label{fondineq}
\forall (\bold{x},\bold{y},t) \in \overline{\Lambda_{L}^{d}}\times \overline{\Lambda_{L}^{d}}\times(0,\infty),\quad G_{L,\kappa}^{(d)}(\bold{x},\bold{y};t) \leq \sup_{L>0} G_{L,\kappa}^{(d)}(\bold{x},\bold{y};t) = G_{\infty,\kappa}^{(d)}(\bold{x},\bold{y};t).
\end{equation}
Here, $G_{\infty,\kappa}^{(d)}(\cdot\,,\cdot\,;t)$ is defined by \eqref{Mehler}-\eqref{multd}. It results from \eqref{fondineq} that $\forall L>0$ and $\forall \kappa>0$, the semigroup $\{G_{L,\kappa}(t)\}_{t>0}$ is a trace-class operator on $L^{2}(\Lambda_{L}^{d})$, and moreover, see e.g. \cite[Lem. A.4]{BSa}:
\begin{equation}
\label{trace0}
\mathrm{Tr}_{L^{2}(\Lambda_{L}^{d})} \left\{ G_{L,\kappa}(t)\right\} \leq \mathrm{Tr}_{L^{2}(\mathbb{R}^{d})}\left\{G_{\infty,\kappa}(t)\right\} = \mathrm{e}^{- E_{\infty,\kappa}^{(\bold{0})} t}\left(1 - \mathrm{e}^{- \kappa t}\right)^{-d}.
\end{equation}

From the foregoing and under the conditions of \eqref{grandpotential}, the G-C potential can be rewritten as:
\begin{equation*}
\Omega_{L,\kappa}(\beta,z) = \frac{1}{\beta} \mathrm{Tr}_{L^{2}(\Lambda_{L}^{d})}\left\{\ln\left(\mathbbm{1} - z G_{L,\kappa}(\beta)\right)\right\},
\end{equation*}
and the operator inside the trace is defined via the Dunford functional calculus. Clearly, $\Omega_{L,\kappa}(\beta,\cdot\,)$ is a $\mathcal{C}^{\infty}$-function on $(0,\mathrm{e}^{\beta E_{L,\kappa}^{(\bold{0})}})$ (remind that $E_{L,\kappa}^{(\bold{0})} := \inf \sigma(H_{L,\kappa})$). In fact, we can prove more:

\begin{lema}
\label{analy}
$\forall d \in \{1,2,3\}$, $\forall L \in (0,\infty)$, $\forall \kappa>0$ and $\forall \beta>0$, $\Omega_{L,\kappa}(\beta,\cdot\,)$ has an analytic continuation to the domain $\mathcal{D}:= \mathbb{C}\setminus [\mathrm{e}^{\beta E_{L,\kappa}^{(\bold{0})}},\infty)$. In the following, we denote it by $\hat{\Omega}_{L,\kappa}(\beta,\cdot\,)$.
\end{lema}

The proof of Lemma \ref{analy} is standard, the main arguments can be found in \cite{ABN}. See also \cite{C1}.\\

Denote by $\mathcal{B}(r)$ an open ball in $\mathbb{C}$ centered at the origin and having the radius $r>0$. When restricting to the domain $\mathcal{B}(\mathrm{e}^{\beta E_{\infty,\kappa}^{(\bold{0})}}) \subset \mathcal{D}$, one gets a very convenient representation of the analytic continuation of $\Omega_{L,\kappa}(\beta,\cdot\,)$ involving the semigroup $\{G_{L,\kappa}(\beta)\}_{\beta >0}$. In particular:

\begin{lema}
\label{conv}
$\forall d \in \{1,2,3\}$, $\forall L \in (0,\infty)$, $\forall \kappa>0$, $\forall \beta>0$ and $\forall z \in \mathcal{B}(\mathrm{e}^{\beta E_{\infty,\kappa}^{(\bold{0})}})$:
\begin{equation}
\label{paPhi-L}
\hat{\Omega}_{L,\kappa}(\beta,z) = - \frac{1}{\beta} \sum_{l=1}^{\infty} \frac{z^{l}}{l} \mathrm{Tr}_{L^{2}(\Lambda_{L}^{d})}\left\{G_{L,\kappa}(l \beta)\right\}.
\end{equation}
\end{lema}

The proof of Lemma \ref{conv} follows the strategy used to prove \cite[Prop. 2 (i)]{ABN}, see also
\cite[pp. 4]{C1} for further details.
In view of \eqref{paPhi-L}, introduce $\forall d \in \{1,2,3\}$, $\forall \kappa>0$, $\forall \beta>0$ and $\forall z \in \mathcal{B}(\mathrm{e}^{\beta E_{\infty,\kappa}^{(\bold{0})}})$:
\begin{equation*}
\hat{\Omega}_{\infty,\kappa}(\beta,z) := - \frac{1}{\beta} \sum_{l=1}^{\infty} \frac{z^{l}}{l} \mathrm{Tr}_{L^{2}(\mathbb{R}^{d})}\left\{G_{\infty,\kappa}(l \beta)\right\}.
\end{equation*}

Next, let us turn to the thermodynamic limit of the G-C potential. Here is the main result:
\begin{proposition}
\label{limith}
$\forall d \in \{1,2,3\}$, $\forall 0<\kappa_{1} < \kappa_{2} < \infty$, $\forall 0< \beta_{1} < \beta_{2} < \infty$ and for any compact subset $K \subset \mathcal{B}(\mathrm{e}^{\beta_{1} E_{\infty,\kappa_{1}}^{(\bold{0})}})$:
\begin{equation*}
\lim_{L \uparrow \infty} \hat{\Omega}_{L,\kappa}(\beta,z) = \hat{\Omega}_{\infty,\kappa}(\beta,z),
\end{equation*}
uniformly in $(\kappa,\beta,z) \in [\kappa_{1},\kappa_{2}]\times[\beta_{1},\beta_{2}] \times K$.
\end{proposition}

Because of the Weierstrass theorem, one has as a corollary of Proposition \ref{limith}:

\begin{corollary}
\label{coro}
$\forall d \in \{1,2,3\}$, $\forall \kappa>0$ and $\forall \beta>0$, $z \mapsto \hat{\Omega}_{\infty,\kappa}(\beta,z)$ is analytic on $\mathcal{B}(\mathrm{e}^{\beta E_{\infty,\kappa}^{(\bold{0})}})$.
Moreover $\forall 0<\kappa_{1} < \kappa_{2} < \infty$, $\forall 0<\beta_{1}<\beta_{2}<\infty$ and for any compact subset $K \subset \mathcal{B}(\mathrm{e}^{\beta_{1} E_{\infty,\kappa_{1}}^{(\bold{0})}})$:
\begin{equation*}
\forall m \in \mathbb{N}^{*},\quad \lim_{L \uparrow \infty} \frac{\partial^{m} \hat{\Omega}_{L,\kappa}}{\partial z^{m}} (\beta,z) = \frac{\partial^{m} \hat{\Omega}_{\infty,\kappa}}{\partial z^{m}} (\beta,z),
\end{equation*}
uniformly in $(\kappa,\beta,z) \in [\kappa_{1},\kappa_{2}]\times[\beta_{1},\beta_{2}]\times K$.
\end{corollary}

\begin{remark}
From Proposition \ref{limith} along with Corollary \ref{coro}, one has in particular $\forall \kappa>0$, $\forall \beta>0$ and $\forall z \in (0,\mathrm{e}^{\beta E_{\infty,\kappa}^{(\bold{0})}})$ the following pointwise convergences:
\begin{gather*}
\Omega_{\infty,\kappa}(\beta,z) := \lim_{L \uparrow \infty} \Omega_{L,\kappa}(\beta,z) = - \frac{1}{\beta} \sum_{l=1}^{\infty} \frac{z^{l}}{l} \mathrm{Tr}_{L^{2}(\mathbb{R}^{d})}\left\{G_{\infty,\kappa}(l \beta)\right\},\\
\overline{N}_{\infty,\kappa}(\beta,z) := -\beta z \frac{\partial \Omega_{\infty,\kappa}}{\partial z}(\beta,z) = \lim_{L \uparrow \infty} -\beta z \frac{\partial \Omega_{L,\kappa}}{\partial z}(\beta,z).
\end{gather*}
\end{remark}

The proof of Proposition \ref{limith} leans on the below estimate which is the main subject of \cite{BSa}:

\begin{lema}
\label{Mimp}
$\forall d \in \{1,2,3\}$ there exists a constant $C_{d}>0$ and $\forall 0<\kappa_{0}<1$ there exists a $\mathpzc{L}_{\kappa_{0}}>0$ s.t. $\forall L \in [\mathpzc{L}_{\kappa_{0}},\infty)$, $\forall \kappa \in [\kappa_{0},\infty)$ and $\forall t>0$:
\begin{multline}
\label{cherc}
\left\vert \mathrm{Tr}_{L^{2}(\Lambda_{L}^{d})}\left\{G_{L,\kappa}(t)\right\} - \mathrm{Tr}_{L^{2}(\mathbb{R}^{d})}\left\{G_{\infty,\kappa}(t)\right\}\right\vert \\ \leq C_{d} \left(1+\sqrt{\kappa}\right)(1+\kappa)^{d} (1+t)^{3(d+ \frac{1}{2})} \left(2\sinh\left(\frac{\kappa}{2} t\right)\right)^{-d} \mathrm{e}^{-\frac{\kappa}{32} \frac{L^{2}}{4} \tanh\left(\frac{\kappa}{2}t\right)}.
\end{multline}
\end{lema}

\noindent \textbf{Proof of Proposition \ref{limith}}. $\forall L \in (0,\infty)$ and $\forall (\kappa,\beta,z) \in [\kappa_{1},\kappa_{2}]\times[\beta_{1},\beta_{2}]\times K$, introduce:
\begin{equation*}
\forall M \in \mathbb{N}^{*},\quad \mathscr{Q}_{L,\kappa,M}(\beta,z) := \frac{1}{\beta} \sum_{l=1}^{M} \frac{z^{l}}{l} \left\vert \mathrm{Tr}_{L^{2}(\mathbb{R}^{d})}\left\{G_{\infty,\kappa}(l\beta)\right\} - \mathrm{Tr}_{L^{2}(\Lambda_{L}^{d})}\left\{G_{L,\kappa}(l\beta)\right\}\right\vert.
\end{equation*}
Let $\mathpzc{L}=\mathpzc{L}_{\kappa_{1}}$ s.t. $\forall L \geq \mathpzc{L}_{\kappa_{1}}$ the estimate in \eqref{cherc} holds. Then $\forall L \in [\mathpzc{L},\infty)$ and $\forall z \in K$, one has:
\begin{equation*}
\mathscr{Q}_{L,\kappa,M}(\beta,z) \leq  C_{d} \left(1+\sqrt{\kappa}\right) (1+\kappa)^{d} \frac{(1 + \beta)^{3(d+\frac{1}{2})}}{\beta \left(1-\mathrm{e}^{-\kappa \beta}\right)^{d}}   \mathrm{e}^{- \frac{\kappa}{32} \frac{L^{2}}{4} \tanh\left(\frac{\kappa}{2} \beta\right)} \left(\sum_{l=1}^{M} \left(\vert z\vert \mathrm{e}^{-\beta E_{\infty,\kappa}^{(\bold{0})}}\right)^{l} l^{3d+\frac{1}{2}}\right),
\end{equation*}
for another constant $C_{d}>0$. Since $\forall (\kappa,\beta) \in[\kappa_{1},\kappa_{2}]\times[\beta_{1},\beta_{2}]$ one has $\sup_{z \in K} \vert z\vert \mathrm{e}^{-\beta E_{\infty,\kappa}^{(\bold{0})}} < 1$, then from the above estimate there exists another constant $C_{d}= C_{d}(\kappa_{1},\kappa_{2},\beta_{1},\beta_{2},K)>0$ s.t.
\begin{equation*}
\lim_{L \uparrow \infty}  \sup_{\kappa \in [\kappa_{1},\kappa_{2}]} \sup_{\beta \in [\beta_{1},\beta_{2}]} \sup_{z \in K} \lim_{M \uparrow \infty} \mathscr{Q}_{L,\kappa,M}(\beta,z) \leq C_{d} \lim_{L \uparrow \infty} \mathrm{e}^{-\frac{\kappa_{1}}{32} \frac{L^{2}}{4} \tanh\left(\frac{\kappa_{1}}{2}\beta\right)} =0. \tag*{\qed}
\end{equation*}


We end this section by proving:\\

\noindent \textbf{Proof of \eqref{locmu}.}
Let us show that:
\begin{equation}
\label{liminfsup}
\overline{\mu}_{\infty,\kappa} \leq \overline{\mu}_{\infty,\kappa}^{\mathrm{inf}} \leq   \overline{\mu}_{\infty,\kappa}^{\mathrm{sup}} \leq \overline{\mu}_{\infty,\kappa},\quad \textrm{with\,\,\,\, $\overline{\mu}_{\infty,\kappa}^{\mathrm{inf}} := \liminf_{L \uparrow \infty} \overline{\mu}_{L,\kappa}$\,\, and\,\,  $\overline{\mu}_{L,\kappa}^{\mathrm{sup}}:=\limsup_{L \uparrow \infty} \overline{\mu}_{L,\kappa}$}.
\end{equation}
We prove the first inequality in \eqref{liminfsup}. Suppose the contrary, i.e., $\overline{\mu}_{\infty,\kappa}^{\mathrm{inf}} < \overline{\mu}_{\infty,\kappa}$. Then there exists $\eta>0$ and a divergent sequence $\{L_{n}\}_{n\geq 1}$ s.t. $\lim_{n \uparrow \infty} \overline{\mu}_{L_{n},\kappa} =  \overline{\mu}_{\infty,\kappa}^{\mathrm{inf}}$ and $\overline{\mu}_{L_{n},\kappa} \leq \overline{\mu}_{\infty,\kappa} - \eta$ $\forall n\geq 1$. Now by using that the map $\mu \mapsto \overline{N}_{L_{n},\kappa}(\beta,\mu)$ is increasing on $(-\infty,E_{\infty,\kappa}^{(\bold{0})})$, then:
\begin{equation*}
\nu = \nu_{L_{n},\kappa}(\beta, \overline{\mu}_{L_{n},\kappa}) \leq \nu_{L_{n},\kappa}(\beta, \overline{\mu}_{\infty,\kappa} - \eta) \quad \forall n\geq 1.
\end{equation*}
Afterwards, since $\{\nu_{L_{n},\kappa}(\beta,\cdot\,)\}_{n\geq 1}$ converges uniformly on compacts w.r.t. $\mu$ to $\nu_{\infty,\kappa}(\beta,\cdot\,)$ as a result of Corollary \ref{coro}, then by using that $\mu \mapsto \nu_{\infty,\kappa}(\beta,\mu)$ is strictly increasing on $(-\infty,E_{\infty,\kappa}^{(\bold{0})})$:
\begin{equation*}
\nu =  \nu_{\infty,\kappa}(\beta, \overline{\mu}_{\infty,\kappa}^{\mathrm{inf}}) \leq \nu_{\infty,\kappa}(\beta, \overline{\mu}_{\infty,\kappa} - \eta) < \nu_{\infty,\kappa}(\beta, \overline{\mu}_{\infty,\kappa}) = \nu.
\end{equation*}
This contradiction yields $\overline{\mu}_{\infty,\kappa} \leq \overline{\mu}_{\infty,\kappa}^{\mathrm{inf}}$. The last inequality in \eqref{liminfsup} can be proved similarly. \qed

\subsection{Appendix 2 - Some useful identities/inequalities.}
\label{AppendixB}

Here, we collect some miscellaneous inequalities/identities involving the hyperbolic functions we use in this paper. Most of them can be found in \cite[Sec. 4.5]{AS}. For any real $x \geq 0$:
\begin{gather}
\label{Ek1}
1 \leq \cosh(x) \leq \mathrm{e}^{x}, \\
\label{Ek2}
x \leq \sinh(x) \leq \frac{1}{2} \mathrm{e}^{x}, \\
\label{Ek3}
0 \leq \tanh(x) \leq 1,\\
\label{Ek4}
\frac{1}{x} \leq \coth(x) := \frac{1}{\tanh(x)} \leq \frac{1+x}{x},\quad x>0.
\end{gather}

\indent For any reals $x \geq 0$ and $p,q>0$:
\begin{gather}
\label{Ink1}
\frac{x}{1+x} < 1- \mathrm{e}^{-x} < x,\\
\label{Ink2}
x \leq \mathrm{e}^{x} - 1 \leq x \mathrm{e}^{x},\\
\label{redexp}
x^{p} \mathrm{e}^{-q x} \leq \left(\frac{2 p}{\mathrm{e} q}\right)^{p} \mathrm{e}^{-\frac{q}{2} x}.
\end{gather}

\end{document}